\documentclass[iop]{emulateapj}

\usepackage{graphicx}
\usepackage{amsmath,booktabs,threeparttable,url}
\usepackage[colorlinks,citecolor=blue]{hyperref}
\usepackage{CJKutf8}
\usepackage{etex}
\newcommand{\cntext}[1]{\begin{CJK*}{UTF8}{bsmi}#1\end{CJK*}}

\slugcomment{Accepted for publication in ApJ}

\usepackage{natbib}
\begin{document}

\title{TWO-DIMENSIONAL CORE-COLLAPSE SUPERNOVA SIMULATIONS WITH THE ISOTROPIC DIFFUSION SOURCE APPROXIMATION FOR NEUTRINO TRANSPORT}

\author{Kuo-Chuan Pan (\cntext{潘國全})$^{1}$, Matthias Liebend\"{o}rfer$^{1}$, 
Matthias Hempel$^{1}$ and Friedrich-Karl Thielemann$^{1}$}
\affil{$^1$Physik Department, Universit\"{a}t Basel, Klingelbergstrasse 82, CH-4056 Basel, Switzerland; kuo-chuan.pan@unibas.ch}


\begin{abstract}
 
The neutrino mechanism of core-collapse supernova is investigated via non-relativistic, two-dimensional (2D), 
neutrino radiation-hydrodynamic simulations. 
For the transport of electron flavor neutrinos,
we use the interaction rates defined by Bruenn (1985) and
the isotropic diffusion source approximation (IDSA) scheme, 
which decomposes the transported particles into trapped particle and streaming particle components. 
Heavy neutrinos are described by a leakage scheme. 
Unlike the ``ray-by-ray'' approach in some other multi-dimensional supernova models, 
we use cylindrical coordinates and solve the trapped particle component in multiple dimensions, 
improving the proto-neutron star resolution and the neutrino transport in angular and temporal directions. 
We provide an IDSA verification by performing 1D and 2D simulations with 15 and 20 $M_\odot$ progenitors 
from Woosley et al.~(2007) and discuss the difference of our IDSA results with those existing in the literature. 
Additionally, we perform Newtonian 1D and 2D simulations from prebounce core collapse to 
several hundred milliseconds postbounce 
with 11, 15, 21, and 27 $M_\odot$ progenitors from Woosley et al.~(2002) with the HS(DD2) equation of state. 
General relativistic effects are neglected.  
We obtain robust explosions with diagnostic energies $E_{\rm dig} \gtrsim 0.1- 0.5$~B for all considered 2D models within approximately $100-300$ milliseconds 
after bounce and find that explosions are mostly dominated by the neutrino-driven convection, 
although standing accretion shock instabilities are observed as well.
We also find that the level of electron deleptonization during collapse dramatically affect the postbounce evolution,
e.g.~the ignorance of neutrino-electron scattering during collapse will lead to a stronger explosion. 

\end{abstract}

\keywords{hydrodynamics  --- instabilities ---neutrinos --- supernovae: general}


\section{INTRODUCTION}

After nearly five decades from the first attempt to obtain a neutrino-driven explosion by \cite{1966ApJ...143..626C},
the explosion mechanism of core-collapse supernovae (CCSNe) remains elusive.
It is generally believed that neutrino transport and convection are important ingredients to achieve a successful explosion
(see recent reviews in \citealt{2012ARNPS..62..407J,2013RvMP...85..245B,2015arXiv150101334F}).   
However, modeling CCSNe with Boltzmann transport in three dimensions
is numerically expensive and too time consuming with current computing resources,    
since the neutrino radiation hydrodynamics with Boltzmann transport is essentially a seven-dimensional problem: 
three spatial coordinates, two angular degrees of freedom, neutrino energy and time. 
In addition, there are three types of neutrino species and their anti-particles
that would require a solution of the Boltzmann equation. 

CCSN simulations with Boltzmann transport have been studied in 1D  
\citep{1993ApJ...405..669M, 2001PhRvD..63j3004L, 2004ApJS..150..263L, 2005ApJ...629..922S}, 
in 2D \citep{2004ApJ...609..277L,2008ApJ...685.1069O,2011ApJ...728....8B} 
and recently in 3D with low resolutions and fixed background profiles \citep{2015ApJS..216....5S}.
However, these 2D and 3D works have ignored the velocity-dependent terms 
and decoupled these from the energy groups for simplicity, 
and the spatial resolutions are not sufficient to achieve a correct turbulence cascade.
Therefore, approximate methods for the neutrino transport in multi-dimensional simulations are still necessary 
at this moment. 

Simple approximatation schemes include the light-bulb scheme \citep{2008ApJ...688.1159M, 2012ApJ...755..138H, 2013ApJ...765...29C}, 
neutrino leakage \citep{1996A&A...306..167J, 2003MNRAS.342..673R, 2011ApJ...730...70O, 2014ApJ...785..123C},
and gray transport \citep{1999ApJ...516..892F, 2006A&A...457..963S}. 
In these schemes, the neutrino transport is relatively cheap 
and therefore, it is possible to perform high resolution simulations in 3D 
(with effective angular resolutions $\lesssim 1^\circ$),
better describing the turbulence and convection behind the shock 
and the standing accretion shock instabilities (SASI, \citealt{2003ApJ...584..971B}). 
However the neutrino transport in these schemes is possibly over-simplified 
because it does not really follow the transport of the neutrino distributions 
and therefore, cannot describe the neutrino heating self-consistently. 

A more sophisticated but still approximated scheme for the neutrino transport is called the Moment scheme:
The multi-group flux limited diffusion (MGFLD) scheme \citep{2013ApJ...767L...6B, 2015ApJ...800...10D} 
takes the zeroth angular moment of neutrino moment expansions (i.e.~energy density). 
The M1 moment scheme additionally evolves the momentum density and considers the higher moment closure 
with an analytic form 
\citep{2000MNRAS.317..550P, 2012ApJ...755...11K,2013ApJ...762..126O,
2014MNRAS.445.3169O, 2014arXiv1411.7058O, 2015arXiv150106330K} 
or by a variable Eddington tensor  
\citep{2000ApJ...539..865B, 2002A&A...396..361R, 2003ApJ...592..434T, 2006A&A...447.1049B, 
2012ApJ...756...84M, 2015ApJ...801L..24M}. 
However, as reported by \cite{2015arXiv150106330K}, the M1 scheme in 3D general relativity (GR) is still very expensive 
and difficult to run in long term postbounce simulations with high resolutions (an effective angular resolution $\lesssim 4^\circ$). 

Another approximated transport scheme is the isotropic diffusion source approximation (IDSA, \citealt{2009ApJ...698.1174L}). 
In the IDSA, the distribution function is separated into an opaque trapped particle component 
and a transparent streaming particle component.  
The two components are linked by a source term.
Therefore, the transport equation becomes a diffusion problem in the opaque region 
and a ray tracing problem in the transparent region \citep{2009ApJ...698.1174L}. 
Multi-dimensional simulations with the IDSA have been performed by 
\cite{2011ApJ...738..165S, 2014arXiv1406.6414S} in 2D, 
and \cite{2014ApJ...786...83T, 2012ApJ...749...98T} in 3D.
However, most of these multi-dimensional studies with the IDSA or with other transport schemes do not really 
solve the neutrino transport in multi-dimensions.
Instead, they consider the angular distribution to be several 1D problems, i.e.~apply the ray-by-ray (RbR) approach.  
\cite{2015ApJS..216....5S} and \cite{2015ApJ...800...10D} have pointed out that the RbR approach 
may artificially exaggerate the neutrino flux variations in the angular and temporal components, 
since the temporal variation of the neutrino fluxes in convective regions is greatly ignored in the RbR approach.  
Additionally, the RbR approximation overestimates the heating of accretion luminosity on its own ray
and underestimates the heating in neighbor rays.

In this Paper, we present two-dimensional CCSN simulations with the IDSA for neutrino transport in cylindrical coordinates, 
which is in principle easy to extend to 3D  and 
has better resolution and boundary conditions for the description of the proto-neutron star (PNS) 
than simulations in spherical coordinates.  
We solve the trapped particle component in multi-dimensions to improve 
the neutrino transport in angular and temporal directions.
Furthermore, we study the neutrino transport effects during collapse 
by comparing our IDSA solver with a parametrized deleptonization scheme from \cite{2005ApJ...633.1042L}.
We find that the postbounce explosion dynamics is sensitive to the detailed neutrino interactions before core bounce, 
such as neutrino-electron scattering (NES) and electron-capture rates.
Additional effects such as GR and magnetic fields are also crucial factors on studying CCSN explosion mechanism, 
in particular for fast-rotating and more massive progenitors,  
but we leave these parts for future work.   
In the following section, the numerical methods and the IDSA implementations are described. 
A verification of our IDSA implementation and a comparison 
with other neutrino transport schemes is shown in Section~\ref{sec_verification}. 
In Section~\ref{sec_results}, we apply our IDSA implementation to multiple progenitors, 
using a new SN equation of state (EOS).
In Section~\ref{sec_conclusions}, we summarize our results and conclude. 
A discussion of the different EOS is presented in Appendix~A.

%
%
\section{NUMERICAL METHODS AND MODELS} \label{sec_methods}
We describe the numerical code and the corresponding setup of our simulations in
\ref{sec_code}. The method and implementation of the IDSA for neutrino transport is demonstrated in \ref{sec_neutrino}.  
We present the investigated EOS and supernova progenitors in Sections~\ref{sec_eos} and \ref{sec_progenitors}.


%
%
%
\begin{deluxetable*}{lc}
\tabletypesize{\scriptsize}
\tablecolumns{2}
\tablecaption{Neutrino Interactions included in the Simulations \label{tab_interactions}}
\tablewidth{0pt}
\tablehead{ 
\colhead{Neutrino Interaction} & \colhead{Reference} }
\startdata
1. $\nu_e + n \rightleftharpoons  e^- +p$ & \cite{1985ApJS...58..771B}\\
2. $\bar{\nu}_e + p \rightleftharpoons e^+ + n$ & \cite{1985ApJS...58..771B}\\
3. $\nu_e + A \rightleftharpoons  e^- + A' $ & \cite{1985ApJS...58..771B}\\
4. $\nu + \alpha/A/N \rightleftharpoons  \nu + \alpha/A/N $ & \cite{1985ApJS...58..771B}\\
5. $\nu + e^\pm \rightleftharpoons \nu +e^\pm $ & \cite{1993ApJ...410..740M, 2005ApJ...633.1042L}$^\dagger$ \\
\enddata
\tablecomments{$n$= free neutrons, $p$= free protons, $N$= free neutrons or protons, $A$= nuclei besides $\alpha$ particles, $\alpha$=alpha particles, 
$\nu_e$= electron type neutrinos, $\bar{\nu}_e$= anti-electron type neutrinos,  $\nu$= all type of neutrinos,
$e^-$=electron, and $e^+$= positron.}
\tablenotetext{$\dagger$}{    
In this Paper, we include the neutrino-electron scattering by means of using the 
parametrized deleptonization in \cite{2005ApJ...633.1042L} in the pre-bounce phase, 
since this reaction is only important during core collapse.}
\end{deluxetable*}

\subsection{Numerical Code And Initial Setup} \label{sec_code}

We use {\tt FLASH}\footnote{\url{http://flash.uchicago.edu}} version~4 
\citep{2000ApJS..131..273F, 2008PhST..132a4046D, 2013JCoPh.243..269L} 
to solve the Eulerian equations of hydrodynamics,
\begin{equation}
\frac{\partial \rho}{\partial t}    +   \nabla \cdot \left(\rho \bm{v} \right)=0 
\end{equation}
\begin{equation}
\frac{\partial \rho \bm{v}}{\partial t}    +   \nabla \cdot \left(\rho \bm{v}\bm{v} \right) + \nabla P= -\rho \nabla \Phi 
\end{equation}
\begin{equation}
\frac{\partial \rho E}{\partial t}    +   \nabla \cdot \left[(\rho E + P)\bm{v} \right]= -\rho \bm{v} \cdot \nabla \Phi  
\end{equation}
\begin{equation}
\frac{\partial \rho Y_e}{\partial t}    +   \nabla \cdot \left(\rho \bm{v} Y_e \right)=0 
\end{equation}
\begin{equation}
\frac{\partial \rho Y_l^t}{\partial t}    +   \nabla \cdot \left(\rho \bm{v} Y_l^t \right)=0 
\end{equation}
\begin{equation}
\frac{\partial \left(\rho Z_l^t \right)^{3/4}}{\partial t}    +   \nabla \cdot \left(\bm{v} (\rho Z_l^t)^{3/4} \right)=0 \label{eq_nu_energy}
\end{equation}
where $\rho$ is the gas density, $\bm{v}$ is the velocity vector, $P$ is the gas pressure, 
$E$ is the total specific energy, $\Phi$ is the gravitational potential, 
$Y_e$ is the electron fraction, $Y^t$ is the particle number fraction, 
and $Z$ is the particle mean specific energy. 
The index $l$ labels different species of trapped particles (i.e. $\nu_e$ and $\bar{\nu_e}$).  
A detailed description of the neutrino-transport method will be presented in Section~\ref{sec_neutrino}.  

{\tt FLASH} is a parallel, multidimensional hydrodynamics code based on block-structured Adaptive Mesh Refinement (AMR).
Our simulation setup is essentially similar to what has been implemented in \cite{2013ApJ...765...29C} 
and \cite{2014ApJ...785..123C}, but replacing their radiation transfer solver by an IDSA solver \citep{2009ApJ...698.1174L}.
We use the third-order piecewise parabolic method (PPM, \citealt{1984JCoPh..54..174C}) for spatial reconstruction, 
the HLLC Riemann solver and the Hybrid slope limiter for shock-capture.
The ``hybrid'' Riemann solver is widely used to avoid an odd-even numerical instability 
when the shock is aligned with the grid \citep{1991PhDT........16Q}. 
However, we don't see this instability in our CCSN simulations by comparing simulation results 
with the hybrid Riemann solver. 
The HLLC Riemann solver shows a better turbulent cascade 
based on the implicit large eddy simulations \citep{2015arXiv150103169R}.
Effects from general and special relativity and from magnetic fields are ignored. 

Simulations are performed in 1D spherical and 2D cylindrical coordinates. 
The center of a progenitor star is located at the origin of the simulation box. 
The simulation box includes the inner $10^4$~km in radius of a progenitor star in 1D
and the full $180^\circ$ in cylindrical coordinates with 
$r_{\rm max} = z_{\rm max} = 10^4$~km and $z_{\rm min}= -10^4$~km in 2D.  
The central $r \lesssim 100$~km sphere has the smallest zone size of $0.488$~km 
and the AMR mesh is dynamically adjusted based on the magnitude of the second derivatives of 
gas density, pressure, and entropy.
To save computation time, we apply additional AMR decrements based on the distance to the origin.  
E.g.~the first AMR decrement is enforced at $r \sim 100$~km 
and the second AMR decrement at $r \sim 200$~km, the next at $r \sim 400$~km, and so on. 
The maximum zone size is $62.5$~km.
We employ the ``outflow'' boundary condition at the outer boundaries 
and the ``reflect'' boundary condition at the inner boundaries. 
The gravitational potential is solved by the new improved multipole solver in {\tt FLASH} \citep{2013ApJ...778..181C} 
with a maximum Legendre order, $l_{\rm max} = 16$.

It should be noted that the ``outflow'' boundary condition,  as reported by \cite{2013ApJ...765...29C}, 
causes a zero-gradient mass accretion at the boundary which will overestimate the inflow and 
suppress the explosion at late time. 
We therefore extend our simulation domain to 10,000 km to minimize the effect of boundary conditions. 


\begin{deluxetable*}{ccccccc}
\tabletypesize{\scriptsize}
\tablecaption{Parameters for the Parametrized Deleptonization \label{tab_parameters}}
\tablewidth{0pt}
\tablehead{
\colhead{Progenitor} & \colhead{$s11.0$ (W02)} & \colhead{$s15.0$ (W02)} & \colhead{$s21.0$ (W02)} & \colhead{$s27.0$ (W02)} &\colhead{$s15$ (W07)} & \colhead{$s20$ (W07)}}
\startdata
$\rho_1$ [g~cm$^{-3}$] & $1.5\times 10^8$     & $9.0\times 10^{7}$ & $2.5 \times 10^{8}$ & $2.5 \times 10^{8}$ & $2.2\times 10^{8}$ & $3.0 \times 10^{8}$\\
$\rho_2$ [g~cm$^{-3}$] & $1.2\times 10^{13}$ & $9.0\times 10^{12}$ & $1.0\times 10^{13}$ & $1.0\times 10^{13}$ & $9.5\times 10^{12}$ & $1.0\times 10^{13}$\\
$Y_1$ & 0.5 & 0.5 & 0.5 & 0.5 & 0.5 & 0.5 \\
$Y_2$ & 0.287 & 0.282 & 0.279 & 0.279 & 0.279 & 0.273\\
$Y_c$ & 0.02 & 0.03 & 0.017 & 0.017 & 0.022 & 0.017\\

\enddata
\tablecomments{Parameters used in the fitting formula from \cite{2005ApJ...633.1042L} }
\end{deluxetable*}

\subsection{Neutrino Transport} \label{sec_neutrino}

In the IDSA \citep{2009ApJ...698.1174L, 2013SIAM}, we decompose the distribution function $f$ 
of transported particle species and the neutrino transport operator $D$ 
into a trapped particle component and a streaming particle component. 
We assume that these two components evolve separately. 
With this assumption, we rewrite the transport equation,
$D(f=f^t+f^s) = C$, where $C= C^t+C^s$ is a collision integral, by two equations,   
\begin{eqnarray}  
D(f^t) = C^t - \Sigma \\
D(f^s) = C^s + \Sigma
\end{eqnarray}
where $\Sigma$ is the diffusion source term, which converts trapped particles ($t$) into streaming particles ($s$) and vice versa. 

By using the diffusion limit, the diffusion source term $\Sigma$ in the trapped transport equation could be expressed by 
(see \citealt{2009ApJ...698.1174L} for a more detailed discussion),
\begin{equation}
\Sigma = \min \left\{ \max \left[\alpha + (j+\chi) \frac{1}{2} \int f^s d\mu,0 \right],j \right\}
\end{equation}
where, 
\begin{equation}
\alpha = \nabla \cdot \left(\frac{-1}{3(j+\chi+\phi)} \nabla f^t \right), \label{eq_alpha}
\end{equation}
is a non-local diffusion scalar, $j$ is the emissivity, $\chi$ is the absorptivity, $\phi$ is the opacity, 
and $\mu$ is the angle cosine.
The distribution functions $f^t$ and $f^s$ can be solved by using equations~(8) and (10) in \cite{2009ApJ...698.1174L}.
Once the distribution functions are known, 
the net interaction rates of transported particles can be calculated and further updates of the fluid quantities, 
such as $\dot{v}$, $\dot{Y_e}$, $\dot{E}$, $\dot{Y_l^t}$, and $\dot{Z_l^t}$, can be derived.

We solve the streaming component in 1D with the original IDSA solver,
but solve the trapped component and $\alpha$ in multi-dimensions. 
In each time step, we take the angle-averaged radial bins of fluid and neutrino quantities, 
such as $\rho$, $s$, $T$, $Y_e$, $Y_l^t$, $Z_l^t$, as 1D inputs for the streaming component. 
The radial bins contain 600 zones sampled from the center of the proto-neutron star (PNS), 
defined by the location of maximum density, up to $r=5,000$~km. 
The radial bins are uniformly spaced ($\Delta r = 2$~km) up to a radius of $r = 400$~km.
Beyond 400~km, the zone spacing logarithmically increases.  
The streaming component carries only to the location of neutrino spheres $R_\nu (E)$ 
and the angular integration of the spectral neutrino flux $F_s(E)$. 
The local heating from streaming neutrinos are then determined based on the local neutrino 
interaction rates based on the multi-dimensional hydro quantities. 

It should be noted that the assumption of spherical symmetry in the streaming component 
may calculate the neutrino flux and heating incorrectly when shocks become highly asymmetric,
especially in 2D simulations after the SASI has developed. 
On the other hand, a RbR approach may artificially enhance the asymmetry and
lead to incorrect results as well \citep{2015ApJS..216....5S, 2015ApJ...800...10D}. 
It is important to have multi-dimensional simulations with both approximations because they are likely
to bracket the actual solution: The RbR focusses all heating from the accretion luminosity on its own ray,
while the angular integration distributes it over all directions. In reality, 
a solution in between these extremes is expected.

For the trapped component, we explicitly solve the diffusion source $\Sigma$ and $\alpha$ in multi-dimensions 
and update $f^t$ locally.  
Together with the streaming component, the new multi-dimensional interaction rates can be updated. 
The new neutrino sources are re-evaluated based on multi-dimensional neutrino sources and production rates, 
and then used for the next streaming step.
Since we solve the diffusion part explicitly, a stable solution requires a small neutrino time step
$t_\nu \sim \Delta x^2/(2\lambda c) \sim \Delta x/ 2c$. 
Therefore, in our 1D and 2D simulations, we use a fixed hydro timestep with $t_{\rm hydro} = 10^{-6}$~sec 
and do sub-cycling for the neutrino transport. 
Since the IDSA solver in the streaming component allows larger time step, 
we adopt two types of sub-cycles. One for the streaming component with $t_\nu^s = 5 \times 10^{-7}$~sec, 
and the other for the trapped component with $t_\nu^t = 10^{-7}$~sec. 
Additionally, the neutrino pressure contributes extra momentum and can be expressed by,
\begin{equation}
\frac{\partial \bm{v}}{\partial t} = - \frac{1}{\rho}\nabla \left( \frac{\rho Z^t_l}{3m_{\rm b}}\right). \label{eq_nu_pressure}
\end{equation}
We note that Equations (\ref{eq_nu_energy}) and (\ref{eq_nu_pressure}) include 
all thermodynamically important $O(v/c)$~terms of the Boltzmann transport equation.
These $O(v/c)$~terms have been considered as crucial 
for CCSN modeling \citep{1993ApJ...405..669M, 2009ApJ...698.1174L, 2012ApJ...747...73L}.

Our IDSA solver only includes electron flavor neutrinos.
We use 20 energy bins which are spaced logarithmically from 3~MeV to 300~MeV.
For heavy neutrinos, the energy release is treated by a leakage scheme 
that is based on the local diffusion and the local production rates 
\citep{1998ApJ...507..339H, 2003MNRAS.342..673R, 2014MNRAS.443.3134P}.

Table~\ref{tab_interactions} summarizes all weak interactions that are included in this work. 
All weak interactions in the IDSA solver are using the Bruenn description \citep{1985ApJS...58..771B}. 
Note that the IDSA solver dose not include the NES 
(interaction~5 in Table~\ref{tab_interactions}).
NES is important during the collapse phase 
but provides minor contributions in the postbounce phase \citep{1989ApJ...340..955B} 
(see Section~\ref{sec_cc} for a demonstration). 
\cite{2005ApJ...633.1042L} found a simple relation between the electron deleptonization with the gas density in the collapse phase, where electron fraction and entropy can be parametrized by density and chemical potentials.
This parametrized deleptonization (PD) scheme can take interactions effectively into account that
are only implemented in the Boltzmann solver that is used to determine the parameters for the PD.
We have recalibrated the fitting parameters with different progenitors 
for the PD scheme by running {\tt AGILE-BOLTZTRAN} \citep{2004ApJS..150..263L}, 
since the original parameters in \cite{2005ApJ...633.1042L} are calibrated for the progenitors 
G15 (s15s7b2, \citealt{1995ApJS..101..181W}) and N13 \citep{1988PhR...163...13N}.  
Table~\ref{tab_parameters} summarizes the fitting parameters we use during collapse.   
In this Paper we include the effect of NES by means of using the PD scheme in the collapse phase. 
Since the scheme is independent of the neutrino fractions, 
we update the neutrino fractions and energy through the IDSA solver during collapse. 
After bounce, we turn off the PD and switch to the IDSA solver.  
For simulations without NES, we use the IDSA solver from the beginning of core collapse to postbounce. 

To verify our IDSA implementation in {\tt FLASH}, we provide a code comparison of {\tt FLASH-IDSA} 
with {\tt AGILE-IDSA} and with existing results from the literature in Section~\ref{sec_verification}.


\subsection{Equation of State} \label{sec_eos}

We use the nuclear EOS unit in {\tt FLASH} which incorporates the finite temperature EOS routines 
from \cite{2010CQGra..27k4103O} and \cite{2013ApJ...765...29C}\footnote{\url{http://www.stellarcollapse.org}}. 
The Lattimer \& Swesty EOS (with incompressibility, $K=220$~MeV; \citealt{1991NuPhA.535..331L}) 
and Hempel \& Schaffner-Bielich (HS) DD2 EOS are used in this work. 
The HS(DD2) EOS employs the density-dependent relativistic mean-field interactions of \cite{2010PhRvC..81a5803T}. 
The description of nuclei in supernova matter is based on \citep{2010NuPhA.837..210H}. 
This EoS was first applied in core-collapse supernova simulations by \cite{2014EPJA...50...46F}, 
where further details can be found.
LS220 is one of the most common and well-studied EOS in the supernova community. 
However, it has some deficiencies, for example it is based on the single nucleus approximation for heavy nuclei, 
and considers only the alpha particle as a degree of freedom of all possible light nuclei. 
See \cite{2015arXiv150300518H} for a comparison of predictions for cluster formation 
for the HS(DD2) and the LS220 EOS with experiments, where good agreement was found. 
Furthermore, it was shown by \cite{2013PhRvC..88b5802K,2014EPJA...50...46F}, 
that the neutron matter EOS of LS220 is in disagreement with constraints from Chiral effective field theory. 
Furthermore, its low-density symmetry energy deviates from constraints obtained from finite nuclei, 
see Fig.~9 of \cite{2014arXiv1410.6337H}.
No multi-dimensional simulations have been performed with HS(DD2) at this moment.
We use LS220 for a code verification test in Section~\ref{sec_verification}
and then use DD2 for our main simulations in Section~\ref{sec_results}.
A low-density extension for the EOS is included in the routines from  \cite{2010CQGra..27k4103O}. 
In Appendex~B, we provide a brief discussion of the differences between LS220 and DD2.


\subsection{Supernova Progenitors} \label{sec_progenitors}

\begin{figure}
\begin{center}
\epsscale{1.2}
\plotone{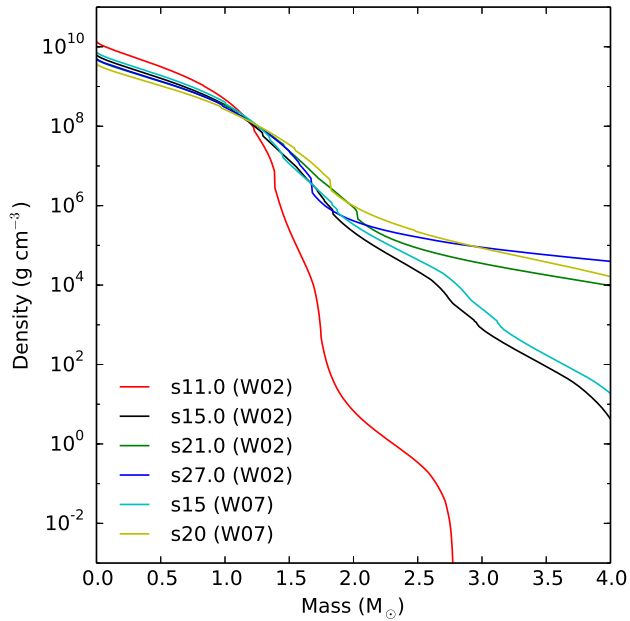}
\end{center}
\caption{\label{fig_progenitors} Density as function of enclosed mass for the four investigated progenitor models from \cite{2002RvMP...74.1015W} and \cite{2007PhR...442..269W}.}
\end{figure}

In this Paper, we consider four different nonrotating, solar-metallicity progenitors, 
s11.0, s15.0, s21.0, and s27.0 from \cite{2002RvMP...74.1015W}\footnote{\url{http://2sn.org/stellarevolution/}}
for our multiple progenitor study. 
We also consider two nonrotating, solar-metallicity progenitors, 
s15 and s20 from \cite{2007PhR...442..269W} for a comparison study.
Figure~\ref{fig_progenitors} shows the initial density distribution of these six progenitors. 
The s11.0 progenitor has the highest core density but the most dilute envelope.   
The s21.0 and s27.0 and s20 progenitors have a similar density distribution and the most massive envelope among the models,
but have a different mass of the iron core and Si/O shell.  
The locations of regions with a high density gradient correspond to the Si/O interface. 
For the same progenitor mass, s15 has a denser core and more massive envelope than s15.0.
Another common progenitor model used in the literature is the s15s7b2 progenitor from \cite{1995ApJS..101..181W}.
It has the same progenitor mass as $s15.0$ and $s15$ but the Si/O interface is located much further inside.  
This difference may make significant changes on the postbounce shock-radius evolutions 
when the shock reaches the interface due to a different mass-accretion history \citep{2014arXiv1406.6414S}.  

To adopt the progenitor models in {\tt FLASH},
we map the one-dimensional density, temperature, electron fraction, and radial velocity
profiles from \cite{2002RvMP...74.1015W} into our 1D/2D grids in {\tt FLASH}.
Other thermodynamic quantities are re-calculated using the EOS solver in {\tt FLASH}.
Neutrino fractions are set to zero at beginning.

%
%

\section{IDSA VERIFICATION} \label{sec_verification}

To verify the IDSA implementation in {\tt FLASH}, we first compare our 1D {\tt FLASH} simulations 
with simulations with {\tt AGILE-IDSA} \citep{2009ApJ...698.1174L} in Section~\ref{sec_cc_idsa}.
{\tt AGILE-IDSA} is a 1D spherically symmetric Lagrangian code 
which is publicly available online\footnote{\url{https://physik.unibas.ch/~liebend}}.
In \cite{2009ApJ...698.1174L}, a nice agreement of {\tt AGILE-IDSA} 
with the GR Boltzmann code {\tt AGILE-BOLTZTRAN} \citep{2004ApJS..150..263L} was shown. 
Since we want to compare our results with the same neutrino transport scheme (IDSA), 
we run additional simulations in {\tt AGILE-IDSA} but turn off the GR correction for the gravitational potential. 
We also turn off the PD during collapse in both codes to test the IDSA solver for neutrino transport before and after bounce. 

In Section~\ref{sec_cc}, we show 1D and 2D {\tt FLASH} simulations 
with the two widely used progenitors, s15 and s20 from \cite{2007PhR...442..269W} 
and discuss the differences compared to other transport schemes.
Furthermore, we run {\tt AGILE-BOLTZTRAN} simulations to demonstrate the importance of NES 
during core collapse and discuss the influence of the utilized neutrino opacities and electron capture rates 
on the shock evolution and neutrino signal.

%
\begin{figure}
\begin{center}
\epsscale{1.2}
\plotone{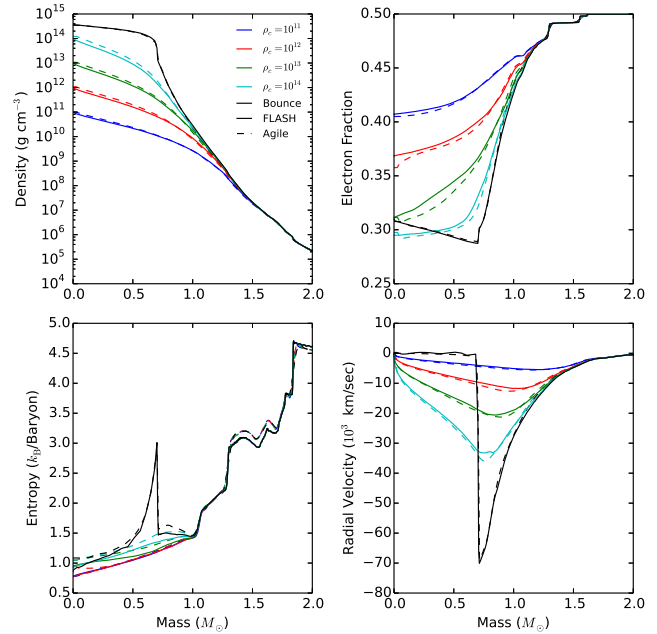}
\end{center}
\caption{\label{fig_agile_collapse} 
1D radial profiles in mass coordinate during the collapse of the $s15.0$ progenitor with LS220 EOS. 
Different colors represent different times when the central density reaches $\rho_c = 10^{11},10^{12}, 10^{13}, 10^{14}$ and at bounce. 
Solid lines show simulation results from {\tt FLASH} and dashed lines show simulation results from {\tt AGILE-IDSA}. 
Note that there are some slight time offsets between the results from {\tt FLASH} and {\tt AGILE-IDSA} 
because their output files do not exactly match the given central densities.  
}
\end{figure} 

\subsection{Code Comparison with AGILE-IDSA} \label{sec_cc_idsa}

We perform 1D CCSN simulations of the four investigated progenitors ($s11.0$, $s15.0$, $s21.0$, and $s27.0$) 
with the LS220 EOS in both {\tt FLASH} and {\tt AGILE-IDSA}. 
Figure~\ref{fig_agile_collapse} shows the radial profiles of density, electron fraction, entropy 
and radial velocity of the progenitor $s15.0$ during collapse. 
It is clear to see that the bounce profiles (black lines in Figure~\ref{fig_agile_collapse}) are nearly identical in the two codes.
Small differences at the center could originate from the fact that {\tt AGILE-IDSA} is a Lagrangian code
with a moving mesh. At the beginning of a simulation, the innermost region of the progenitor star is
much less resolved than with the Eulerian grid of {\tt FLASH}.  
Therefore, the radial profiles in {\tt AGILE-IDSA} evolve slightly slower than those in {\tt FLASH}.  
Furthermore, the bounce time in {\tt FLASH} is about $0.1-12$~ms later than in {\tt AGILE-IDSA}, 
depending on the progenitor star.

%
\begin{figure}
\begin{center}
\epsscale{1.2}
\plotone{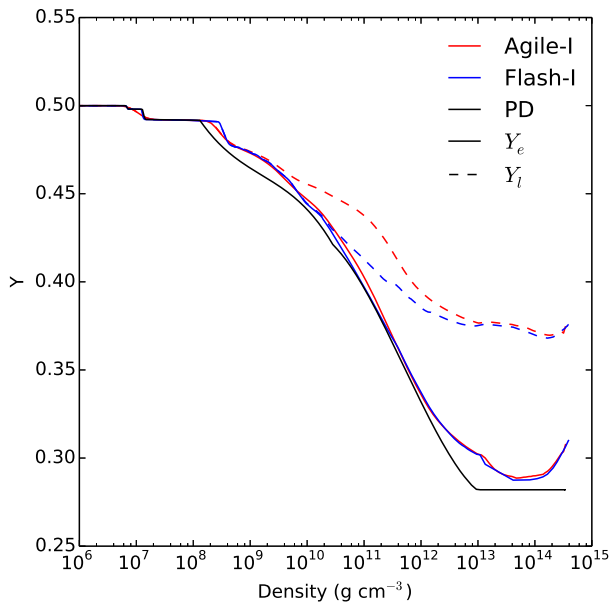}
\end{center}
\caption{\label{fig_agile_delep} 
Electron (solid lines) and lepton (dashed lines) fraction profiles at core bounce in model $s15.0$.
Red lines show simulation data from {\tt Agile-IDSA} and blue lines represent data from {\tt FLASH}.
The black line indicates the electron fraction from the fitted PD scheme (Table~\ref{tab_parameters}).
Both {\tt Agile-IDSA} and {\tt FLASH} are 1D simulations with LS220 EOS and without NES. }
\end{figure}

\begin{figure}
\begin{center}
\epsscale{1.2}
\plotone{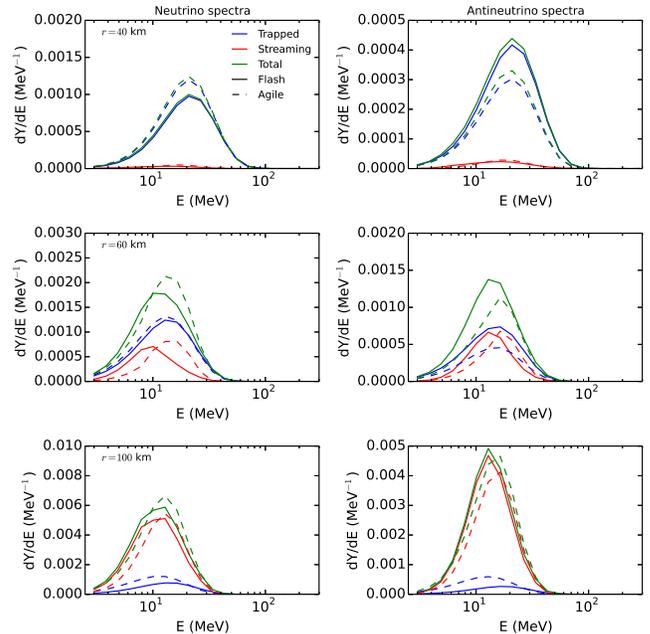}
\end{center}
\caption{\label{fig_agile_spectra} 
Particle spectra at 150~ms postbounce for the trapped particle component (blue lines), 
streaming particle component (red lines), and their sum (green lines). 
Solid lines show spectra from {\tt FLASH} and dashed lines represent spectra from {\tt AGILE-IDSA}.
Left panels show neutrino spectra and right panels show anti-neutrino spectra.
Each row indicates spectra at a different radius. }
\end{figure}

Figure~\ref{fig_agile_delep} shows the electron and lepton fractions of the progenitor $s15.0$ at core bounce. 
It is found that the electron fraction is consistent in both codes, 
but lepton fraction shows a mismatch at $10^{10}\lesssim \rho \lesssim 10^{12}$~g~cm$^{-3}$,
suggesting a lower electron neutrino distribution at low densities region in {\tt FLASH}.  
However, this difference does not alter the postbounce evolution much 
since the neutrinos are simply free streaming in the low density region.  
We note that in our multi-dimensional IDSA solver, we have employed a limiter for the diffusion scalar $\alpha$.
The limiter enforces the trapped component to have a small value, when $r > 1.25 \times R_\nu(E)$, 
to prevent unphysical oscillation on the $\alpha$, 
when the trapped neutrino density goes to zero in the free-streaming regime. 
This limiter leads to small differences of the neutrino luminosity and mean energy, 
but does not affect the hydrodynamic quantities.  
Figure~\ref{fig_agile_spectra} shows the particle spectra of the trapped and streaming components 
at 150~ms postbounce of the progenitor $s15.0$.
We show three different regions where the trapped particle component dominates (at $r=40$~km), 
trapped and streaming particle components are comparable (at $r=60$~km), 
and where the streaming particle component dominates ($r=100$~km).
There are small differences in the particle spectra, but we do not expect exactly identical spectra in the two codes,
since the hydrodynamic parts are different.  
 
%

\begin{figure}
\begin{center}
\epsscale{1.2}
\plotone{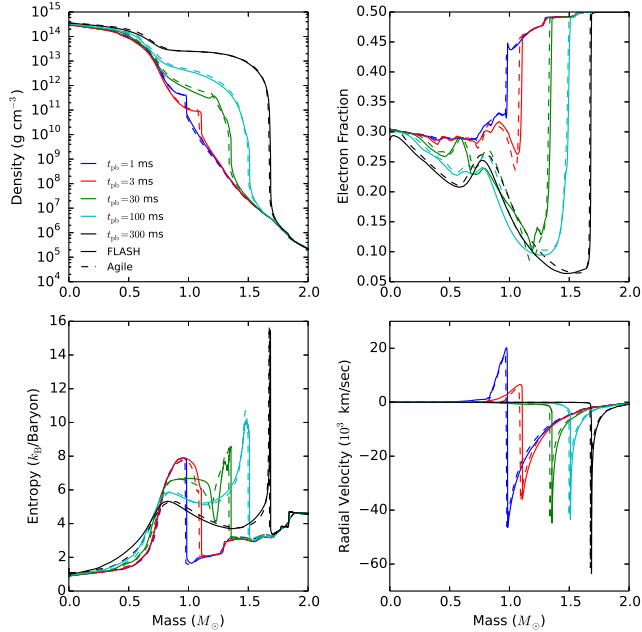}
\end{center}
\caption{\label{fig_agile_postbounce} 
Similar to Figure~\ref{fig_agile_collapse} but for the postbounce evolution. }
\end{figure}

\begin{figure}
\begin{center}
\epsscale{1.2}
\plotone{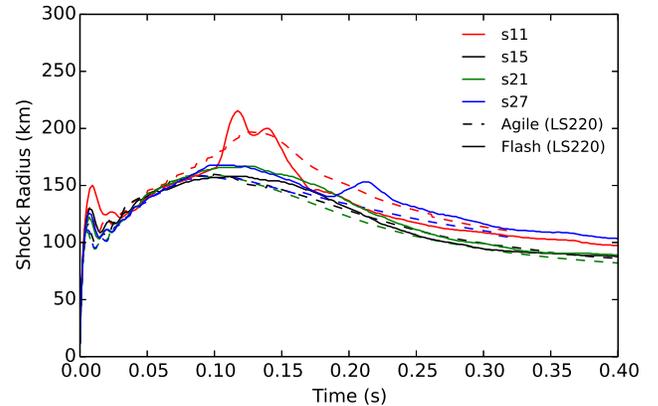}
\end{center}
\caption{\label{fig_agile_rshock} 
Time evolution of the shock radius for the four investigated progenitor models in both {\tt FLASH} (solid lines) 
and {\tt AGILE-IDSA} (dashed lines). Different colors represent different progenitor models. 
All results are 1D with LS220 EOS and without NES. The differences in the models $s11.0$ and $s27.0$ 
after $0.1$~s are due to a different handling of shell interfaces and a large diffusivity of the implicit adaptive mesh in {\tt AGILE-IDSA}. }
\end{figure}

A comparison of radial profiles of the density, electron fraction, entropy, 
and radial velocity of the progenitor $s15.0$ in both codes 
at different postbounce times is shown in Figure~\ref{fig_agile_postbounce}.
We note that although there are some differences in the radial shock location, 
the profiles and shock locations are consistent in the mass coordinate.  
The post bounce shock radius evolution for the four progenitor models are shown in Figure~\ref{fig_agile_rshock}.
Overall, the evolutions of the shock radius are nearly the same 
but show slight differences at $\sim 120$~ms ($\sim 220$~ms) in the model $s11.0$ ($s27.0$) 
when the shock reaches the Si/O interface.

\begin{figure}
\begin{center}
\epsscale{1.2}
\plotone{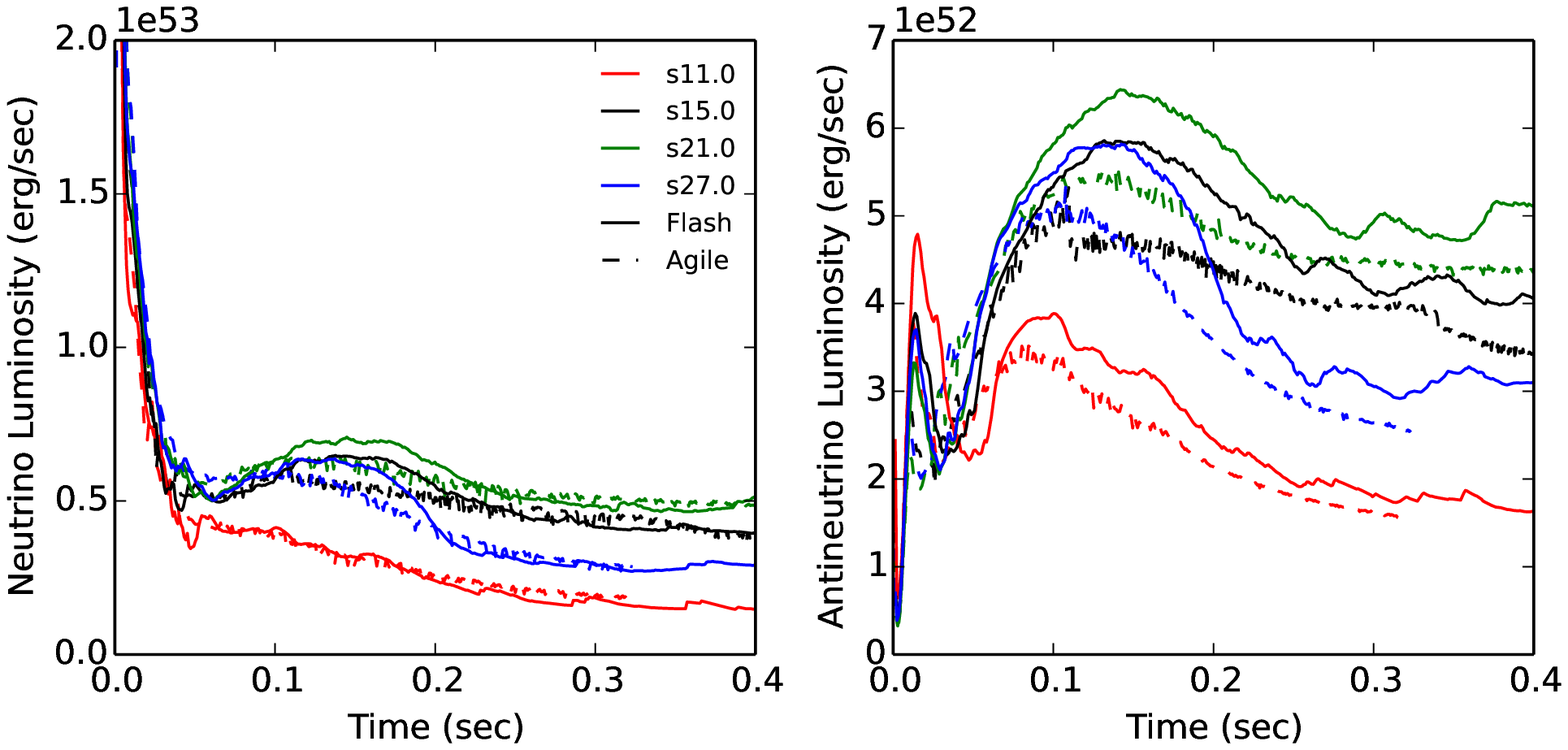}
\plotone{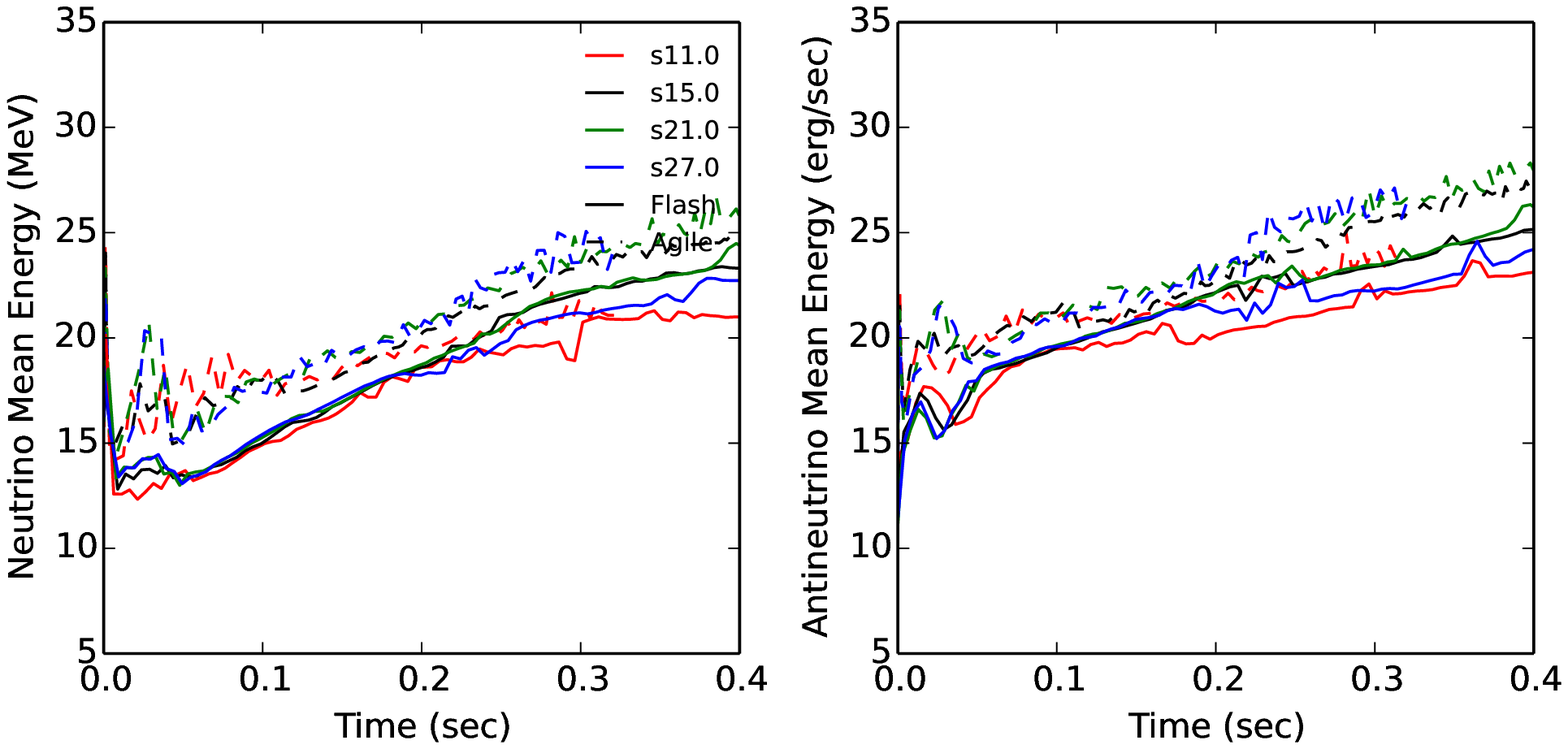}
\end{center}
\caption{\label{fig_agile_luminosity} 
The electron neutrino (left) and electron antineutrino (right) luminosities and mean energies as a function of time. 
The quantities are averaged in a $5$~ms time interval. 
Different colors represent the different progenitor models.  
Solid lines represent the {\tt FLASH} simulations and 
dashed lines indicate the {\tt AGILE-IDSA} simulations.}
\end{figure}

Figure~\ref{fig_agile_luminosity} shows the time evolution of electron-type neutrino luminosity and mean neutrino energy
for the four considered models.
The values are sampled at a radius of 500~km in both codes.
The electron neutrino luminosities and electron anti-neutrino luminosities 
are similar for the two codes around bounce 
and early postbounce.
However, at $\sim 100$~ms post bounce, the neutrino and anti-neutrino luminosities in {\tt FLASH} 
become slightly higher than in {\tt AGILE-IDSA}
for the progenitors $s15.0$, $s21.0$ and $s27.0$.
The maximum difference is an increase of about $\lesssim 15\%$ in {\tt FLASH} at $\sim 150$~ms post bounce.
After $\sim 200$~ms post bounce, the neutrino luminosities are back to the same value in both codes 
but the anti-neutrino luminosity remains slightly higher in {\tt FLASH} simulations. 
For the progenitor $s11.0$, the electron neutrino luminosity is the same in both codes 
but the electron antineutrino luminosity is slightly higher in {\tt FLASH}.
However, we note that the mean neutrino and anti-neutrino energies in {\tt FLASH} 
are about $10-20 \%$ lower than in {\tt AGILE-IDSA}.
This difference could originate from the lower neutrino fraction in the low density region 
in {\tt FLASH} (see Figure~\ref{fig_agile_delep}), since we measure the neutrino mean energy at $r=500$~km.

The kernel of our IDSA solver in {\tt FLASH} is the same as the solver in {\tt AGILE-IDSA}.
The main differences are from the hydrodynamics, 
the explicit solver for the diffusion scalar $\alpha$ in Equation~\ref{eq_alpha},
the detailed treatment of the EOS, and potentially, also the gravity solver. 
The low-density extension for the EOS in {\tt FLASH} and differences in the internal energy shift may also provide slightly differences.  
In principle, we should expect identical results in both codes in 1D.
As presented above, although some little differences have been observed, most features are consistent and in nice agreement.  
\cite{2005ApJ...620..840L} have performed a code comparison of the Boltzmann solver {\tt AGILE-BOLTZTRAN} 
with the variable Eddington factor {\tt VERTEX} code.  
Both codes have sophisticated physics input in spherical symmetry but different implementations.
As pointed out in \cite{2005ApJ...620..840L}, the different grids in the Lagrangian or Eulerian coordinates 
produce late time differences in the shock evolution when the shock runs through shell interfaces. 
Since {\tt FLASH} is also an Eulerian code, similar differences between {\tt FLASH} and {\tt AGILE-IDSA} as we have found here can be expected.

%

%

%

%
\begin{table*}
 \footnotesize
  \begin{center}
  \caption{1D Simulation Results During Core Collapse}
  \label{tab_collapse}
  \begin{tabular}{*{10}{p{.16\linewidth}}}
  	\midrule
	\midrule
	\multicolumn{2}{l}{Property} & \multicolumn{8}{c}{Models} \\
	\multicolumn{2}{l}{} & \multicolumn{2}{c}{$s11.0$} &  \multicolumn{2}{c}{$s15.0$} &  \multicolumn{2}{c}{$s21.0$} &  \multicolumn{2}{c}{$s27.0$} \\ 
	\midrule 
	\multicolumn{2}{l}{Progenitor mass ($M_\odot$)} & \multicolumn{2}{c}{11} &  \multicolumn{2}{c}{15} &  \multicolumn{2}{c}{21} &  \multicolumn{2}{c}{27} \\
	\multicolumn{2}{l}{Progenitor compactness ($\xi_{1.75}$)} & \multicolumn{2}{c}{0.066$^\dag \tnote{a}$} &  \multicolumn{2}{c}{0.54} &  \multicolumn{2}{c}{0.68} &  \multicolumn{2}{c}{0.53} \\
	\multicolumn{2}{l}{Progenitor compactness ($\xi_{2.5}$)} & \multicolumn{2}{c}{0.004} &  \multicolumn{2}{c}{0.15} &  \multicolumn{2}{c}{0.22} &  \multicolumn{2}{c}{0.23} \\
	\\
	\multicolumn{2}{l}{} & \multicolumn{1}{l}{w PD} & \multicolumn{1}{l}{wo PD} & \multicolumn{1}{l}{w PD} & \multicolumn{1}{l}{wo PD} & \multicolumn{1}{l}{w PD} & \multicolumn{1}{l}{wo PD} & \multicolumn{1}{l}{w PD} & \multicolumn{1}{l}{wo PD} \\
	\multicolumn{2}{l}{Abbreviation (1D-)} & \multicolumn{1}{c}{DP11} & \multicolumn{1}{c}{DA11} & \multicolumn{1}{c}{DP15} & \multicolumn{1}{c}{DA15} & \multicolumn{1}{c}{DP21} & \multicolumn{1}{c}{DA21} & \multicolumn{1}{c}{DP27} & \multicolumn{1}{c}{DA27} \\
	\midrule
	\multicolumn{2}{l}{Time at $\rho_c = 10^{11}$g~cm$^{-3}$ (ms)} & \multicolumn{1}{c}{176.3} & \multicolumn{1}{c}{194.1} &  \multicolumn{1}{c}{267.6} & \multicolumn{1}{c}{278.2} &  \multicolumn{1}{c}{269.7} & \multicolumn{1}{c}{284.3} &  \multicolumn{1}{c}{273.0} & \multicolumn{1}{c}{287.2} \\
	\multicolumn{2}{l}{Time at $\rho_c = 10^{12}$g~cm$^{-3}$ (ms)} & \multicolumn{1}{c}{197.6} & \multicolumn{1}{c}{224.2} &  \multicolumn{1}{c}{285.7} & \multicolumn{1}{c}{300.8} &  \multicolumn{1}{c}{286.3} & \multicolumn{1}{c}{304.0} &  \multicolumn{1}{c}{289.4} & \multicolumn{1}{c}{306.9} \\
	\multicolumn{2}{l}{Time at $\rho_c = 10^{13}$g~cm$^{-3}$ (ms)} & \multicolumn{1}{c}{201.9} & \multicolumn{1}{c}{229.3} &  \multicolumn{1}{c}{289.6} & \multicolumn{1}{c}{305.5} &  \multicolumn{1}{c}{290.2} & \multicolumn{1}{c}{308.4} &  \multicolumn{1}{c}{293.2} & \multicolumn{1}{c}{311.4} \\
	\multicolumn{2}{l}{Time at $\rho_c = 10^{14}$g~cm$^{-3}$ (ms)} & \multicolumn{1}{c}{203.0} & \multicolumn{1}{c}{230.7} &  \multicolumn{1}{c}{290.7} & \multicolumn{1}{c}{306.8} &  \multicolumn{1}{c}{291.2} & \multicolumn{1}{c}{309.7} &  \multicolumn{1}{c}{294.3} & \multicolumn{1}{c}{312.7} \\
	\multicolumn{2}{l}{Time at bounce (ms)} & \multicolumn{1}{c}{203.5} & \multicolumn{1}{c}{231.3} &  \multicolumn{1}{c}{291.2} & \multicolumn{1}{c}{307.3} &  \multicolumn{1}{c}{291.7} & \multicolumn{1}{c}{310.2} &  \multicolumn{1}{c}{294.7} & \multicolumn{1}{c}{313.2} \\
	\multicolumn{2}{l}{Bounce, central $\rho$ ($10^{14}$ g~cm$^{-3}$)} & \multicolumn{1}{c}{3.21} & \multicolumn{1}{c}{3.39} &  \multicolumn{1}{c}{3.14} & \multicolumn{1}{c}{3.43} &  \multicolumn{1}{c}{3.08} & \multicolumn{1}{c}{3.35} &  \multicolumn{1}{c}{3.10} & \multicolumn{1}{c}{3.45} \\
	\multicolumn{2}{l}{Bounce, central $Y_e$} & \multicolumn{1}{c}{0.287} & \multicolumn{1}{c}{0.316} &  \multicolumn{1}{c}{0.282} & \multicolumn{1}{c}{0.312} &  \multicolumn{1}{c}{0.279} & \multicolumn{1}{c}{0.310} &  \multicolumn{1}{c}{0.279} & \multicolumn{1}{c}{0.311} \\
	\multicolumn{2}{l}{Bounce, shock position ($M_\odot$)} & \multicolumn{1}{c}{0.54} & \multicolumn{1}{c}{0.77} &  \multicolumn{1}{c}{0.56} & \multicolumn{1}{c}{0.76} &  \multicolumn{1}{c}{0.56} & \multicolumn{1}{c}{0.75} &  \multicolumn{1}{c}{0.56} & \multicolumn{1}{c}{0.75} \\
	\midrule
  \end{tabular}
  \begin{tablenotes}
  	\item[a] $^\dag$ Note that the compactness parameter of the s11.0 progenitor is about ten times smaller than others due to a smaller core mass and light envelope.
  \end{tablenotes}
  \end{center}
\end{table*}


\begin{figure*}
\begin{center}
\epsscale{0.5}
\plotone{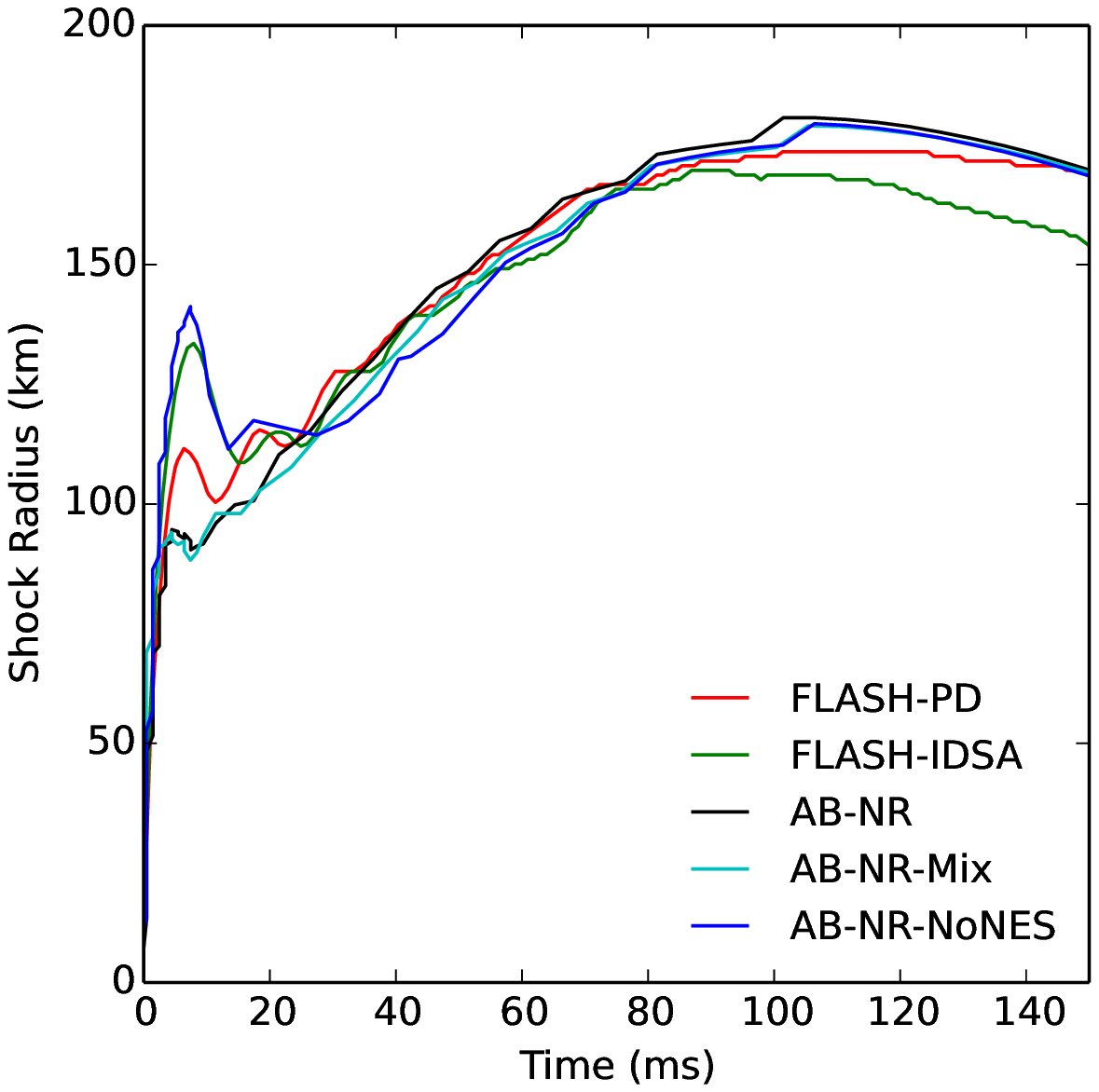}
\plotone{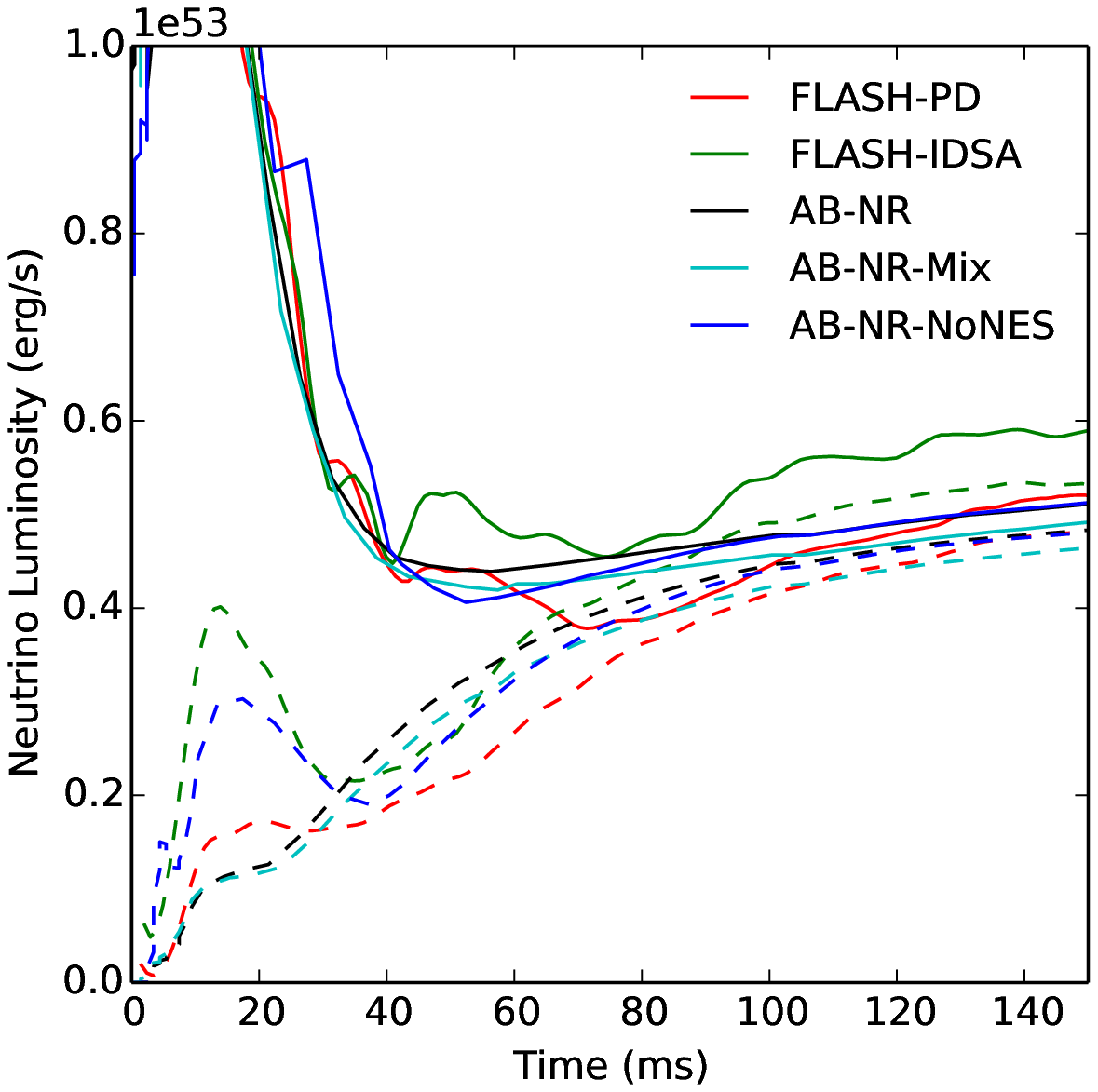}
\end{center}
\caption{\label{fig_cc_1d} 
Time evolution of shock radius (left) and electron-type neutrino luminosities (right). 
Different colors represent different models from {\tt FLASH} and {\tt AGILE-BOLTZTRAN} with different neutrino physics. 
See Section~\ref{sec_cc} for detailed description. }
\end{figure*}

\begin{figure*}
\begin{center}
\epsscale{0.5}
\plotone{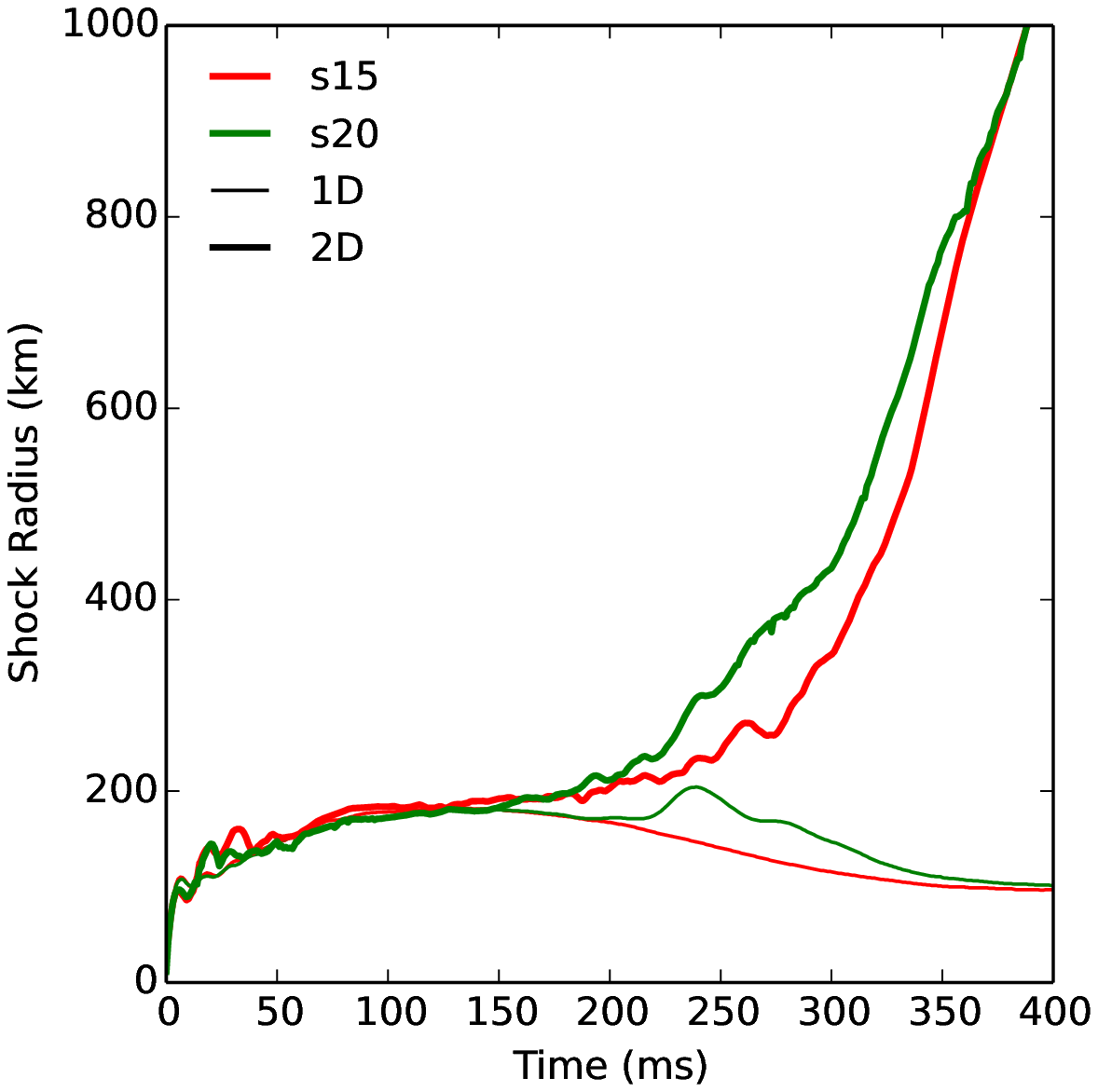}
\plotone{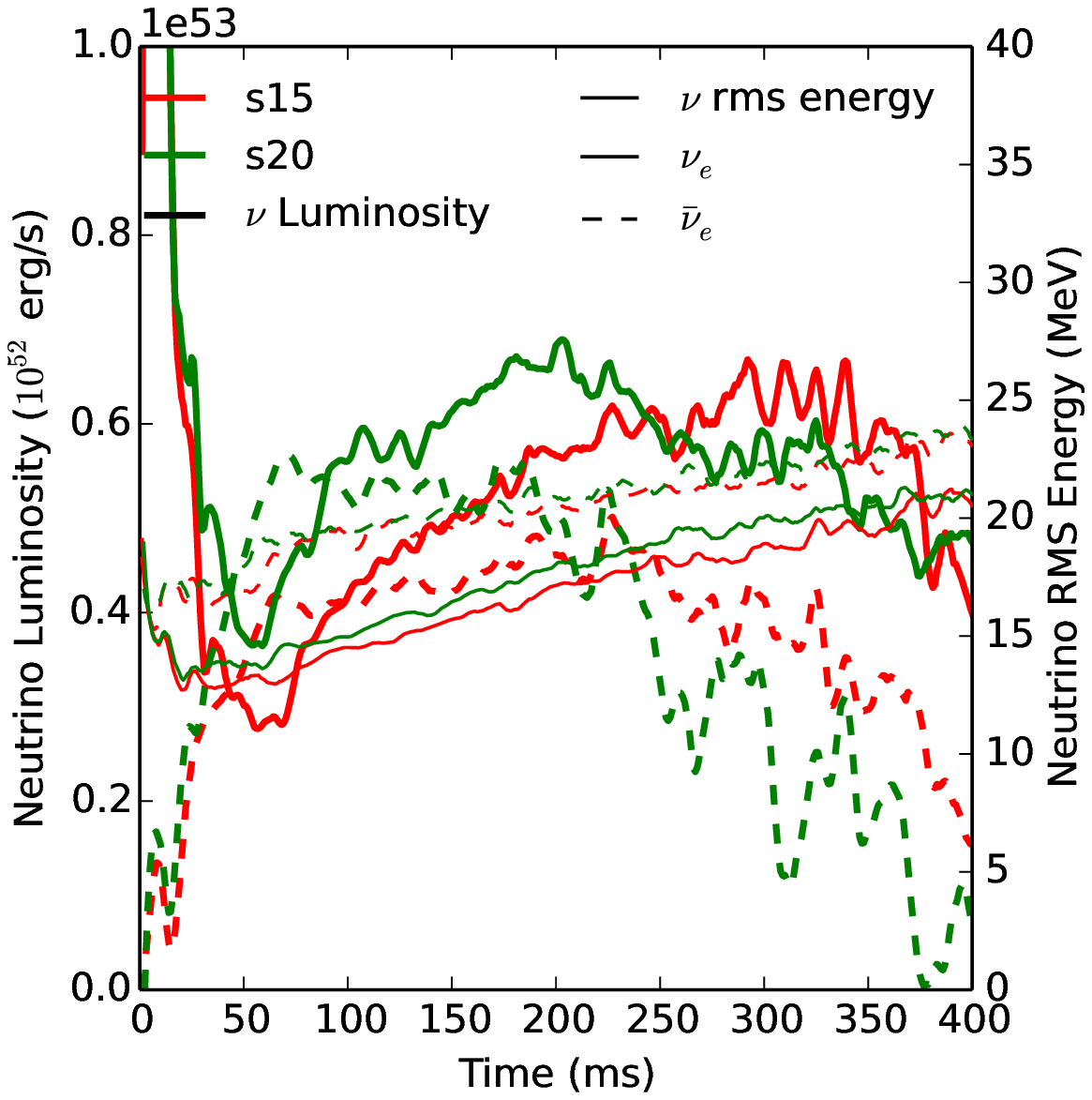}
\end{center}
\caption{\label{fig_cc_2d} 
Time evolution of shock radius (left) and electron-type neutrino signals (right).
Different colors represent different progenitor models. 
In the left panel, thick lines show the evolution of 2D simulations and 
thin lines represent 1D simulations. 
In the right panel, thick lines indicate electron-type neutrino luminosities 
and thin lines show the RMS neutrino mean energies.
Solid lines represent electron neutrino signals 
and dashed lines represent electron antineutrino signals.}
\end{figure*}

\subsection{Code Comparisons for the 15 $M_\odot$ and 20 $M_\odot$ progenitors} \label{sec_cc}

The s15 and s20 progenitors from \cite{2007PhR...442..269W} have been widely studied in the literature, 
e.g. by \cite{2013ApJ...767L...6B}, \cite{2015ApJ...800...10D}, \cite{2014arXiv1406.6414S}, Hanke (2014, PhD Thesis),
and \cite{2015ApJ...808L..42M}.  
In addition, \cite{2012ApJ...747...73L, 2012ApJ...760...94L} have shown that different neutrino opacities 
and approximations could lead to different post-bounce evolutions by using spherically symmetric {\tt AGILE-BOLTZTRAN} simulations
with the s15 progenitor. 
Therefore these two progenitor models are very suitable for comparisons among different SN codes.
It is generally agreed that the {\tt Vertex-Prometheus} \citep{2002A&A...396..361R} 
and the {\tt CHIMERA} \citep{2013ApJ...767L...6B} codes use state-of-the-art physics 
for neutrino transport. 
Therefore, it is worth to perform simulations with these progenitor models to have a direct comparison. 
However, there is still different physics employed in the above mentioned works. 
For instance, {\tt CASTRO} simulations \citep{2015ApJ...800...10D}, 
{\tt ZEUS} simulations \citep{2014arXiv1406.6414S}, and our {\tt FLASH-IDSA} simulations are Newtonian, 
but {\tt AGILE-BOLTZTRAN} simulations (\citealt{2012ApJ...747...73L, 2012ApJ...760...94L}), 
 {\tt CHIMERA} simulations \citep{2013ApJ...767L...6B}, 
and {\tt Vertex-Prometheus} simulations (Hanke, 2014, and \citealt{2015ApJ...808L..42M}) are GR or post-Newtonian. 
Furthermore, \cite{2015ApJ...800...10D} use Shen's EOS while other groups use the LS220 EOS. 
Therefore in this Section, we only give a qualitative discussion.

To evaluate that the PD scheme could effectively represent NES, 
we perform three Newtonian {\tt AGILE-BOLTZTRAN} simulations with the s15 progenitor and the LS220 EOS.   
The first simulation (model AB-NR) includes the NES;
the second simulation (model AB-NR-NoNES) ignores the NES; 
and  the third simulation (model AB-NR-Mix) includes NES only in the collapse phase. 
Figure~\ref{fig_cc_1d} shows the shock-radius and neutrino luminosity evolutions of these three simulations
together with {\tt FLASH-IDSA} simulations with and without PD. 
Ignoring NES (models FLASH-IDSA and AB-NoNES in Figure~\ref{fig_cc_1d}) 
gives a short-period of shock expansion at $\sim 10$~ms postbounce. 
A signature from this can also be seen in the electron anti-neutrino luminosity.  
This behavior is consistent with the model ``N-ReduceOp'' in \cite{2012ApJ...747...73L}, 
and the shock and electron-type neutrino luminosity evolutions of our {\tt FLASH-IDSA} simulation without PD 
is also consistent with our {\tt AGILE-BOLTZTRAN} simulation (AB-NR-NoNES).     

The model AB-NR-Mix is nearly identical to the model AB-NR,
demonstrating that NES is important mainly during the collapse phase (Figure~\ref{fig_cc_1d}).
\cite{2012PhRvD..85h3003F} showed that that NES plays a role in the late time PNS cooling 
and deleptonization ($> 1$~s) after the explosion. However, our simulations focus only on the first few hundred milliseconds.
The PD scheme in model FLASH-PD greatly improves the post-bounce simulation,
making our {\tt FLASH-IDSA} simulations closer to the {\tt AGILE-BOLTZTRAN} simulation with full neutrino reactions. 
Although there is no perfect match, we conclude that the use of the PD scheme in the collapse phase 
could effectively take into account the NES.
It should be noted that heavy neutrinos are treated by a simple leakage scheme 
which could lead to some difference in our IDSA simulations as well.


In Figure~\ref{fig_cc_2d}, we show the shock-radius evolution and neutrino signatures of our
2D {\tt FLASH-IDSA} simulations of the s15 and s20 progenitors from \cite{2007PhR...442..269W}.
To get a fair comparison, we enable the PD scheme during collapse and use the LS220 EOS. 
Both 2D models explode at $\sim 300$~ms postbounce (see Table~\ref{tab_simulations}).  
The explosion time, $t_{400}$ is defined by the time when the averaged shock radius exceeds $400$~km 
and never recedes at the end of the simulation.  
Overall, the shock-radius evolutions are similar to the models B15-WH07 and B20-WH07 in \cite{2013ApJ...767L...6B} 
except that the {\tt CHIMERA} simulations show earlier explosions.
It should be noted that we use the old neutrino interactions from \cite{1985ApJS...58..771B} 
and that our simulations are Newtonian. 
\cite{2012ApJ...756...84M} have shown that GR effects could enlarge the neutrino luminosities 
and therefore make it easier to explode.
The model s20-2007 in Hanke et al.~(2014) and \cite{2015ApJ...808L..42M} shows a similar explosion time as our model s20,
but the shock radius at $\sim 150$~ms shrinks to $\sim 150$~km due to GR effects.  
The reasons for the differences between {\tt Vertex-Prometheus} and {\tt CHIMERA} are still unclear, 
but the overall features for the progenitor s20 are still rather similar. 
However, the model s15-2007 in Hanke et al.~(2014) shows a very different result.
The shock stalls for $\sim 500$~ms and then explodes at around $\sim 600$~ms. 

On the other hand, 2D {\tt CASTRO} and {\tt ZEUS} Newtonian simulations 
by \cite{2015ApJ...800...10D} and \cite{2014arXiv1406.6414S} 
did not obtain explosion with the s15 and s20 progenitors.  
Our 2D simulations show a fast shock expansion after the prompt convection ($\sim 20$~ms, see Figure~\ref{fig_cc_2d}).
This is similar to what was observed in \cite{2015ApJ...800...10D} but somewhat less dramatic.
The prompt convection and fast shock expansion coincide with an oscillation of the electron antineutrino luminosity 
at $10-20$~ms (see Figure~\ref{fig_cc_2d} and Figure~6 of \citealt{2015ApJ...800...10D}).  
These could be caused by the reduced opacity or incomplete neutrino interactions as discussed before. 
Note that \cite{2015ApJ...800...10D} use the Shen EOS which is considered more difficult to lead to explosions than LS220 
\citep{2013ApJ...765...29C, 2013ApJ...764...99S}.

\cite{2014arXiv1406.6414S} also use the IDSA (without PD) but with spherical coordinates and the ``Ray-by-ray'' approach. 
In principle, we should expect similar results, but the non-explosion of s15 and s20 in \cite{2014arXiv1406.6414S} 
suggest that the different hydrodynamics code, geometry, resolutions, and multidimensional neutrino transport approximation 
may also cause significant differences. 
\cite{2014arXiv1406.6414S} use 300 logarithmically spaced radial zones (from 1km to 5,000~km) 
and $1.4^\circ$ angular resolution. This is roughly three times lower than our simulations. 
A~detailed code comparison is therefore necessary.

%
%


\begin{deluxetable*}{clcccccccc}
\tabletypesize{\scriptsize}
\tablecolumns{10}
\tablecaption{Simulation Parameters and Results \label{tab_simulations}}
\tablehead{ \colhead{Progenitor} \vspace{-0.0cm} & Abbreviation$^{\dagger}$ & ${NT_{\rm collapse}}^A$  & \colhead{EOS} & \colhead{${t_{\rm bounce}}^{B}$} &\colhead{${t_{400}}^{C}$} &\colhead{${t_{\rm end}}^{D}$} & \colhead{${E_{\rm diag}}^{E}$} & \colhead{${M_{\rm PNS}}^{F}$} & \colhead{${R_{\rm PNS}}^{G}$}  \\
 & & & & (ms) & (ms) & (ms) & (B) & ($M_\odot$) & (km)
}
\startdata
\cutinhead{1D}
s15 (W07) \vspace{-0.0cm}& 1D-LA15-07 & IDSA & LS220 & 237 & --- & 763 & --- & 1.85 & 29.1 \\
s15 (W07) \vspace{-0.0cm}& 1D-LP15-07 & PD & LS220 & 249 & --- & 523 & --- & 1.77 & 36.1 \\
s20 (W07) \vspace{-0.0cm}& 1D-LP20-07 & PD & LS220 & 322 & --- & 678 & --- & 1.99 & 30.4 \\
s11.0 (W02) \vspace{-0.0cm}& 1D-LA11  & IDSA & LS220 &  206 &  --- & 794 &--- & 1.48 & 27.5 \\
s15.0 (W02) \vspace{-0.0cm}& 1D-LA15 & IDSA & LS220 &  273 &  --- & 727 &--- & 1.84 & 29.5 \\
s21.0 (W02) \vspace{-0.0cm}& 1D-LA21 & IDSA & LS220 & 274 &  --- & 726 &--- & 1.98 & 30.4 \\
s27.0 (W02) \vspace{-0.0cm}& 1D-LA27 & IDSA & LS220 & 283 &  --- & 717 &--- & 1.81 & 31.2 \\
s11.0 (W02) \vspace{-0.0cm}& 1D-DA11 & IDSA & DD2 &  231 &  --- & 766 &--- & 1.47 & 32.2 \\
s15.0 (W02) \vspace{-0.0cm}& 1D-DA15 & IDSA & DD2 & 307 &  --- & 693 &--- & 1.84 & 34.6 \\
s21.0 (W02) \vspace{-0.0cm}& 1D-DA21& IDSA & DD2 & 310 &  --- & 604 &--- & 1.95 & 37.1 \\
s27.0 (W02) \vspace{-0.0cm}& 1D-DA27 & IDSA & DD2 & 313 &  --- & 687 &--- & 1.81 & 36.1 \\
s11.0 (W02) \vspace{-0.0cm}& 1D-DP11 & PD & DD2 &  203 &  --- & 761 &--- & 1.47 & 33.6 \\
s15.0 (W02) \vspace{-0.0cm}& 1D-DP15 & PD & DD2 & 291 &  --- & 709 &--- & 1.84  & 36.6 \\
s21.0 (W02) \vspace{-0.0cm}& 1D-DP21 & PD & DD2 & 292 &  --- & 708 & --- & 1.98 & 36.6 \\
s27.0 (W02) \vspace{-0.0cm}& 1D-DP27 & PD & DD2 & 295 &  --- & 705 & --- & 1.81 & 37.1 \\
\cutinhead{2D}
s15 (W07) \vspace{-0.0cm}& 2D-LP15-07 & PD & LS220 & 249 & 312 & 524 & 0.298 & 1.69 & 37.1 \\
s20 (W07) \vspace{-0.0cm}& 2D-LP20-07 & PD & LS220 & 324 & 284 & 490 & 0.347 & 1.86 & 40.5 \\
s15.0 (W02) \vspace{-0.0cm}& 2D-LA15 & IDSA & LS220 & 274 & 209 & 311 & 0.464 & 1.60 & 46.8 \\
s15.0 (W02) \vspace{-0.0cm}& 2D-LA15low$^{\dagger \dagger}$ & IDSA & LS220 & 274 & 210 & 369 & 0.523 & 1.62 & 43.5 \\
s11.0 (W02) \vspace{-0.0cm}& 2D-DA11 & IDSA & DD2 & 232 & 86   & 374 & 0.821 & 1.31 & 42.3 \\
s15.0 (W02) \vspace{-0.0cm}& 2D-DA15 & IDSA & DD2 & 308 & 186 & 417 & 0.506 & 1.65 & 43.5 \\
s21.0 (W02) \vspace{-0.0cm}& 2D-DA21 & IDSA & DD2 & 311 & 189 & 411 & 0.635 & 1.75 & 44.8 \\
s27.0 (W02) \vspace{-0.0cm}& 2D-DA27 & IDSA & DD2 & 314 & 162 & 429 & 0.248 & 1.66 & 42.3 \\
s11.0 (W02) \vspace{-0.0cm}& 2D-DP11 & PD & DD2 & 204 & 170 & 376 & 0.267 & 1.37 & 47.4 \\
s15.0 (W02) \vspace{-0.0cm}& 2D-DP15 & PD & DD2 & 292 & 275 & 456 & 0.255 & 1.71 & 45.4 \\
s21.0 (W02) \vspace{-0.0cm}& 2D-DP21 & PD & DD2 & 292 & 282 & 484 & 0.417 & 1.82 & 44.8 \\
s27.0 (W02) \vspace{-0.0cm}& 2D-DP27 & PD & DD2 & 295 & 199 & 484 & 0.157 & 1.70 & 43.5 \\

\enddata
\tablenotetext{A}{ Neutrino transport scheme during the collapse phase \vspace{-0.0cm}}
\tablenotetext{B}{ Bounce time \vspace{-0.0cm}}
\tablenotetext{C}{ Explosion time after bounce \vspace{-0.0cm}}
\tablenotetext{D}{ Termination time after bounce \vspace{-0.0cm}}
\tablenotetext{E}{ Diagnostic explosion energy at the end of the simulation \vspace{-0.0cm}}
\tablenotetext{F}{ PNS mass at the end of the simulation \vspace{-0.0cm}}
\tablenotetext{G}{ PNS radius (determined at the average radius corresponding to $\rho=10^{11}$~g~cm$^{-3}$) at the end of the simulation \vspace{-0.0cm}}
\tablenotetext{$\dagger$}{The model abbreviations are defined by the model's dimensionality, EOS, neutrino transport scheme, and progenitor mass. See Section~\ref{sec_results} for a detailed description. \vspace{-0.0cm}}
\tablenotetext{$\dagger \dagger$}{  
The effective angular resolution in this model is a factor of two lower.
\vspace{-0.0cm}}

\end{deluxetable*}


\section{MULTI-PROGENITOR STUDY} \label{sec_results}

We perform 1D and 2D simulations with $s11.0$, $s15.0$, $s21.0$ and $s27.0$ 
progenitor models from \cite{2002RvMP...74.1015W}.
Simulations start from the prebounce core collapse to several hundred milliseconds postbounce 
with and without the PD in the collapse phase. The former is important to effectively take NES into account. 
Table~\ref{tab_collapse} shows the core properties of these four progenitors during collapse based on 1D simulations.
A summary of all performed simulations is shown in Table~\ref{tab_simulations}.
The model abbreviations in Tables~\ref{tab_collapse} and \ref{tab_simulations} 
are defined by a set of letters and numbers:
The first two characters define the dimension of the model (1D or 2D);
the first letter after the hyphen denotes the EOS of the model (L for LS220 and D for DD2);
the second letter shows the transport scheme during the collapse (A for IDSA and P for PD);
and the last two numbers specify the mass of the investigated progenitor model.
A ``'-07'' at the end shows progenitor models from \cite{2007PhR...442..269W}, 
otherwise they are from \cite{2002RvMP...74.1015W}.
For instance, model 1D-DA15 means a 1D simulation of the $s15.0$ progenitor with DD2 EOS 
and using the IDSA in the collapse phase (i.e.~effectively without NES). 
When we refer to models DA, we consider all models with ``DA'' in its abbreviations.

%
%
\subsection{Stellar Collapse and Core Bounce}

Simulations are started from the non-rotating, solar-metallicity pre-supernova progenitors 
from \cite{2002RvMP...74.1015W} without artificial perturbation.    
\cite{2013ApJ...778L...7C} and \cite{2014arXiv1409.4783M} show that 
small perturbations on the Si/O interface during collapse could amplify post-shock turbulence 
and turn a model that failed to explode towards a successful explosion. 
In addition, \cite{2015arXiv150302199C} performed 3D simulations of the final few minutes of iron core growth in a massive star
that includes Si shell burning.  
The results suggest that the non-spherical progenitor structure may have a strong impact on neutrino-driven explosions.  
In our models, we do not include these non-spherical features, 
and therefore the 2D simulations during collapse show nice agreement with 1D simulations in all models. 
However, we will show that spherical variations due to different levels of electron deleptonization 
or neutrino reactions during collapse may also have a significant impact on neutrino-driven explosions.

\begin{figure}
\begin{center}
\epsscale{1.2}
\plotone{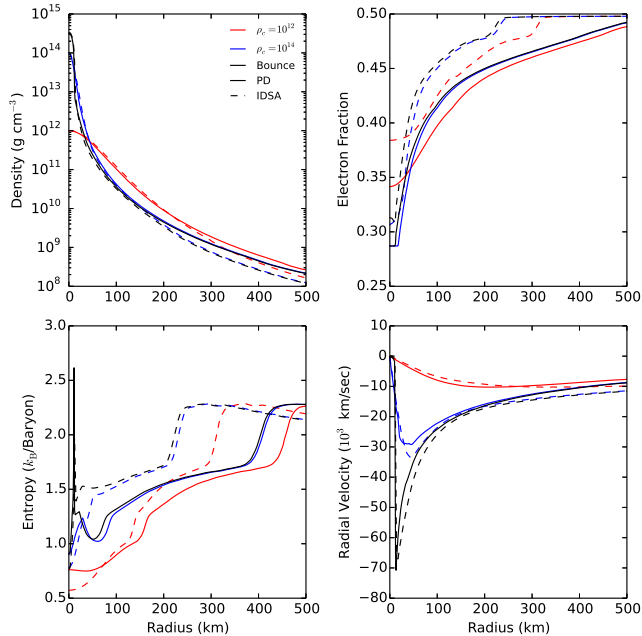}
\end{center}
\caption{\label{fig_collapse_s11} 
Angle averaged radial profiles of progenitor $s11.0$ at different times during core collapse.
All models are using the DD2 EOS. 
Different colors indicate different averaged profiles at certain central densities and bounce.
The solid lines show simulations with the PD, 
and the dashed line show simulations with the IDSA.
 }
\end{figure}

\begin{figure}
\begin{center}
\epsscale{1.2}
\plotone{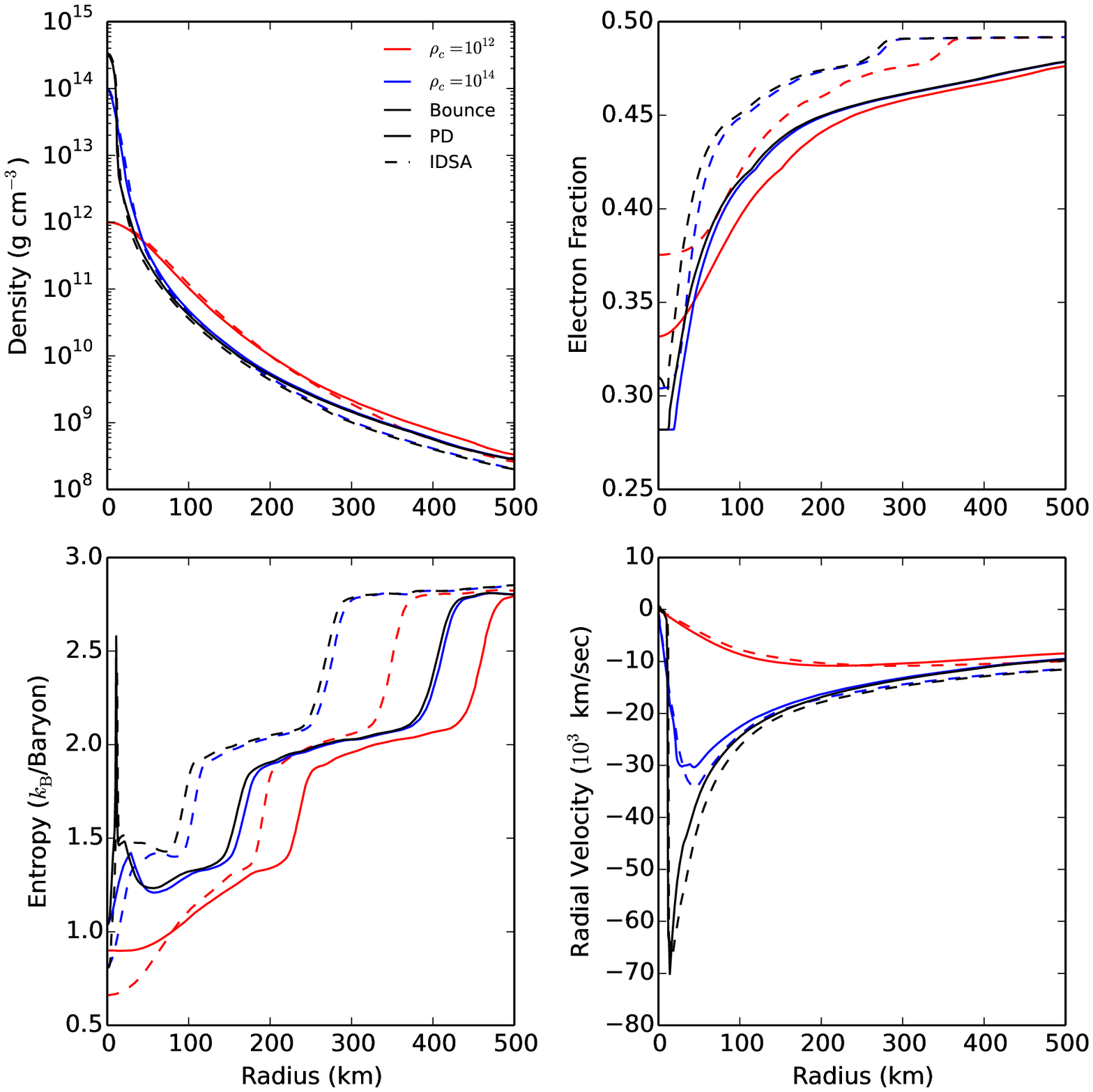}
\end{center}
\caption{\label{fig_collapse_s15} 
Similar to Figure~\ref{fig_collapse_s11} but for progenitor $s15.0$.
 }
\end{figure}

\begin{figure}
\begin{center}
\epsscale{1.2}
\plotone{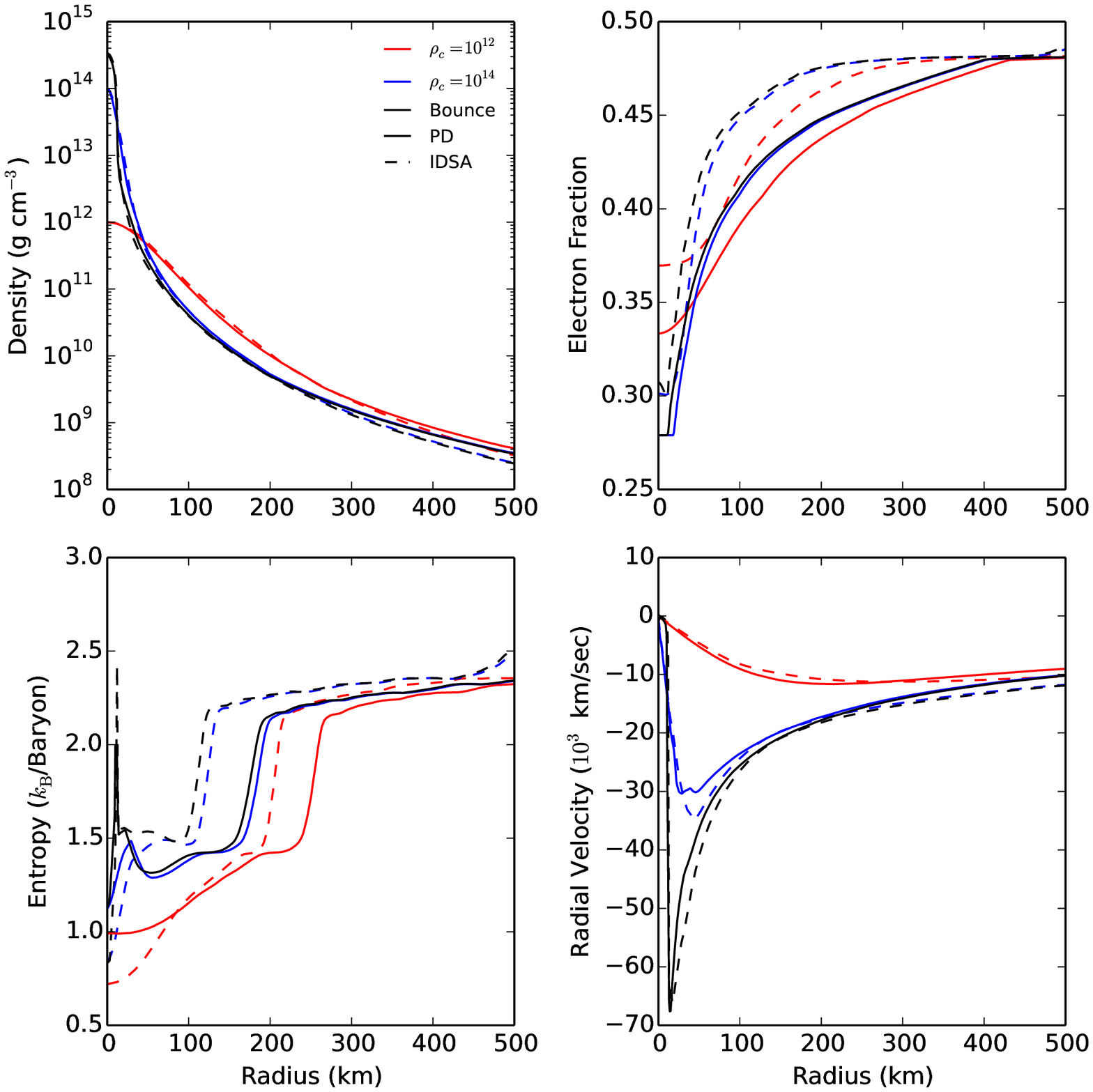}
\end{center}
\caption{\label{fig_collapse_s21} 
Similar to Figure~\ref{fig_collapse_s11} but for progenitor $s21.0$.
 }
\end{figure}

\begin{figure}
\begin{center}
\epsscale{1.2}
\plotone{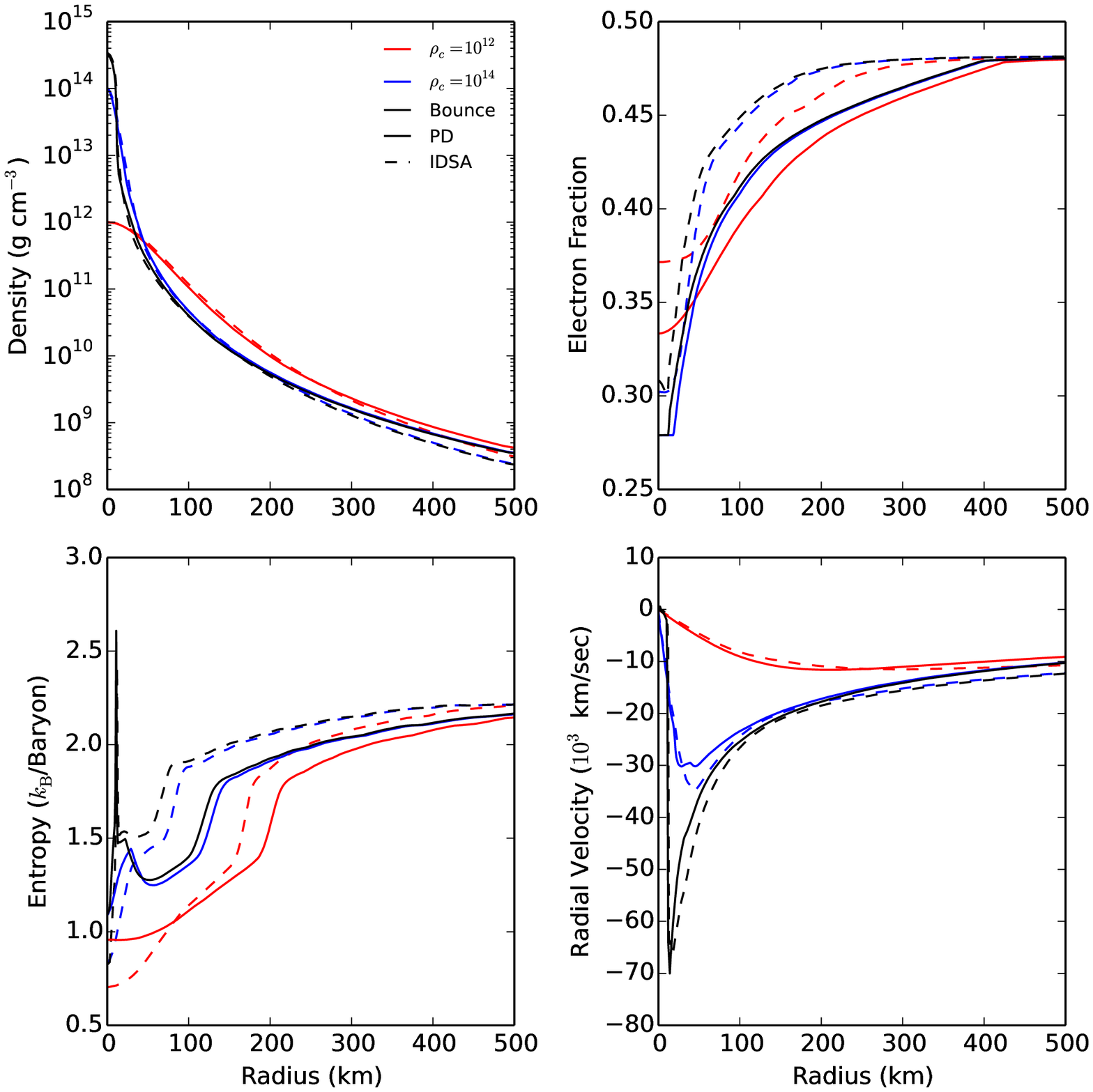}
\end{center}
\caption{\label{fig_collapse_s27} 
Similar to Figure~\ref{fig_collapse_s11} but for progenitor $s27.0$.
 }
\end{figure}

Figures~\ref{fig_collapse_s11}-\ref{fig_collapse_s27} show the radial density, electron fraction, entropy, 
and radial velocity 
evolutions of 2D models with HS(DD2) EOS based on the IDSA or the PD at different time during the collapse phase.
The 1D data are not shown because they are barely distinguishable from the 2D data before bounce.  
In the bounce profiles, the models DP show a slightly lower central density 
but a higher density distribution outside of the iron core, due to an earlier bounce time.

The core evolutions of different progenitors behave qualitatively similar during collapse, 
but show quantitive differences for the central electron fraction, central density, and bounce shock locations. 
Table~\ref{tab_collapse} summarizes important quantities for the four different progenitors during the collapse phase. 
Due to a higher initial central density of the progenitor $s11.0$, 
the central density of $s11.0$ reaches $\rho_c = 10^{11}$~g~cm$^{-3}$ $\sim 100$~ms earlier than other progenitors. 
Once the core densities in different progenitors reach the same $\rho_c$, 
the more massive progenitors collapse faster than other progenitors. 

In addition, since the electrons are highly degenerate and can only gain energy,  
this means that neutrinos lose their energy through NES  and therefore escape more easily, 
accelerating the collapse process \citep{1990RvMP...62..801B}.   
Therefore models DP (with effective NES) collapse faster than models DA (without NES) 
and have a lower $Y_e$ (more $e-$captures), $Y_l$ ($+\nu-$escape) and $\rho_c$ at core bounce.

Core bounce is defined here by the first time 
when the maximum density in the core exceeds $2\times 10^{14}$~g~cm$^{-3}$ 
and the maximum peak entropy is above $3$~k$_{\rm B}$~baryon$^{-1}$.
At bounce, the bounce shock emerges at $\sim 0.55 M_\odot$ 
(defined by the enclosed mass within the shock front at bounce) in models DP 
and at $\sim 0.75 M_\odot$ in models DA. 
Since the core mass is proportional to $Y_e^2$ \citep{1982ASIC...90...53Y}, 
models DA have higher core mass than models DP at bounce.
The highest infall velocity at bounce is about $\sim 70,000$~km~s$^{-1}$ in all progenitor models. 
It should be noted that the Si/O interface, which corresponds to the high entropy gradient region 
at about $100-400$~km in Figures~\ref{fig_collapse_s11}-\ref{fig_collapse_s27}, 
is at different radii for models DA and DP due to different bounce times,
since the bounce time directly sets the location of the Si/O interface. 
These difference will strongly influence the shock evolution after bounce.

%
%

%
%

%
%

%
%

%
%

%
%
\begin{figure*}
\begin{center}
\epsscale{1.0}
\plotone{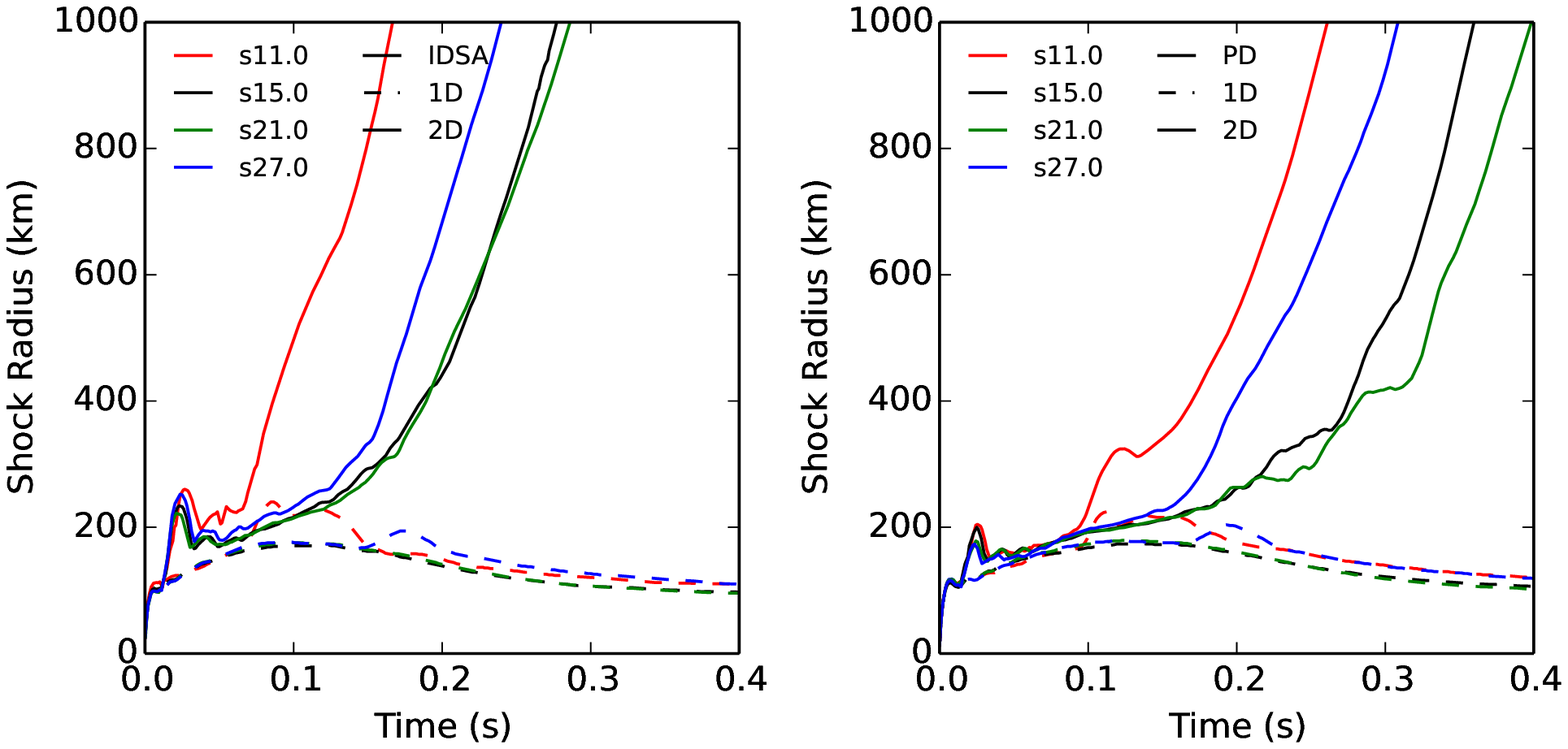}
\end{center}
\caption{\label{fig_2d_shock_radius} 
Average shock radius vs.~postbounce time for different progenitors and neutrino transport approximations. 
All models are using the DD2 EOS. 
Different colors indicate different progenitor models.
The left panel represents simulations without the PD, 
and the right panel represents simulation with an effective inclusion of NES by the PD.   
The solid lines show 2D simulations, and the dashed lines show 1D simulations. } 
\end{figure*}

\subsection{Shock Expansion and Instabilities} \label{sec_shocks}

Figure~\ref{fig_2d_shock_radius} shows the angle averaged shock radius as a function of time 
in our 1D and 2D simulations with the DD2 EOS.
The shock radius of a sample ray is defined here by the highest radius where the entropy is above 
$s_{\rm min} = 4.5$~$k_B$~baryon$^{-1}$ ($s_{\rm min} = 6$~$k_B$~baryon$^{-1}$ 
for progenitor models $s11.0$ and $s27.0$) 
after the postbounce time was reached $t_{\rm pb} = 50$~ms. 
In 2D simulations, we sample 180 radial rays.   
Before $50$~ms postbounce, the shock front is defined by the minimum radial infall velocity.   

1D and 2D models behave very similarly in the first few milliseconds 
until the bounce shock passes through the neutrino sphere at $\sim 10$~ms.
At that time, a prompt entropy- and electron-driven convection occurs and makes the 2D simulations different from 1D. 
Later on, this prompt convection causes a fast shock expansion, and 
changes the shock radius from $r_{\rm sh} \sim 100$~km 
to $r_{\rm sh} \sim 200$~km at $t_{\rm pb} \sim 20$~ms (see Figure~\ref{fig_2d_shock_radius}). 
While the prompt convection is caused by the negative entropy and $Y_e$ gradient, 
it should be noted that this early fast shock expansion during $t_{\rm pb} \sim 10-20$~ms 
may be amplified by our incomplete neutrino interactions and the old neutrino opacities, 
and by the grid-effect in cylindrical coordinates in consideration 
with the multidimensional treatment of neutrino diffusion, 
since a Rayleigh-Taylor bubble along the diagonal axis is observed at that time. 
A similar fast shock expansion is observed in \cite{2015ApJ...800...10D} as well, 
but is not obvious in any other 2D simulation with spherical coordinates.  
However, unlike the results in \cite{2015ApJ...800...10D}, 
the shock radius in our simulations shrinks back to $r_{\rm sh}\sim 150$~km 
after this prompt convection at $t_{\rm pb} \sim 50$~ms. 

%
%
\begin{figure*}
\begin{center}
\epsscale{0.5}
\plotone{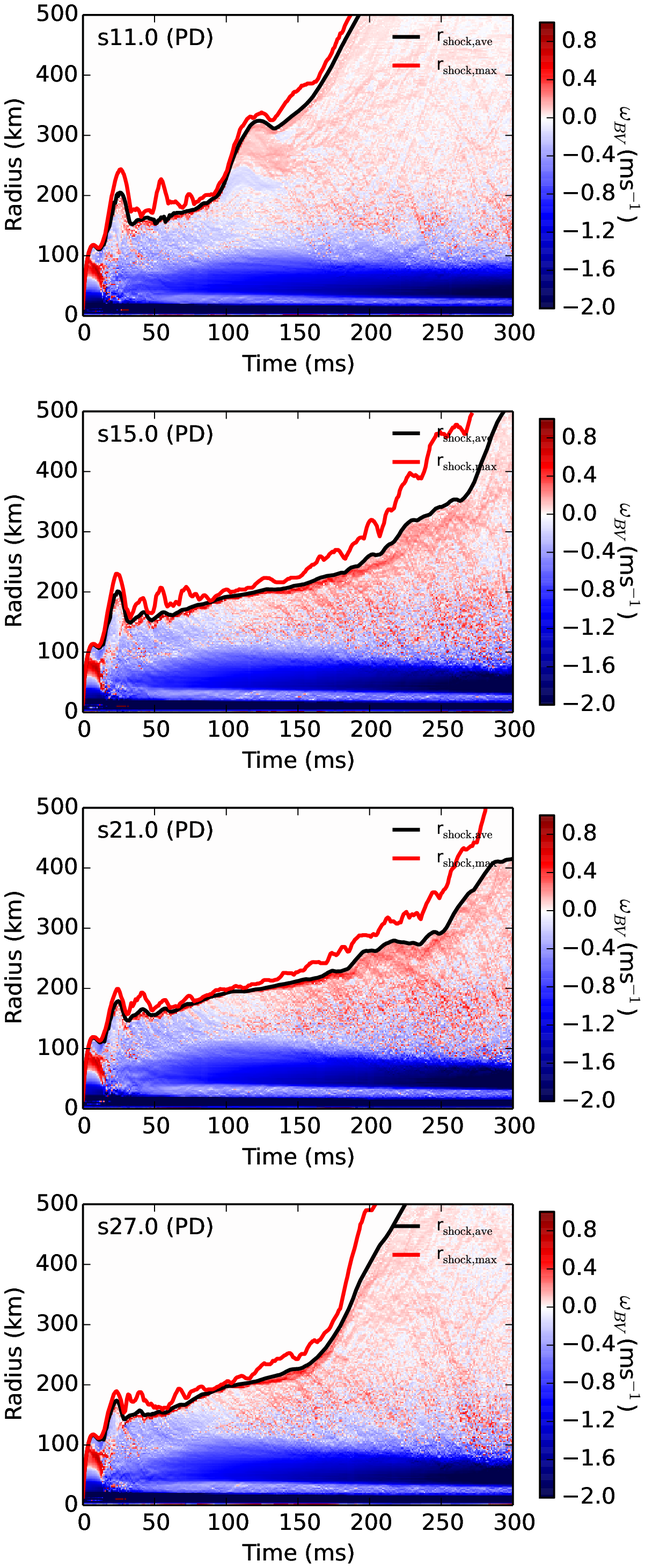}
\plotone{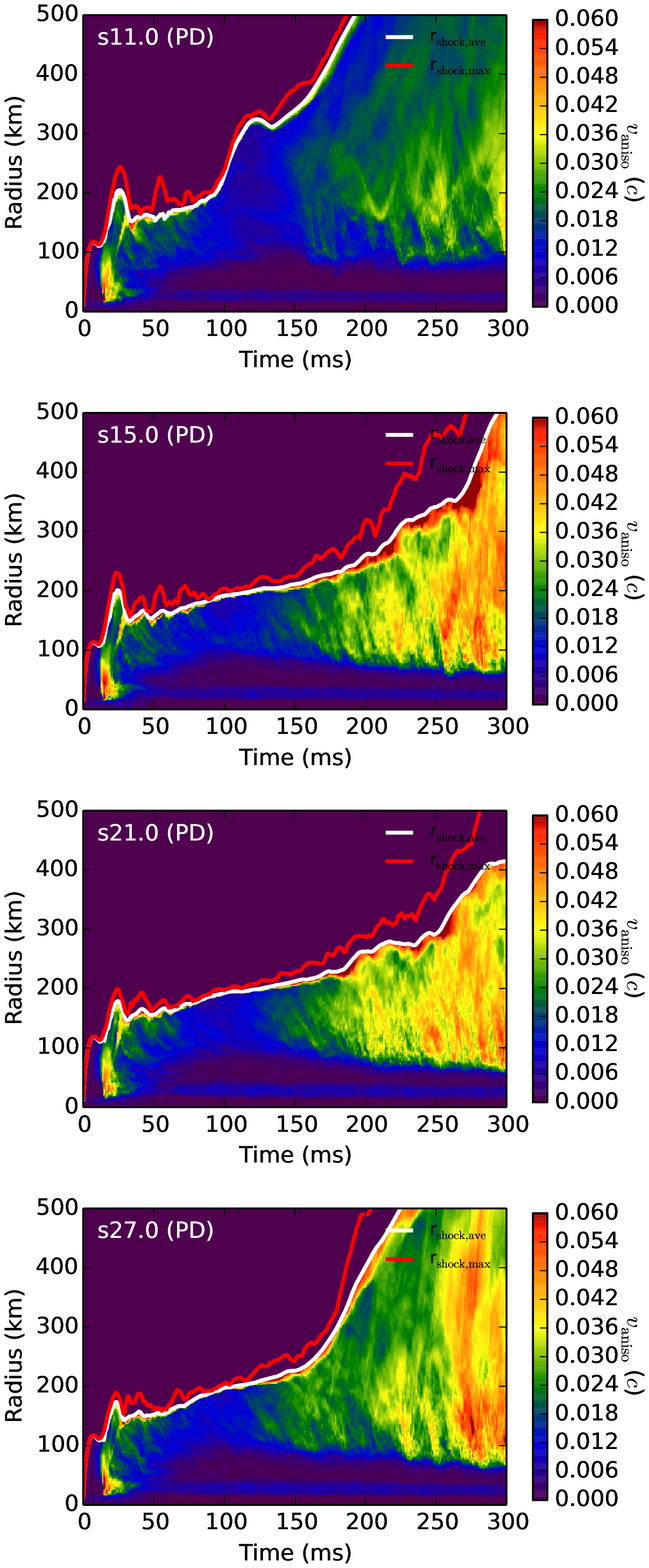}
\end{center}
\caption{\label{fig_bvw_van} 
The color maps show the time evolution of the angle-averaged 
Brunt-V\"{a}is\"{a}l\"{a} (BV) frequency $\omega_{BV}$ in units of ms$^{-1}$ (left) 
and the anisotropic velocity $v_{aniso}$ in units of the speed of light c (right) in our models 2D-DP.
The red lines show the time evolution of the maximum shock radius and 
the black (left) and white (right) lines show the time evolution of the averaged shock radius. }
\end{figure*}

The prompt and late convection can be understood from the local stability analysis 
via the Ledoux criterion \citep{1947ApJ...105..305L},
\begin{equation}
C_L = -\left( \frac{\partial \rho}{\partial P}\right)_{s,Y_l}\left[ \left( \frac{\partial P}{\partial s} \right)_{\rho,Y_l} \left( \frac{d s}{d r} \right) +  \left( \frac{\partial P}{\partial Y_l} \right)_{\rho, s} \left( \frac{d Y_l}{d r} \right) \right],
\end{equation}
where we assume $Y_l \sim Y_e$ for simplicity. 
The Brunt-V\"{a}is\"{a}l\"{a} (BV) frequency, $\omega_{BV}$, describes the  
linear growth frequency for convection. 
Follow the definition of \citep{2006A&A...447.1049B, 2013ApJ...768..115O}, one obtains
\begin{equation}
\omega_{BV} = {\rm sign}(C_L) \sqrt{\left| \frac{C_L}{\rho} \frac{d \Phi}{d r} \right|}, 
\end{equation}
where $\Phi$ is the local gravitational potential and the approximation $d\Phi/dr \sim -GM(r)/r^{-2}$ was used.
Once convection is active, another useful quantity to describe the strength of convention is the anisotropic velocity.
We follow the definition from \cite{2012ApJ...749...98T} and define the anisotropic velocity as  
\begin{equation}
v_{\rm aniso} = \sqrt{\frac{\left< \rho \left[ (v_r - \left< v_r \right>_{4\pi})^2 + v_\phi^2 \right] \right>_{4 \pi}}{\left< \rho \right>_{4\pi}}}.
\end{equation} 

In Figure~\ref{fig_bvw_van}, we show the evolution of the angle-averaged BV frequency 
and anisotropic velocity of our models DP. 
The BV frequencies are positive and high when the shock break through the neutrinosphere 
and the prompt convection happens at $\sim 20$~ms.  
Starting from the core regions, the anisotropic velocities become very strong after the prompt convection has developed, 
and therefore drive the fast-shock expansion at $\sim 20$~ms.
However, the convection stops and the anisotropic velocity returns to small values at $\sim 50$~ms. 
Models DA behave similar but on a different time scale due to different shock evolutions.  
We note that our estimate of $C_L$ may be incorrect at small radii, where neutrinos are trapped, 
due to the ignorance of the contribution from neutrinos in $Y_l$, but the overall features should be the same. 

%
%
\begin{figure}
\begin{center}
\epsscale{1.2}
\plotone{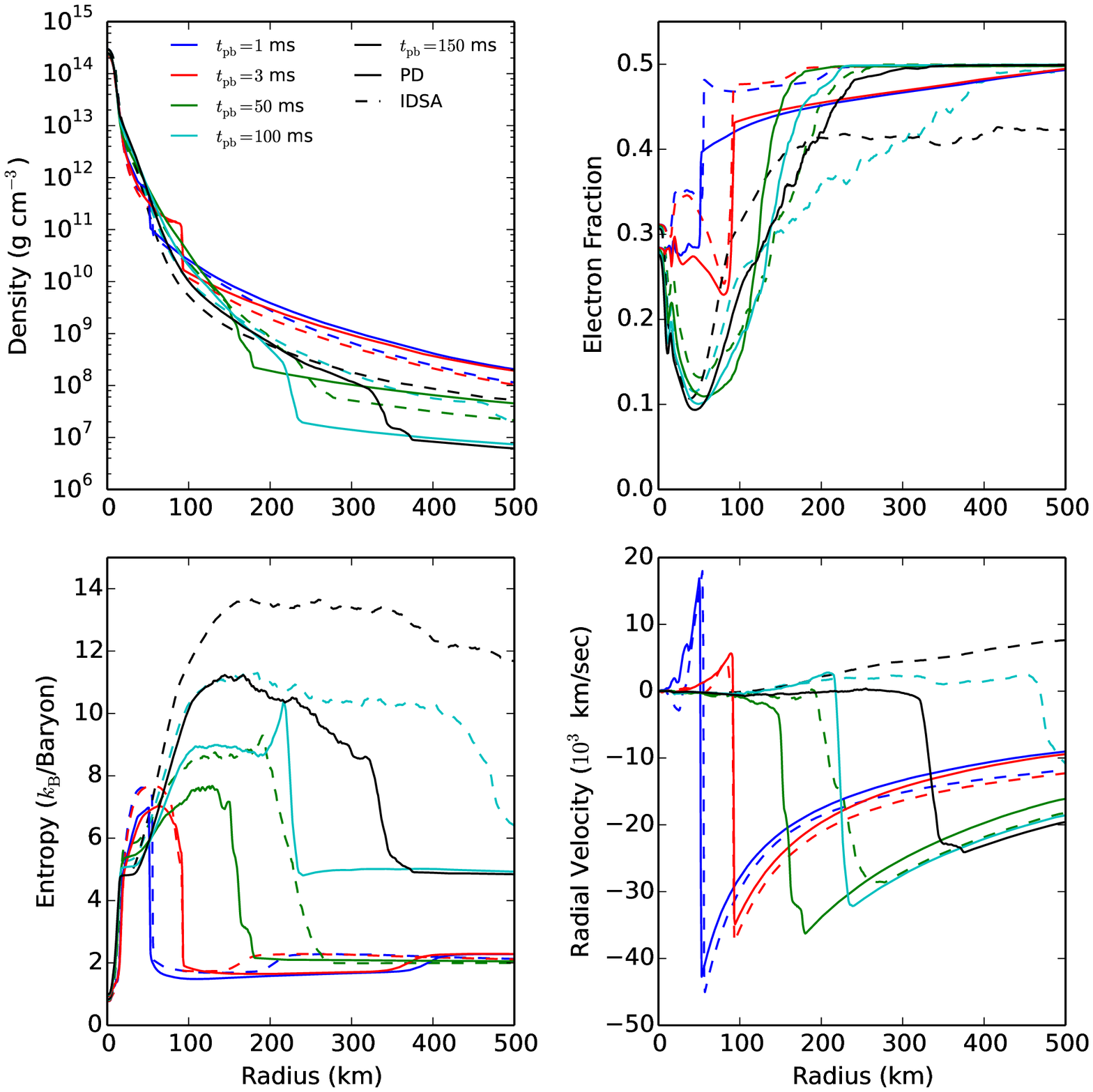}
\end{center}
\caption{\label{fig_postbounce_s11} 
Angle averaged radial profiles of progenitor $s11.0$ at different postbounce times.
All models are evolved in 2D and are using the DD2 EOS. 
Different colors indicate different postbounce times.
The solid lines show simulations with the PD, and the dashed lines show simulations without the PD.
}
\end{figure}

%
%
\begin{figure}
\begin{center}
\epsscale{1.2}
\plotone{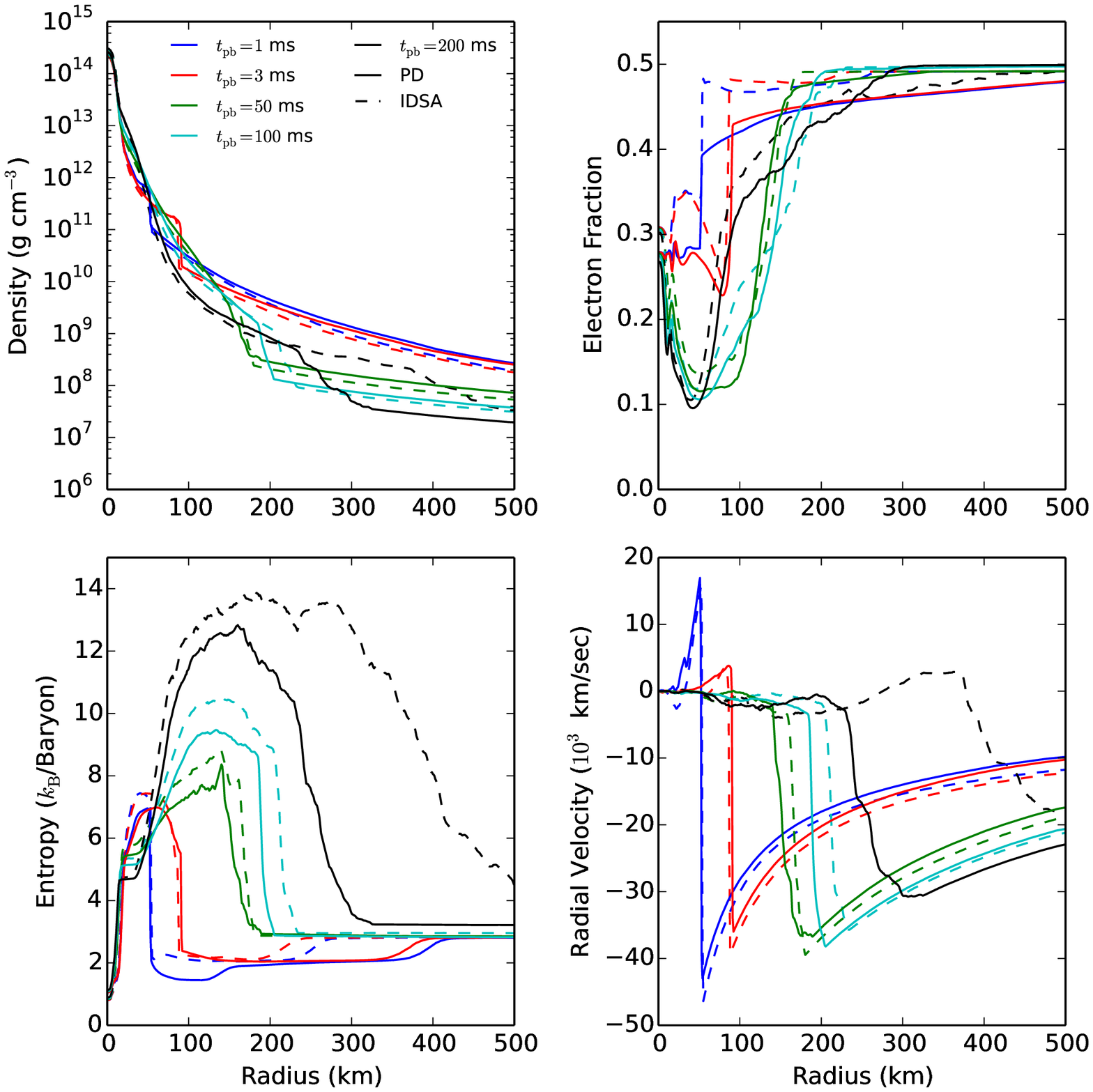}
\end{center}
\caption{\label{fig_postbounce_s15} 
Similar to Figure~\ref{fig_postbounce_s11} but for progenitor $s15.0$.
 }
\end{figure} 

%
%
\begin{figure}
\begin{center}
\epsscale{1.2}
\plotone{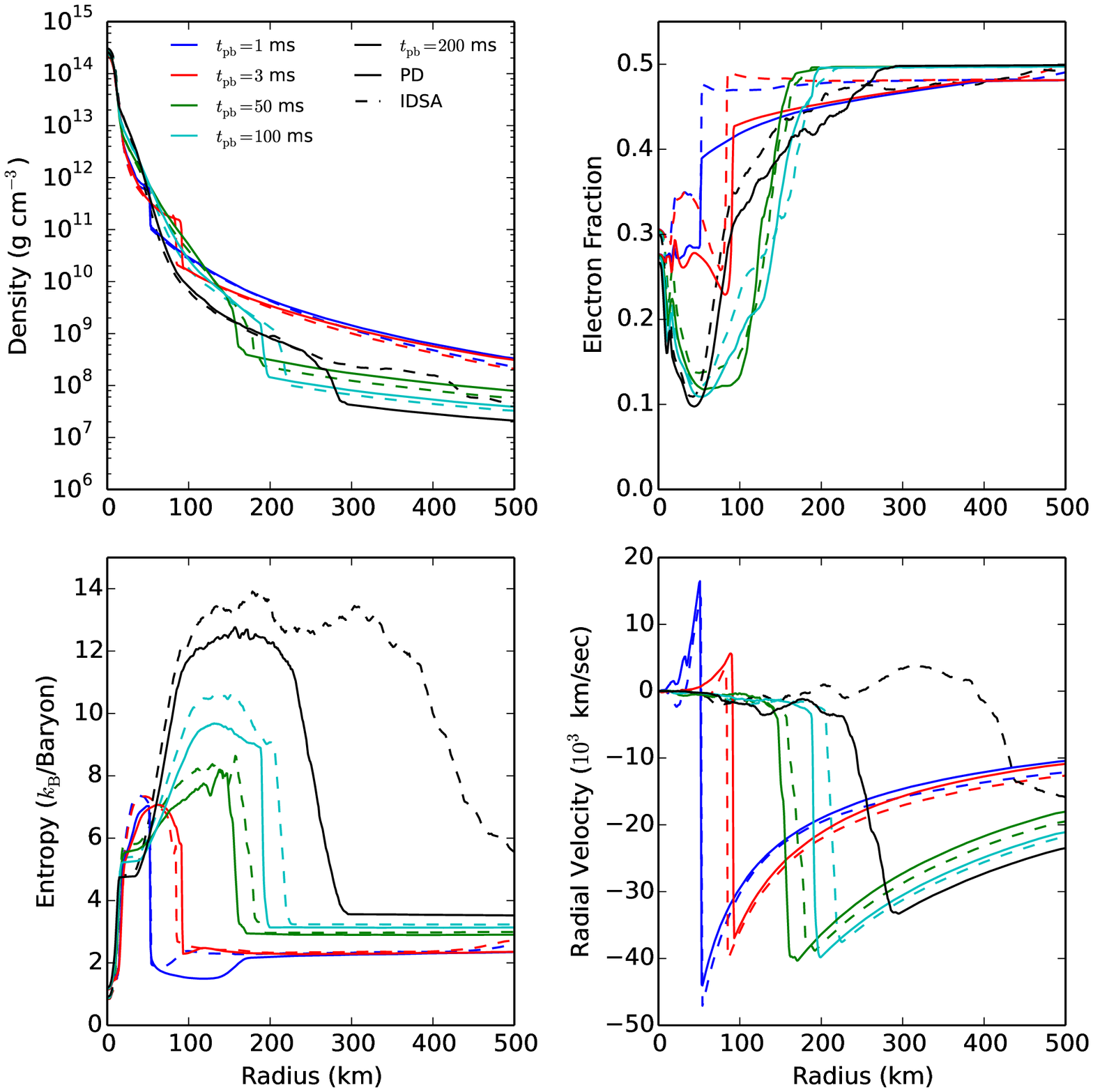}
\end{center}
\caption{\label{fig_postbounce_s21} 
Similar to Figure~\ref{fig_postbounce_s11} but for progenitor $s21.0$.
 }
\end{figure} 

%
%
\begin{figure}
\begin{center}
\epsscale{1.2}
\plotone{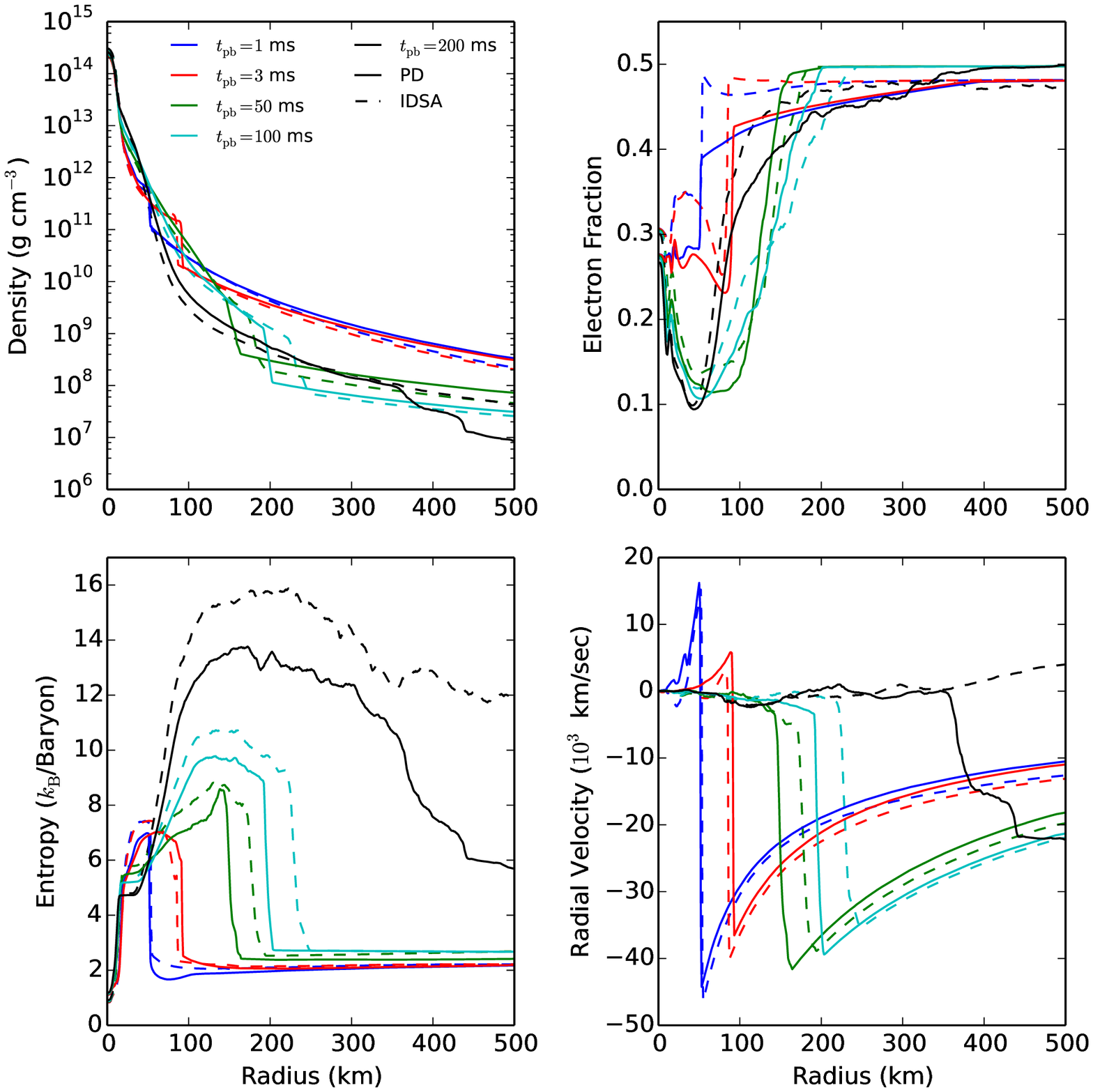}
\end{center}
\caption{\label{fig_postbounce_s27} 
Similar to Figure~\ref{fig_postbounce_s11} but for progenitor $s27.0$.
 }
\end{figure}

For a given progenitor model, the bounce shocks have a similar shock strength in both models DA and DP. 
The difference of shock radius in models DA and DP during the prompt convection 
is mainly due to the difference of the bounce time, 
since different bounce time leads to a different shock strength 
when the bounce shock reaches the location of the negative entropy gradient.
Since convection is suppressed in 1D, we don't observe this prompt convection in 1D models.
However, the prompt convection in 2D models leads to a larger shock radius than in 1D models at $t_{\rm pb} \sim 50$~ms.

After $t_{\rm pb} \sim 50$~ms, the shock stalls at $r_{\rm sh} \sim 150 - 200$~km 
for another $\sim 50 - 100$~ms (except for model 2D-DA11).
It is no surprise that all 1D models fail to explode due to the lack of convection.
However, at first it seems  surprising that all our 2D models explode within $\sim 100-300$ milliseconds postbounce 
regardless of the progenitor mass and the inclusion of NES during collapse. 
But on second thought, one has to consider that our simulations are Newtonian 
(comparing with \citealt{2001ApJ...560..326B, 2001PhRvD..63j3004L, 2012ApJ...756...84M}) 
and do not include the most recent electron capture rates \citep{2003PhRvL..90x1102L, 2003PhRvL..91t1102H}.
In general, Newtonian simulations disfavor explosions compared 
to general relativistic simulations \citep{2012ApJ...756...84M}, 
and also the old neutrino interactions and opacities could make our simulations too optimistic with respect to explosions.   
Furthermore, we are using the DD2 EOS, which has more realistic nuclear matter properties than the LS220 EOS 
that was used in the above mentioned simulations. 
In Appendix~B, we show that DD2 leads in our simulations to more favorable conditions for explosions than LS220.
Table~\ref{tab_simulations} summarizes the essential parameters of our 1D and 2D simulations at the end of the simulations.  
Model 2D-DA11 is the fastest explosion model that explodes already at $t_{400} = 86$~ms by neutrino-driven convection.  
For the other DA models, the explosion time is $t_{400} \sim 160-190$~ms. 
In addition, because of the effective NES, models DP evolve more slowly and explode at a late
$t_{400} \sim 200-280$~ms ($t_{400} = 170$~ms for model 2D-DP11, 
see Table~\ref{tab_simulations} for detailed information).

In model 2D-DP11, the shock radius first expands to $r_{\rm sh} \gtrsim 300$~km at $\sim 100$~ms postbounce 
when it reaches the Si/O interface, but temporarily drops back to $r_{\rm sh} \lesssim 300$~km 
at $\sim 130$~ms postbounce before it explodes.  
Progenitor $s15.0$ and $s21.0$ explode later than progenitor $s11.0$ and $s27.0$ 
but at a similar time in both models 2D-DA and 2D-DP.
In Figures~\ref{fig_postbounce_s11}-\ref{fig_postbounce_s27}, 
we compare the average radial profiles of density, entropy, electron fraction, and
radial velocity at 1, 3, 50, 100, and 200~ms postbounce for models DA and DP. 
We terminate the simulations at $\sim 700-800$~ms postbounce in 1D, and $\sim 300-500$~ms postbounce in 2D.

\begin{figure}
\begin{center}
\epsscale{1.3}  
\plotone{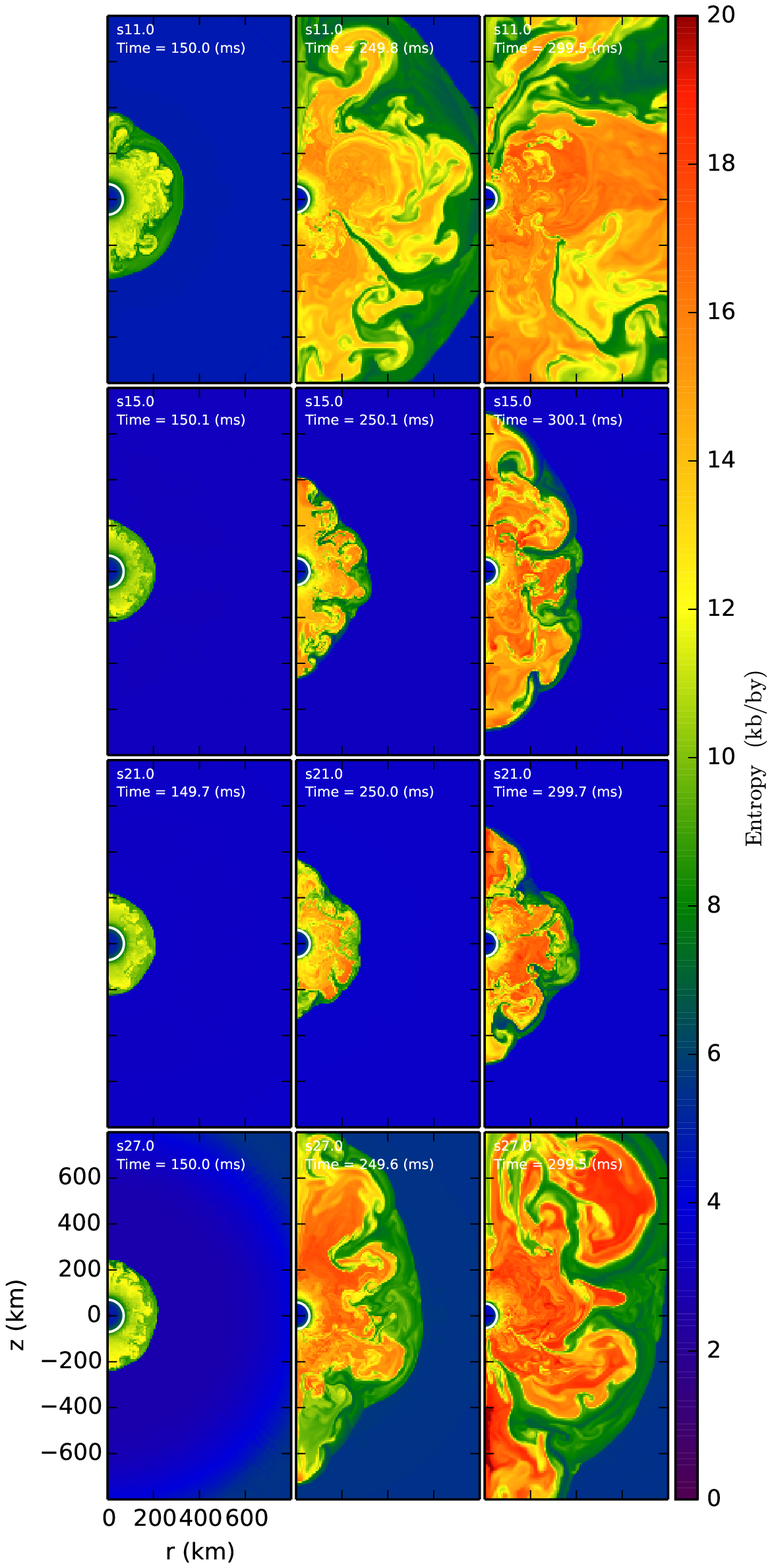}
\end{center}
\caption{\label{fig_2d_entropy_slices} The entropy distributions of our 2D models 
with the PD and the DD2 EOS (See Table~\ref{tab_simulations}). 
Each frame shows a section of the domain spanning 800~km. 
The color scale indicates the entropy in $k_{\rm B}/$baryon.}
\end{figure}

\begin{figure}
\begin{center}
\epsscale{1.2}
\plotone{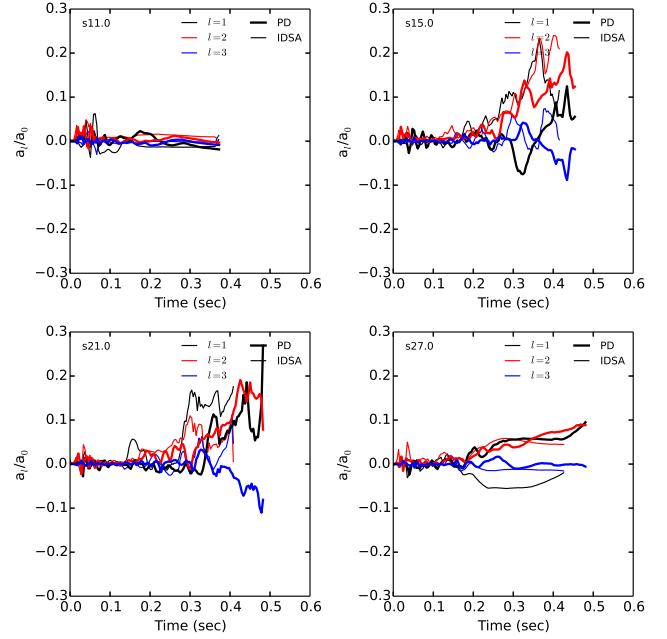}
\end{center}
\caption{\label{fig_2d_sasi} 
First, second, and third coefficients $a_1$, $a_2$, and $a_3$ of the spherical decomposition of the shock radius 
into Legendre polynomials, normalized to the average shock radius ($a_0$) for different progenitors.}
\end{figure}

\begin{figure*}
\begin{center}
\epsscale{1.0}
\plotone{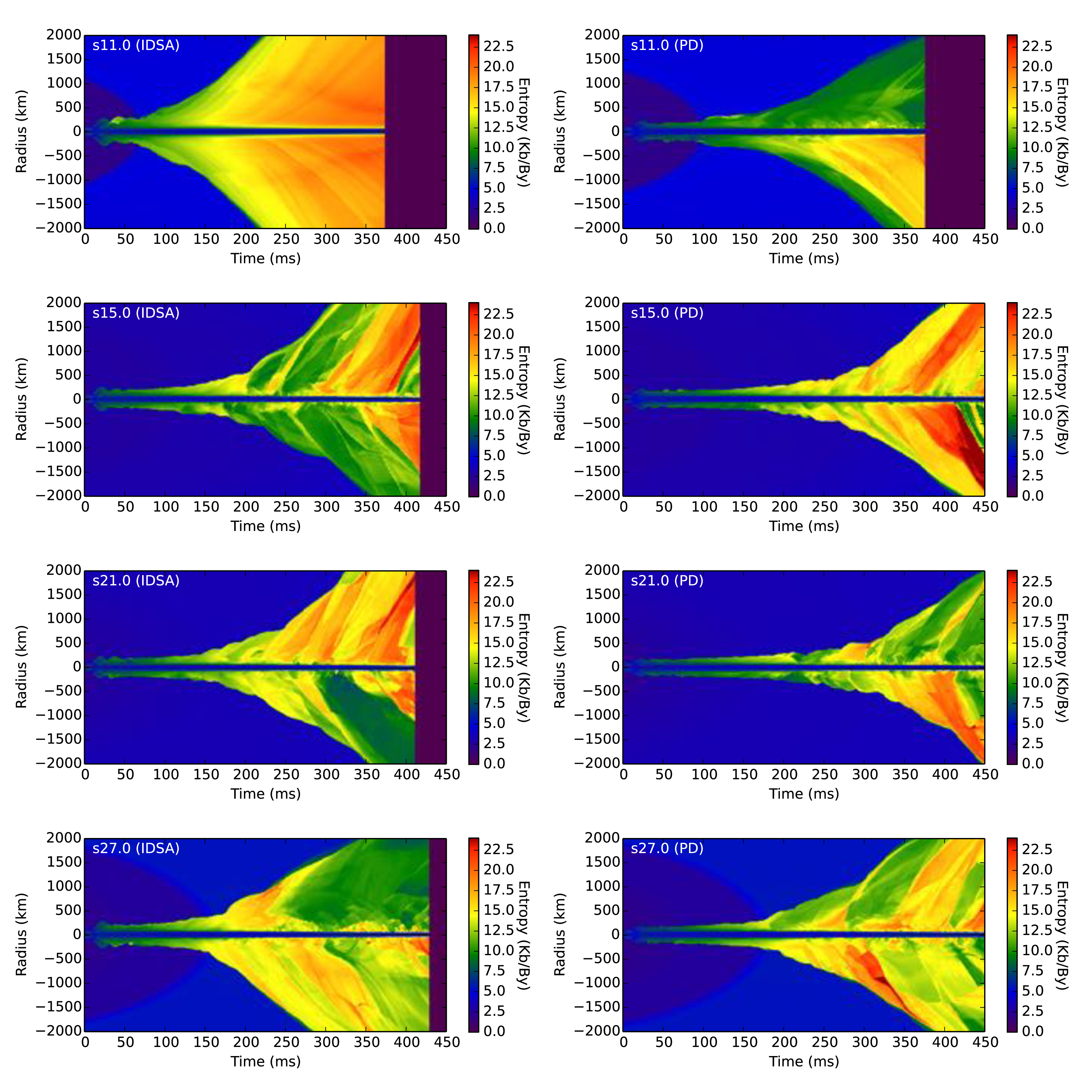}
\end{center}
\caption{\label{fig_2d_entr_timeevo} 
Entropy along the north and the south poles as a function of time for different progenitor masses 
and for models DA (left) and DP (right) in Table~\ref{tab_simulations}. }
\end{figure*}

The 2D entropy distribution of models 2D-DP are shown in Figure~\ref{fig_2d_entropy_slices} 
at 150, 250, and 300~ms postbounce. 
At $\sim 150$~ms, the convection is getting stronger (see Figure~\ref{fig_bvw_van}) but the shock radius distribution is still spherically symmetric.
Later on, at $\sim 200$~ms the SASI starts in models 2D-DP15, 21, and 27 
but the models of progenitor $s11.0$ show mainly convection without SASI. 
The SASI activities can be seen in Figure~\ref{fig_2d_sasi}, where  
we show the normalized coefficients $a_l$ of the decomposition of the shock radius $r_{\rm sh}(\theta)$ 
into Legendre polynomials $P_l$ \citep{2012ApJ...761...72M}. $a_l$ can be calculated by:
\begin{equation}
a_l = \frac{2l+1}{2}\int_0^\pi r_{sh}(\theta) P_l d \cos \theta \label{eq_sasi},
\end{equation}
and $a_0$ corresponds to the averaged shock radius. 
For the progenitor $s11.0$, there is no obvious evidence of SASI activities in both models 2D-DA11 and 2D-DP11.
The amplitude of the normalized coefficients for $l=1$, 2 and 3 modes are small and within the same order of magnitude.  
For progenitors $s15.0$, $s21.0$, and $s27.0$, the SASI activities could be seen and start to grow at $\sim 200$~ms postbounce.
After $\sim 200$~ms postbounce, the dipole ($l=1$) and quadrupole ($l=2$) modes grow to $a_l/a_0 \sim 0.2$ 
in the progenitors $s15.0$ and $s21.0$, and $a_l/a_0 \sim 0.1$ in progenitor $s27.0$.   
The amplitudes do not show significant differences between models DA and DP, 
but the starting time of the high growth rates of the amplitudes corresponds to the time of fast shock expansion.
SASI activities could also be seen in Figure~\ref{fig_2d_entr_timeevo} for the entropy distribution along the north and south poles. 

To verify the code convergence, we have also performed a low-resolution run 
by reducing the angular resolution by a factor of 2 (model 2D-LA15low). 
We find that the explosion time $t_{400}$ is 1ms delayed and the shock expansion evolves slightly slower 
than for the standard resolution run (model 2D-LA15). 
Overall, we find no significant differences between the low-resolution run and the standard run.

%
%

%
%

%
%


%
%

%
%
\begin{figure}
\begin{center}
\epsscale{1.2}
\plotone{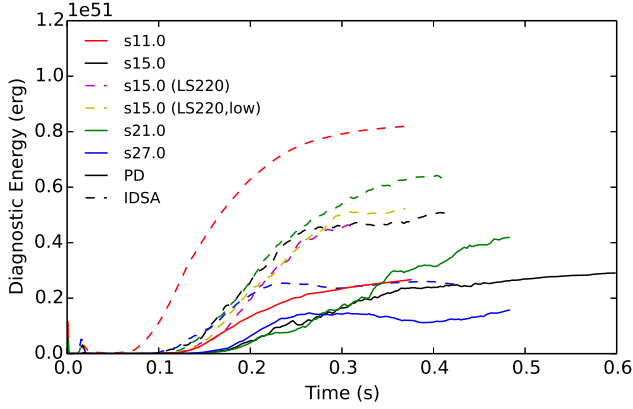}
\end{center}
\caption{\label{fig_2d_explosion}
The diagnostic explosion energies defined in Section~\ref{sec_results} as a function of time.
Different colors represent the different progenitor models. 
Solid lines represent the models 2D-DP and 
dashed lines indicate the models 2D-DA in Table~\ref{tab_simulations}.
The magenta and yellow lines use LS220 EOS, while other lines use DD2 EOS. }
\end{figure}

%
%
\begin{figure}
\begin{center}
\epsscale{1.2}
\plotone{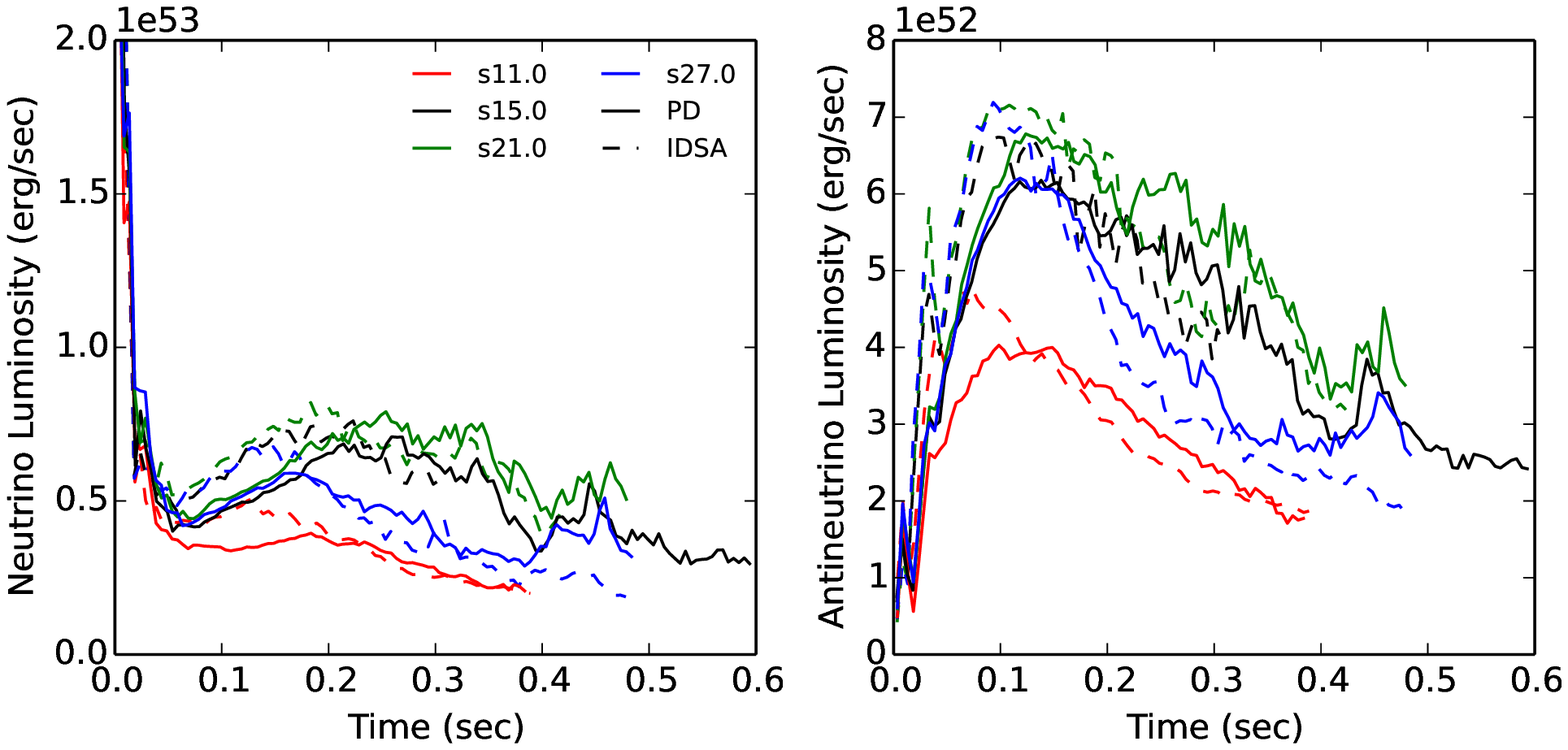}
\plotone{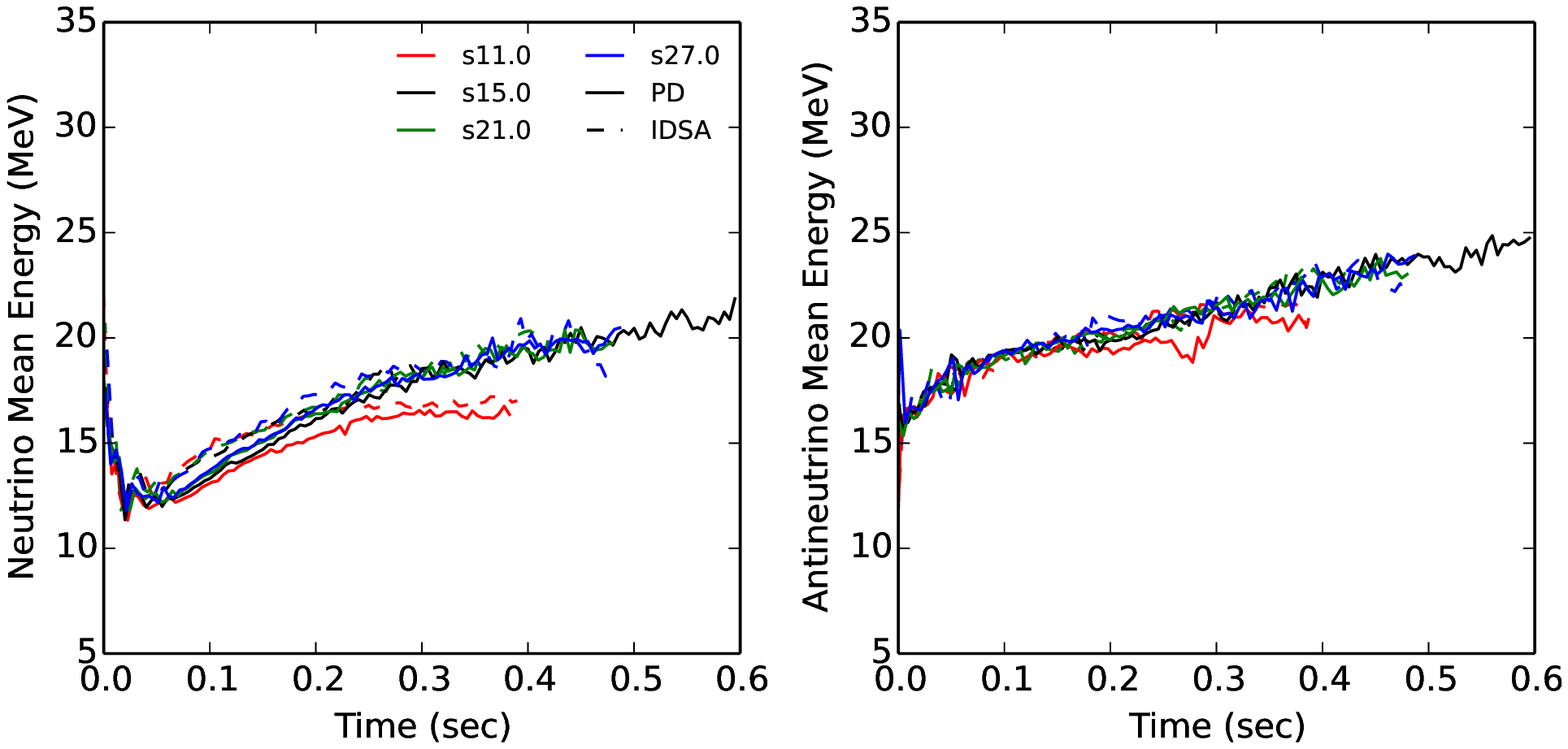}
\end{center}
\caption{\label{fig_2d_luminosity} 
Similar to Figure~\ref{fig_agile_luminosity} 
but for the models DP (solid lines) and the models DA (dashed lines) in Table~\ref{tab_simulations}.}
\end{figure}

%
%
\begin{figure}
\begin{center}
\epsscale{1.3}  
\plotone{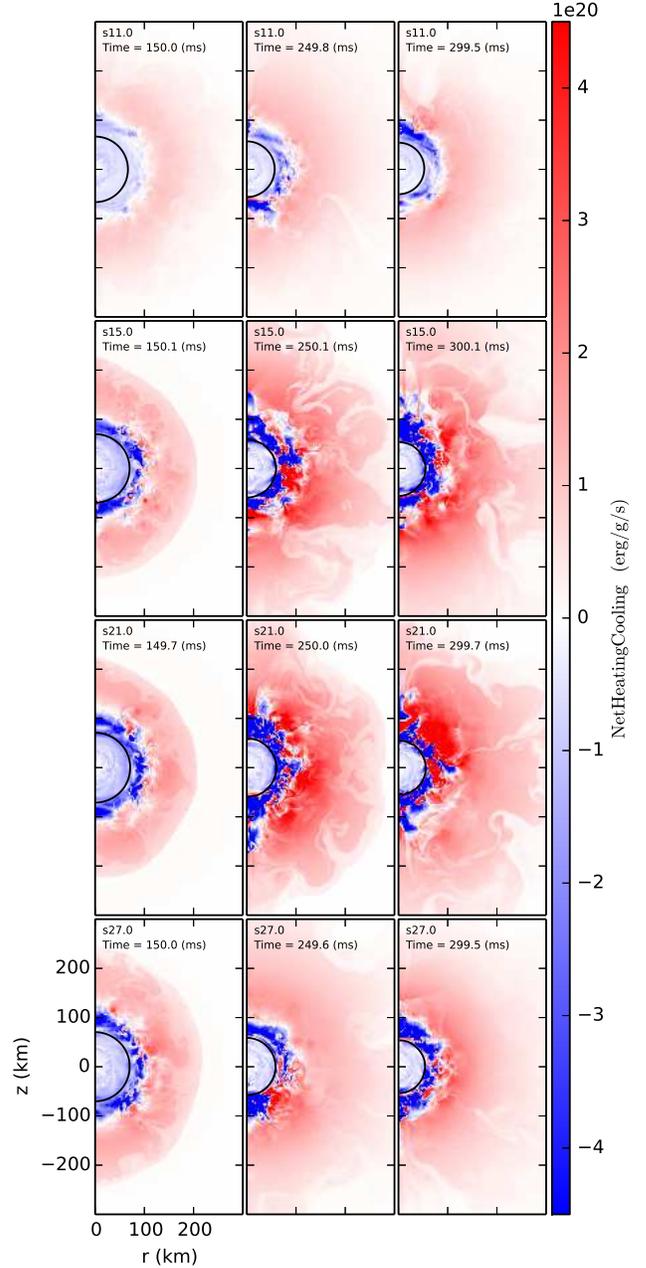}
\end{center}
\caption{\label{fig_2d_heating_slices} The net heating (red) and cooling (blue) distributions of our models 2D-DP (See Table~\ref{tab_simulations}). The solid black lines show the radius of the PNS.}
\end{figure}

%
%
\begin{figure}
\begin{center}
\epsscale{1.2}
\plotone{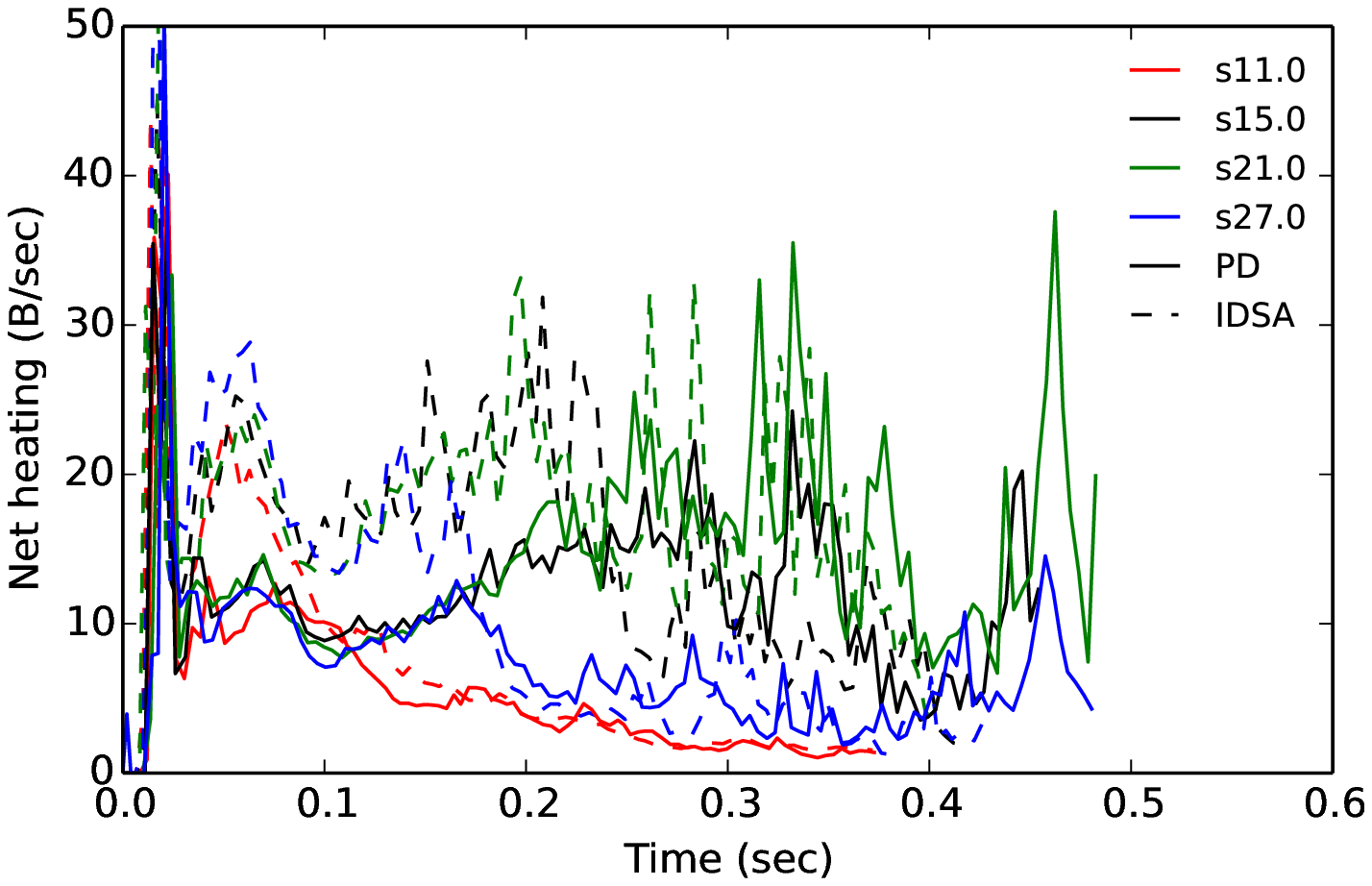}
\plotone{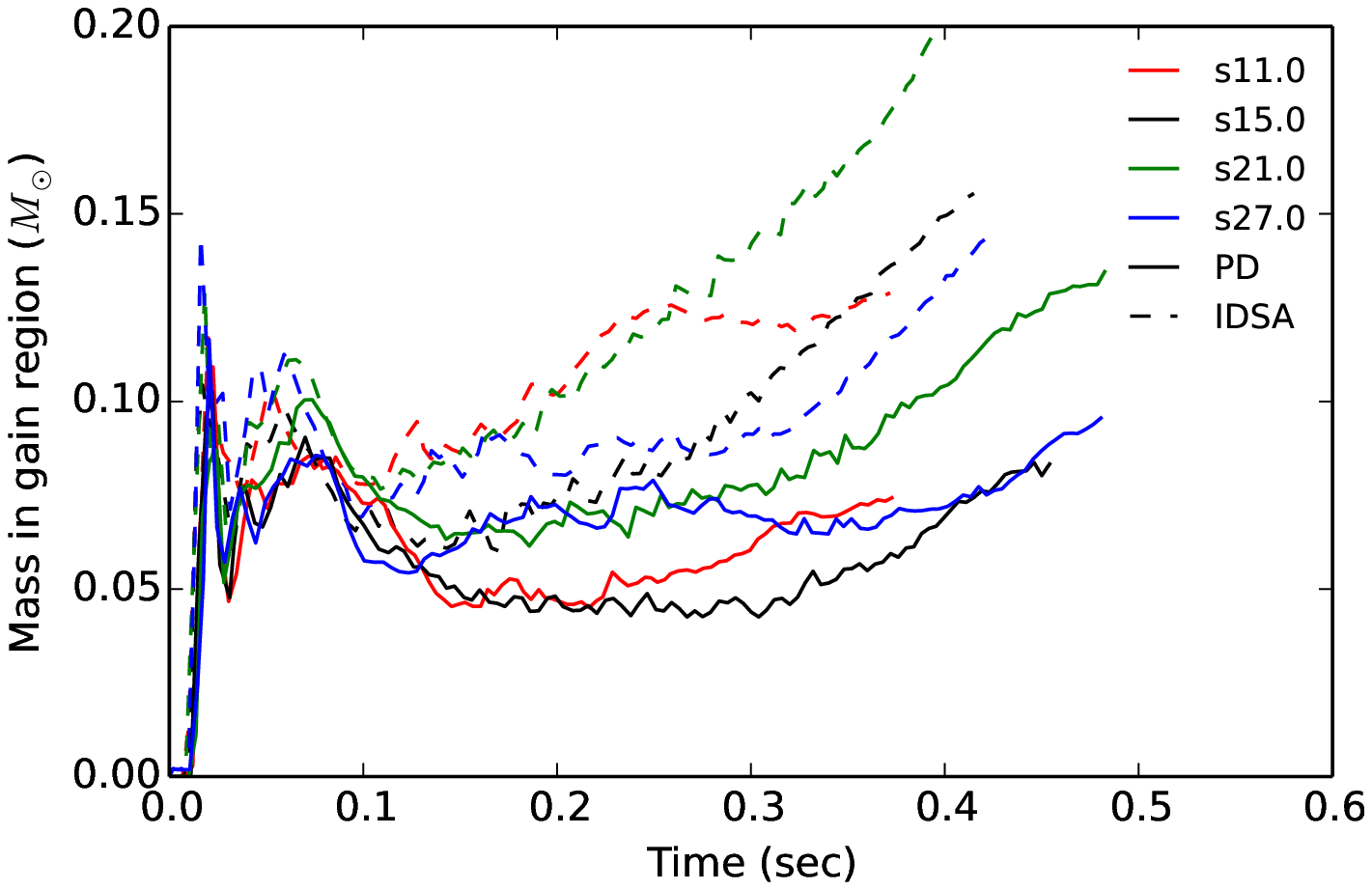}
\plotone{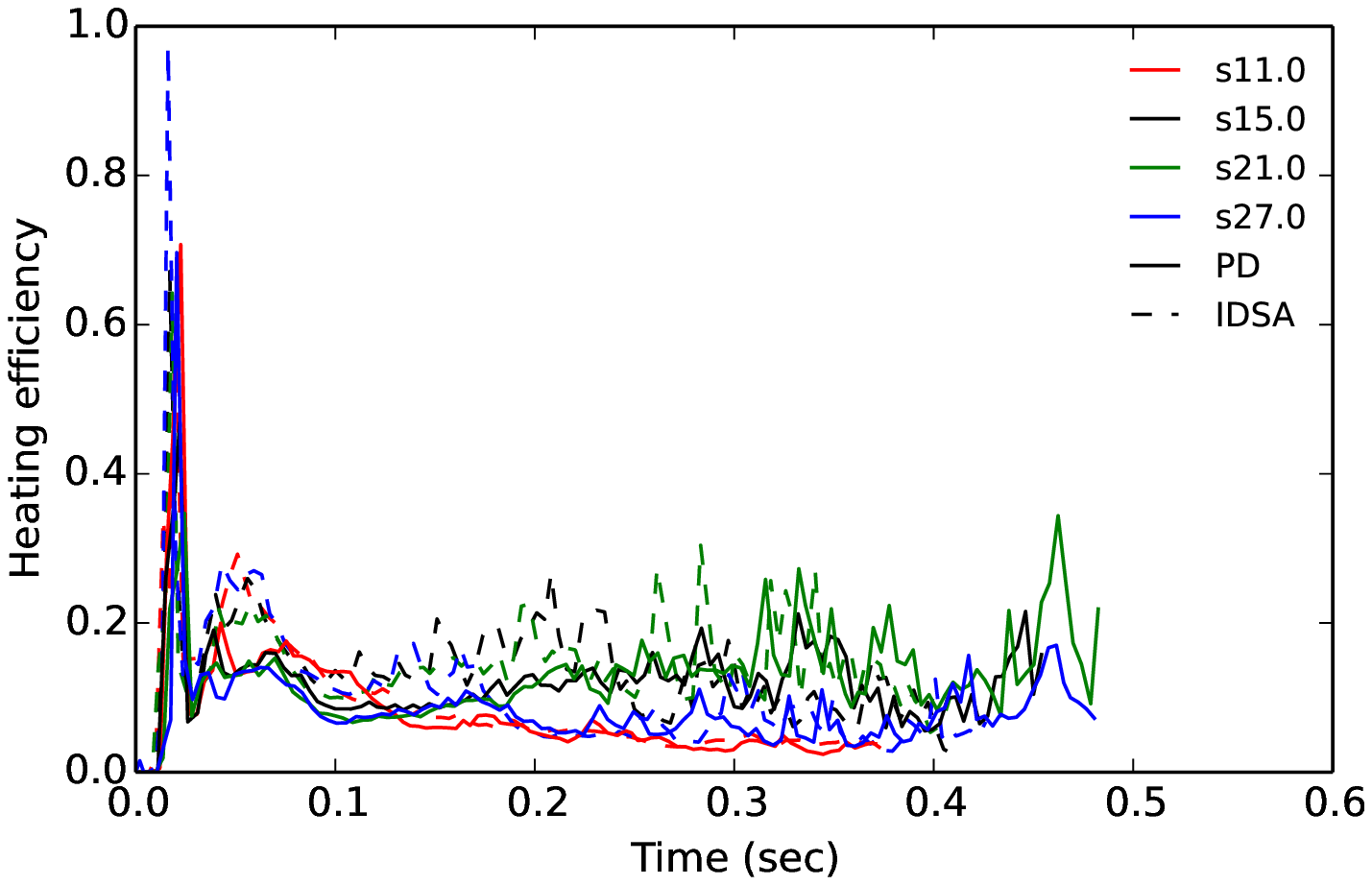}
\end{center}
\caption{\label{fig_2d_gain_mass} 
The integral quantities of net heating (top), mass in the gain region (middle), and heating efficiency (bottom) as functions of time.
See Section~\ref{sec_results} for the definition of these integral quantities. 
Different colors represent different progenitor models. Solid lines represent the models 2D-DP and 
dashed lines indicate the models 2D-DA in Table~\ref{tab_simulations}. }
\end{figure}

\subsection{Neutrino Heating and Explosion Energy}

The energy released in neutrinos is about $10^{53}$~erg in a typical CCSN \citep{1966ApJ...143..626C}. 
To power the observed kinetic energy of a CCSN ($\sim 10^{51}$~erg $\equiv 1 B$),
the baryonic matter has to absorb $\sim 1\%$ of the neutrino energy \citep{1990ApJ...350L..33B}. 
However, the explosion energies in most published 2D models are still lower than the standard $1B$ 
(except for those of \citealt{2013ApJ...767L...6B, 2014arXiv1409.5779B}).
The real explosion energy is not straightforward to calculate and the definition may differ from group to group.
Figure~\ref{fig_2d_explosion} shows the ``diagnostic energy'' of our models 2D-DA and 2D-DP. 
The diagnostic energy, $E_{\rm dia}$, is defined by
\begin{equation}
E_{\rm dia} = \int_{e_{\rm tot} >0} \rho e_{\rm tot} dV,
\end{equation} 
where the volume integration is performed over the region where the total specific energy, $e_{\rm tot}$,
\begin{equation}
e_{\rm tot} \equiv (e - e_0) + \frac{1}{2} \bm{v}^2 + \Phi 
\end{equation}
is positive, 
$e$ represents the specific internal energy (thermal energy plus binding energy),
and $\frac{1}{2} \bm{v}^2$ and $\Phi$ are the specific kinetic and gravitational energies, respectively. 
$e_0$ is the reference energy value that is defined by the minimum specific internal energy at the beginning of the simulation, 
which leads to a negligible diagnostic explosion energy at beginning.
We have checked that alternative values for the reference energy, that are suggested in the literature, 
do not have a significant effect on our final results.
Depending on progenitor and simulation domain, $e_0$ is about $\sim 10^{17}$~erg~g$^{-1}$. 
We find that the diagnostic energy is insensitive to $e_0$ if we make it a few times larger or smaller.    
Note that this expression does not take into account properly the different binding energy contributions 
from different nuclei to thermal energy. Therefore the so-called recombination energy is not included.
$e$ corresponds only to the thermal energy that actually should be used in $E_{\rm dia}$, 
if the composition was dominated by $^{12}C$, or if the composition has a similar average binding energy like $^{12}C$.
The diagnostic explosion energies (Figure~\ref{fig_2d_explosion}) of our models 2D-DP are between $0.1 B$ and $0.4 B$
at $\sim 400$~ms postbounce and are still increasing at the end of our simulations. 
The models 2D-DA, which ignored the NES in the collapse phase, have a much stronger explosion energy, 
$E_{\rm exp} \sim 0.2 - 0.8 B$ at $\sim 400$~ms postbounce.
This result is consistent with the 1D results in \cite{1989ApJ...341..385B}.
It should be noted that the only difference between models 2D-DA and 2D-DP is the deleptonization method during collapse.
The postbounce physics employed are identical in both models 2D-DA and 2D-DP,  
suggesting that the initial conditions at bounce may dramatically affect the postbounce evolutions. 
Furthermore, note that it is difficult to compare the results of models DP with other groups 
because there are differences in the input physics and methods.  
Additionally, we show the diagnostic explosion energies of models 2D-LA15 and 2D-LA15low 
for a comparison with LS220 EOS and with low resolution. 
We find that the model with LS220 EOS leads to a slight lower diagnostic explosion energy due to a delay of the explosion.
The low resolution model also gives a similar results, 
suggesting that the fast explosion in our models is not due to insufficient resolution (Figure~\ref{fig_2d_explosion}).

\cite{2012ApJ...757...69U, 2014arXiv1406.2415N}, and \cite{2015arXiv150102845P} suggest that 
there is a correlation of the compactness parameter with the explosion energy. 
We do see this trend in our simulations except for the progenitor $s11.0$, 
if we define the compactness parameter at an enclosed mass of $1.75 M_\odot$,
i.e. $\xi_{1.75} \equiv \frac{1.75}{R(M=1.75M_\odot)/1000{\rm km}}$ \citep{2011ApJ...730...70O}.
In our simulations, the progenitor $s11.0$, which has the lowest compactness parameter,  
has the second highest diagnostic explosion energy (see Tables~\ref{tab_collapse} and \ref{tab_simulations}). 
However this trend in our simulations will disappear if we use $\xi_{2.5}$. 
It should be noted that we only have four progenitor models in our simulations,  
and the correlation found in \cite{2012ApJ...757...69U} and \cite{2015arXiv150102845P} 
show a huge scatter between different models,
indicating that the compactness parameter oversimplifies the complexity of the explosion mechanism.

Figure~\ref{fig_2d_luminosity} shows the neutrino luminosity and mean energy of electron type neutrinos 
and anti-neutrinos as functions of time.  
The neutrino mean energy, $E_\nu$ is defined by 
$E_\nu = \frac{\int \epsilon^3 f(\nu, \epsilon) d\epsilon}{\int  \epsilon^2 f(\nu, \epsilon) d\epsilon}$,
where $f$ is the distribution function. 
The neutrino luminosities in models DA after $\sim 50$~ms grow faster than in models DP  
but the peak luminosities after $\sim 100$~ms are similar in both DA and DP models.  
The neutrino mean energies show a similar trend but the differences are less pronounced. 
We note that the neutrino luminosities and mean energies in our simulations are of a similar order of magnitude  
as the typical results in the literature \citep{2012ARNPS..62..407J}.  
Therefore, the high explosion energies in our simulations require a larger mass in the gain region or
a higher heating efficiency than other investigators. 

Figure~\ref{fig_2d_heating_slices} presents the 2D net heating and cooling rates in models 2D-DP.
The red color shows the heating region and the blue color indicates the cooling region.   
The PNS radius (defined by the average radius of density $\rho = 10^{11}$~g~cm$^{-3}$) 
is plotted as well in Figure~\ref{fig_2d_heating_slices}. 
The gain region is defined as the post-shock region with positive net heating. 
In Figure~\ref{fig_2d_gain_mass}, we show the integrated quantities of net heating ($Q_{\rm net}$), 
total mass in the gain region, and the heating efficiency as functions of time. 
The heating efficiency is defined by $\eta_{\rm heat} = Q_{\rm net}/ (L_{\nu_e,{\rm gain}} + L_{\bar{\nu}_e,{\rm gain}})$, 
since the heating originates mainly from the electron type neutrinos. 

The early prompt convection shows a very high heating efficiency peak at $\sim 20$~ms, 
which could be associated with a grid-effect in cylindrical coordinates. 
However, this prompt heating does not affect the explosion energy directly 
since matter is still bound at that time. 
The shock expansion and the early convection at $t_{\rm pb} \sim 50$~ms enlarge the gain radius 
and the mass in the gain region. 
After $t_{\rm pb} \sim 100-200$~ms, the mass in the gain region continues to increase in time. 
We noted that the masses in the gain region ($\sim 0.1 M_\odot$) in our simulations are higher 
than the values that are reported by other groups, 
explaining the high diagnostic explosion energy in our simulations.

The models 2D-DA11 and 2D-DP11 have the lowest net heating and heating efficiency among the models 
(except during the prompt convection period), but have the second highest mass in the gain region and diagnostic energy. 
This is probably due to the fast shock expansion (see Figure~\ref{fig_2d_shock_radius}) 
and low mass-accretion in the progenitor $s11.0$.


%
%
%
%

\begin{figure}
\begin{center}
\epsscale{1.2}
\plotone{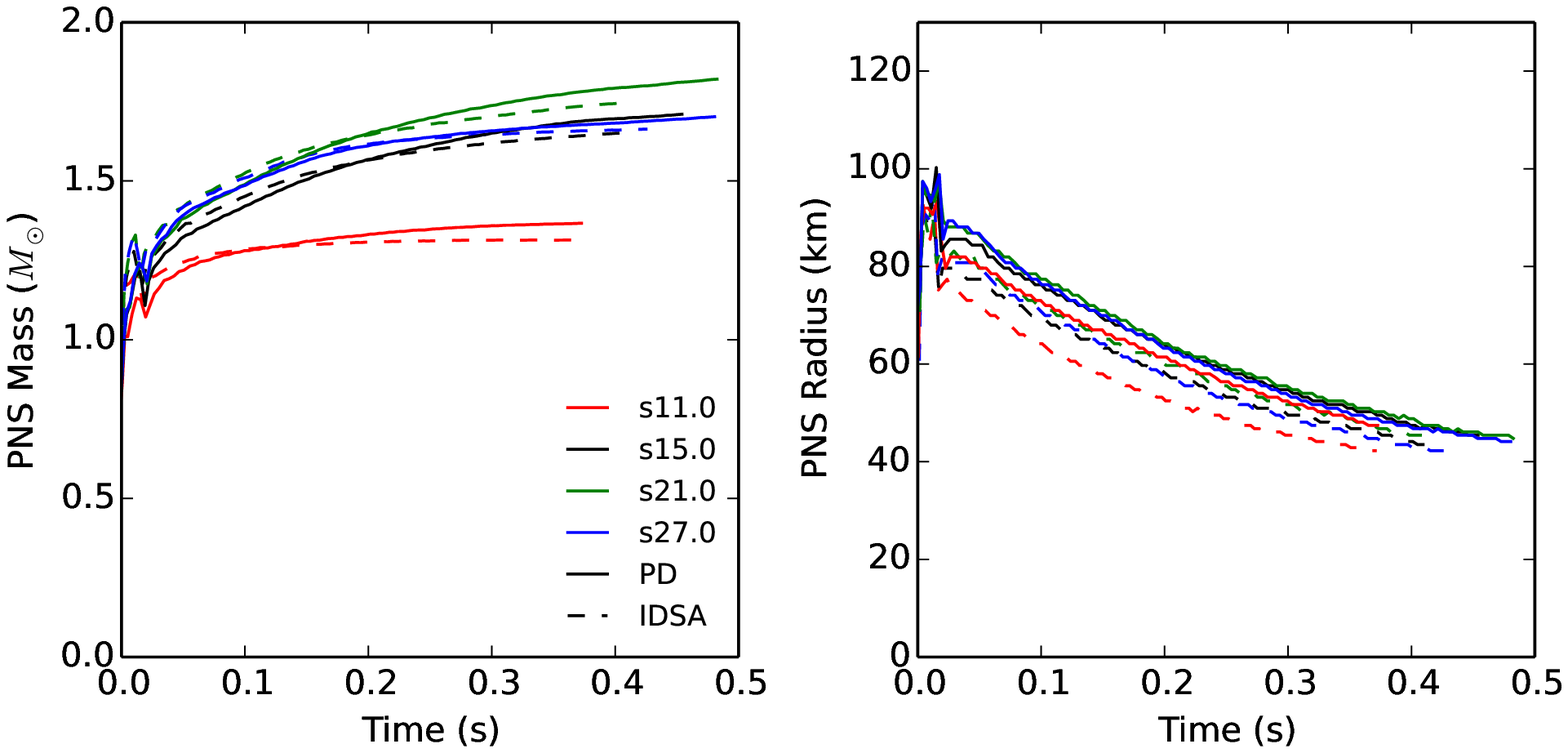}
\end{center}
\caption{\label{fig_2d_pns} The time evolution of the PNS mass (left) and radius (right). 
The PNS mass and radius are determined at the average radius corresponding to $\rho = 10^{11}$~g~cm$^{-3}$.
Different colors represent different progenitor stars. 
Solid lines show the models 2D-DP and dashed lines show the models 2D-DA in Table~\ref{tab_simulations}.}
\end{figure}

%
%
\begin{figure}
\begin{center}
\epsscale{1.2}
\plotone{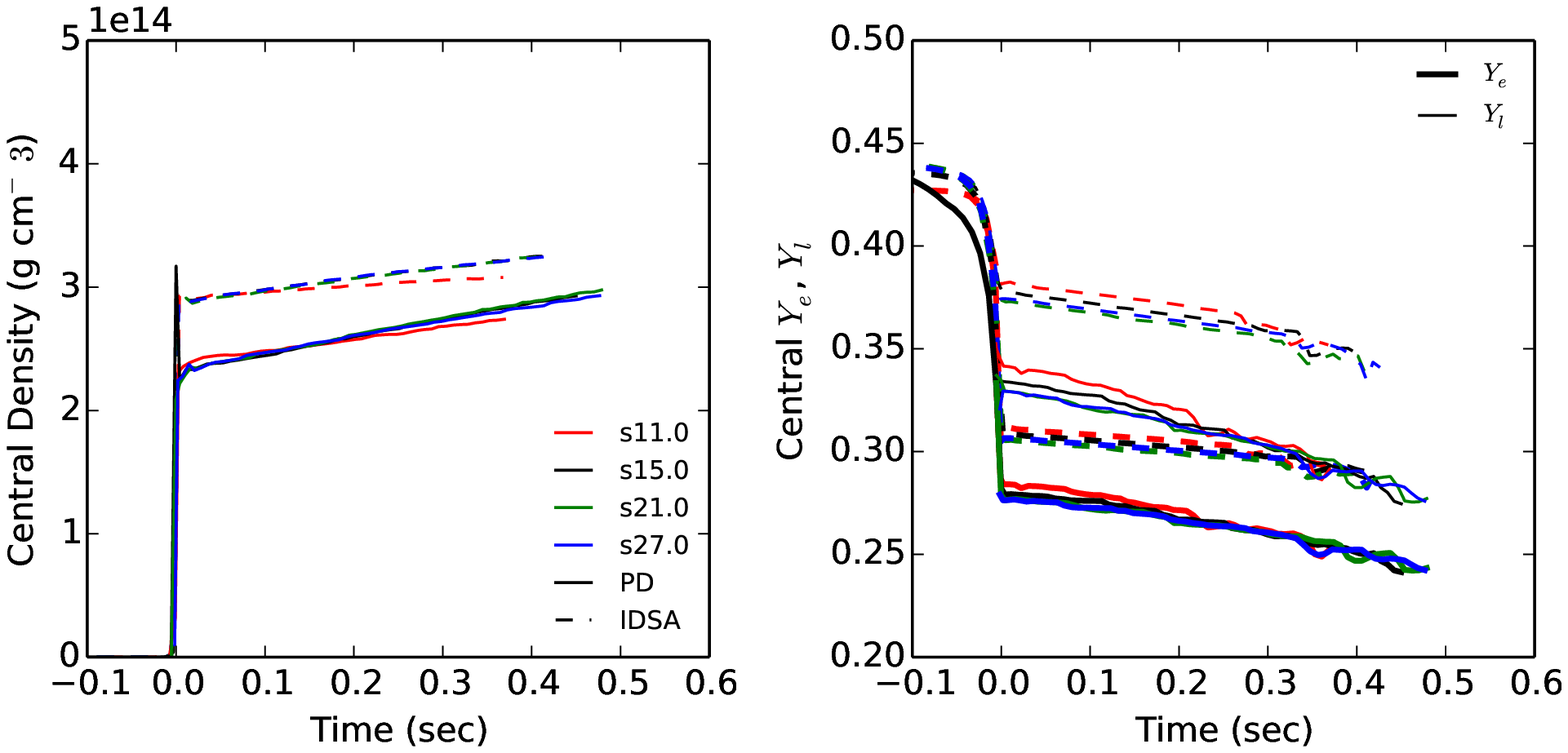}
\plotone{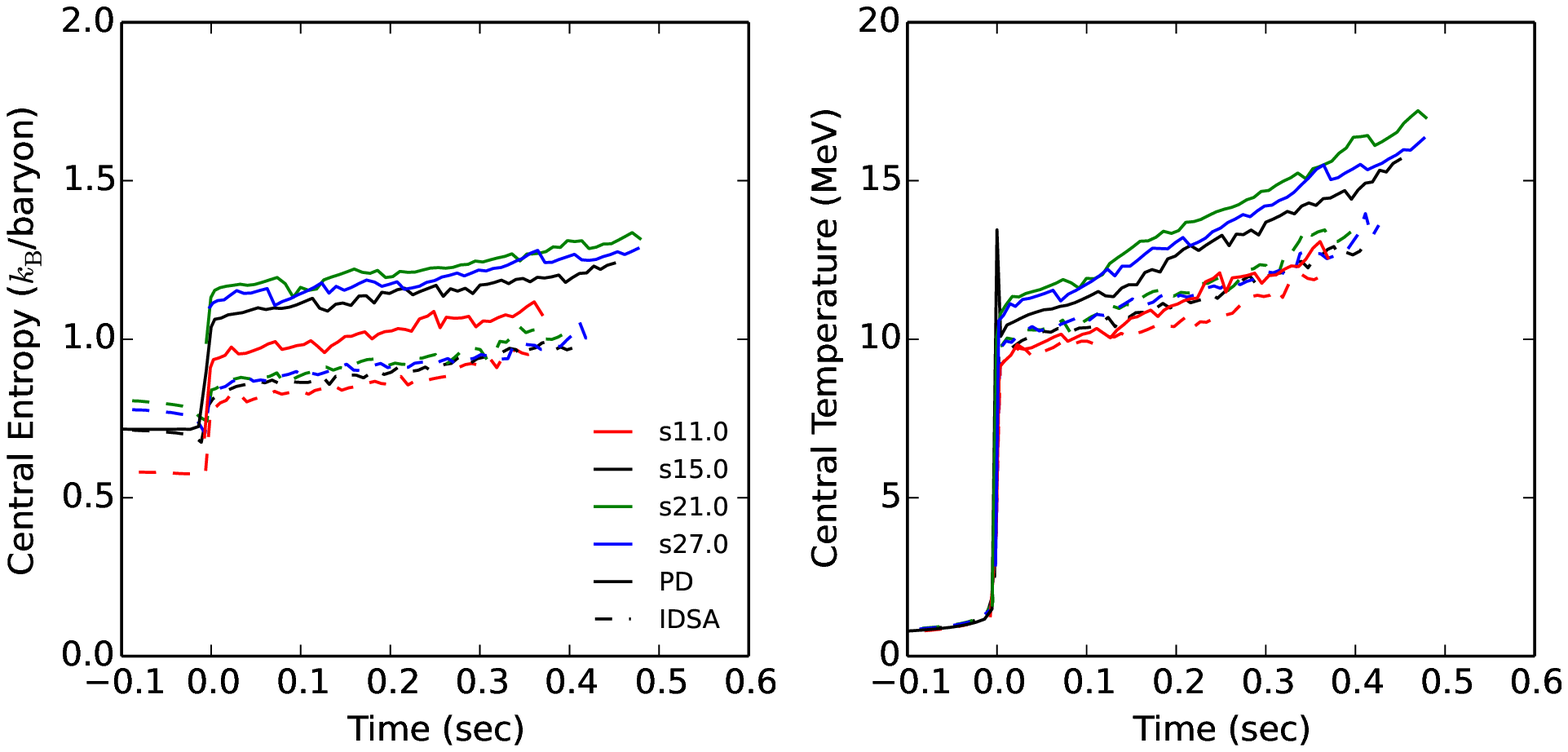}
\end{center}
\caption{\label{fig_2d_central_values}
The time evolution of central density ($\rho_c$, upper left), central electron fraction ($Y_e$) and lepton fraction ($Y_l$, upper right), central entropy (lower left), and central temperature (lower right).
Different colors represent different progenitor stars. 
Solid lines show the models 2D-DP and dashed lines show the models 2D-DA in Table~\ref{tab_simulations}.
In the upper right panel, the thin lines represent the central lepton fraction 
and the thick lines represent the central electron fraction. }
\end{figure}

\subsection{Proto Neutron Stars}

During its evolution, the PNS shrinks due to the neutrino cooling behind the gain radius. 
In Figure~\ref{fig_2d_pns}, we show the mass and radius evolutions of the PNSs after bounce. 
An initial oscillation of the PNS is taking place at $\sim 10$~ms, when the shock breaks through the neutrino sphere.
Later on, mass continuously accretes onto the PNS for several hundred milliseconds so that it
reaches $\sim 1.3 - 2.0 M_\odot$ at the end of the simulations.  
At that time, the PNS radii have shrunken to $\sim 40$~km. 
The mass and radii of the PNSs at the end of the simulations are summarized in Table~\ref{tab_simulations}.

The convection and SASI activities help the PNS to accrete mass during the setup of an explosion. 
Accretion funnels are created along the high entropy convective plumes.
This can be seen in Figure~\ref{fig_2d_entropy_slices}. 
Therefore, the progenitor $s11.0$ which has the less SASI activities and accretion funnels, 
shows the lowest PNS mass increment
and the growth rate of the PNS mass is the lowest as well. 
Furthermore, the progenitor $s11.0$ has the lowest density distribution outside the iron core, 
which gives the lowest mass accretion rate.   
After $\sim 400$~ms, the mass increment of the PNS becomes small 
but the mass of the PNS is still increasing at the end of the simulations.  
We find that although the growth rate of the PNS mass in models 2D-DA and 2D-DP are different, 
the final PNS mass in models 2D-DP are only slightly larger ($< 5\%$), 
and the radii of PNSs in models 2D-DA are slightly lower than that in models 2D-DP,
giving a more compact PNS core in models 2D-DA. 
This can also be related to the earlier explosion times of models DA.
 
Figure~\ref{fig_2d_central_values} shows the central density, central electron, lepton fraction, central entropy 
and central temperature as functions of time.
Models 2D-DA show higher central densities and higher central $Y_e$ (higher electron pressures),
but lower central entropies and central temperatures than the models 2D-DP,
giving a more compact PNS core.
However, the density in the outer layer of the PNS is higher and more convective in models 2D-DP than that in models 2D-DA.
Therefore, models 2D-DP have a higher PNS mass and radius (see Figure~\ref{fig_2d_pns}).

\section{CONCLUSIONS} \label{sec_conclusions}

We have performed self-consistent CCSN simulations for the four nonrotating, solar-metallicity progenitors, 
s11.0, s15.0, s21.0, and s27.0 from \cite{2002RvMP...74.1015W} and s15 and s20 from \cite{2007PhR...442..269W}
using the AMR code {\tt FLASH} with an IDSA for neutrino transport. 
A very good agreement of {\tt FLASH-IDSA} with {\tt AGILE-IDSA} is shown in Section~\ref{sec_cc_idsa}, 
though some small differences still exist. 
In addition, we have provided a comparison of our multi-dimensional IDSA results for the s15 and s20 progenitors 
from \cite{2007PhR...442..269W} with results in the existing literature.
However, a detailed comparison with an exchange of data and collaboration among groups would be necessary 
to understand remaining different results with regard to the underlying physics.

The new SN EOS HS (DD2) is employed in this study and a comparison 
with the standard LS220 EOS is discussed in Appendix~A.
It is found that the DD2 EOS shows a faster shock expansion, and a  
higher mass in the gain region, while the neutrino luminosities are similar,
causing the DD2 EOS to lead to slightly earlier and stronger explosions than the LS220. 

We have presented two sets of simulations which compared the neutrino transport with the IDSA 
(but ignoring NES; abbreviated DA) 
with the parametrized deleptonization scheme (which includes NES effectively; abbreviated DP) during collapse.
The results show clearly that the treatment of neutrino weak interactions 
and the level of deleptonziation during the collapse phase have a significant impact on the neutrino-driven explosions.  
All our 2D models explode within $\sim 100-300$~ms. 
Models without NES explode much easier, stronger and faster than models with NES, 
The diagnostic explosion energy in our simulations are around $0.1 - 0.5 B$ 
in models 2D-DP (see Table~\ref{tab_simulations}) 
and around $0.2 - 0.8$~B in models 2D-DA.  
Our explosion energies are likely to decrease when we include general relativistic effects 
and better electron-capture rates in future models. 

In our simulations, we do not use the typical RbR approach for the neutrino transport. 
Instead, we solve the diffusion part in multi-dimensions, 
improving the neutrino transport in angular and temporal directions.
With this approach, the prompt convection after the core bounce causes a fast shock expansion ($t_{\rm pb}\sim 50$~ms),
enlarging the gain region at late time, and therefore helps to increase the explosion energy at late time. 
However, this prompt convection may be amplified by grid-effects in cylindrical coordinates. 
To address this issue, a full 3D simulation will be necessary 
or a detailed comparison with simulations in spherical coordinates. 
Furthermore, we have found SASI activities at late time in progenitors $s15.0$, $s21.0$, and $s27.0$, 
but in most cases, the explosions are mainly due to neutrino-driven convection. 
Future work will relax the constraints imposed by axisymmetry.  
Models in 3D will permit to study a more realistic turbulence cascade and rotation.


\acknowledgments
We thank the anonymous referee for his/her valuable comments and suggestions.
This work was supported by the European Research Council (ERC; FP7) 
under ERC Advanced Grant Agreement N$^\circ$~321263~-~FISH, 
by the PASC High Performance Computing Grant DIAPHANE, 
and by the Swiss National Science Foundation (SNF).
The Basel group is member of the COST Action New Compstar.
Part of this work was inspired during the MICRA2013 meeting in ECT*.
We thank Albino Perego for the development of the leakage scheme for heavy neutrinos.
We thank Takami Kuroda, Rub\'{e}n Cabez\'{o}n, Kei Kotake, Tomoya Takiwaki, Ko Nakamura, 
and Yudai Suwa for useful discussions. 
K.C.P.~acknowledges Sean Couch for releasing his supernova setup in the public version of {\tt FLASH}. 
{\tt FLASH} was developed largely by the DOE-supported ASC/Alliances Center for Astrophysical  
Thermonuclear Flashes at the University of Chicago. 
The simulations have been carried out at the CSCS Monte-Rosa under grant No.~412.
Analysis and visualization of simulation data were completed using the analysis toolkit {\tt yt}
\citep{2011ApJS..192....9T}.


%

\section*{APPENDIX A\\ A COMPARSION BETWEEN DD2 AND LS220}

%
\begin{figure}
\begin{center}
\epsscale{1.2}
\plotone{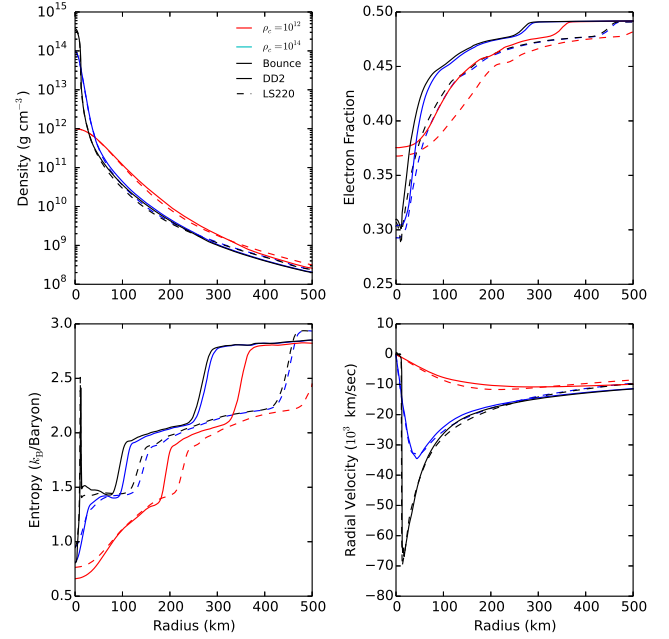}
\end{center}
\caption{\label{fig_eos_collapse} 
Similar to Figure~\ref{fig_collapse_s11} but for models 2D-DA15, and 2D-LA15 in Table~\ref{tab_simulations}.}
\end{figure}

%
\begin{figure}
\begin{center}
\epsscale{1.2}
\plotone{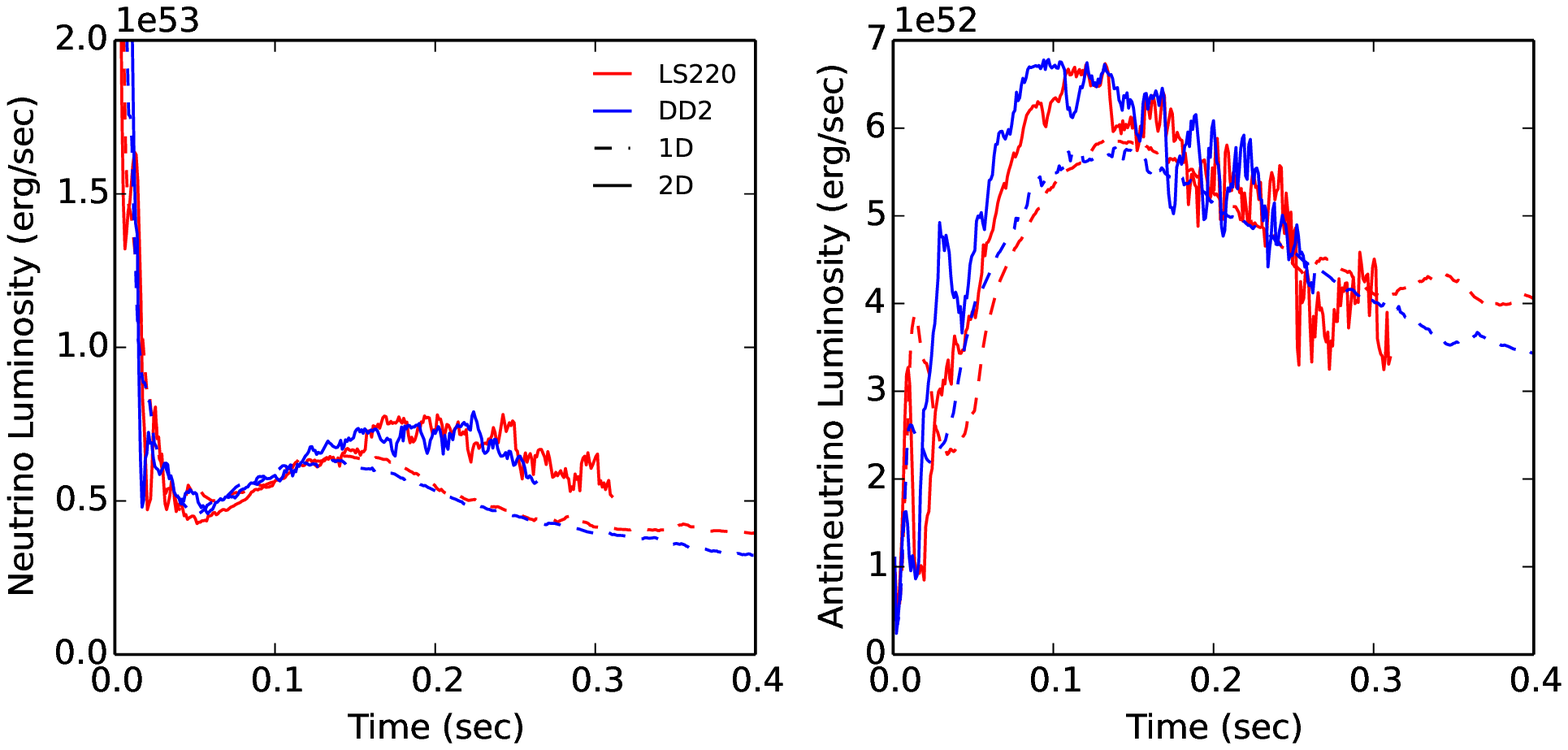}
\plotone{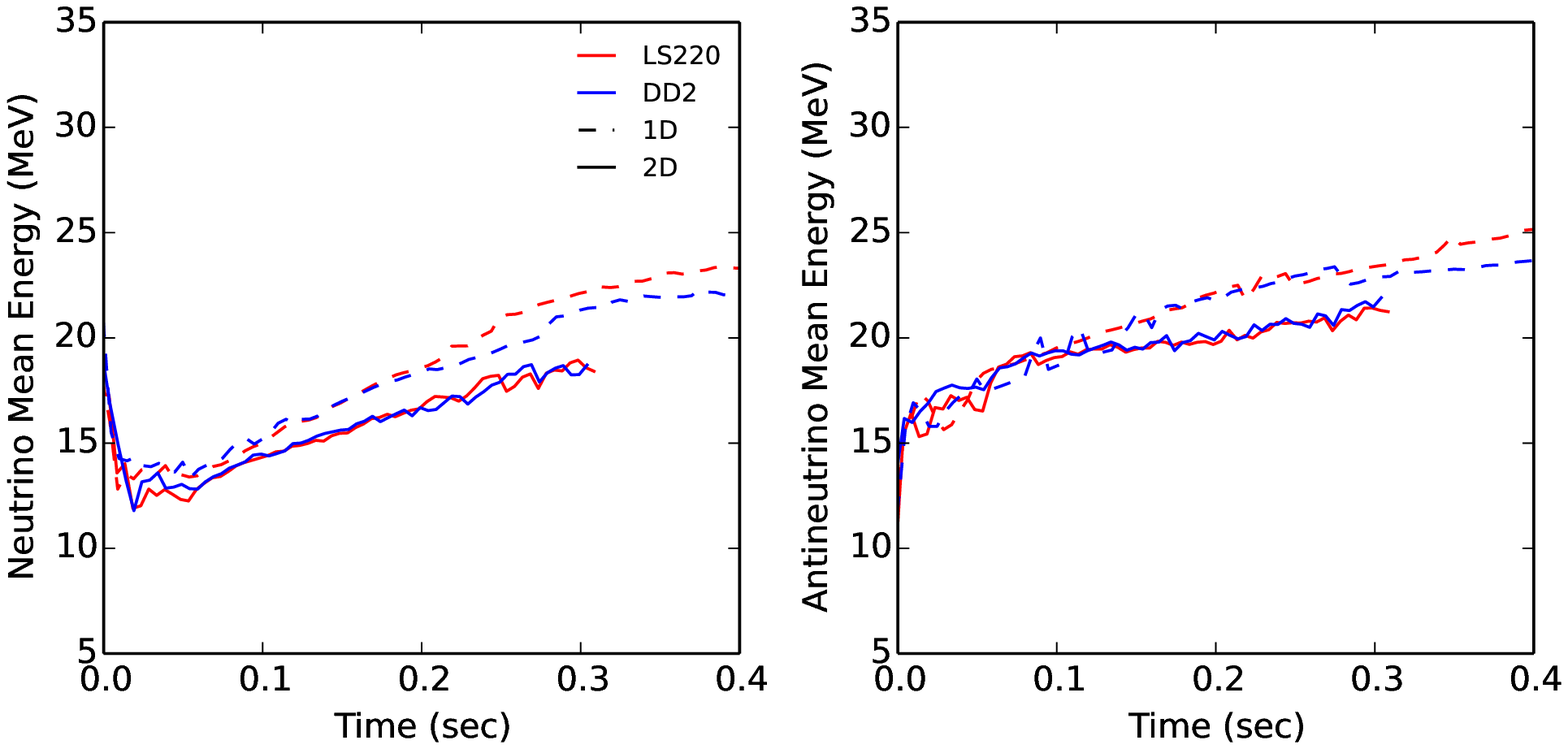}
\end{center}
\caption{\label{fig_eos_luminosity} 
Similar to Figure~\ref{fig_2d_luminosity} but for models 1D-DA15, 1D-LA15, 2D-DA15, and 2D-LA15 in Table~\ref{tab_simulations}.}
\end{figure} 

Several simulations in this paper use the new DD2 EOS,
while most multi-dimensional simulations found in the literatures use LS220. 
We therefore briefly describe the qualitative differences between the simulations with the DD2 
and the LS220 EOS in this section. 
We perform 1D and 2D simulations of the progenitor $s15.0$ with DD2 and LS220 EOS 
(models 1D-DA15, 2D-DA15, 1D-LA15 and 2D-LA15 in Table~\ref{tab_simulations}).
Hereafter, we use DA15 to refer to both models 1D-DA15 and 2D-DA15, and similarly 
we use LA15 to refer to the models 1D-LA15 and 2D-LA15. 
Their main features are summarized in Table~\ref{tab_simulations}). 
The PD scheme is turned off in order to study the impact of the EOS on the neutrino transport. 

%
\begin{figure}
\begin{center}
\epsscale{1.2}
\plotone{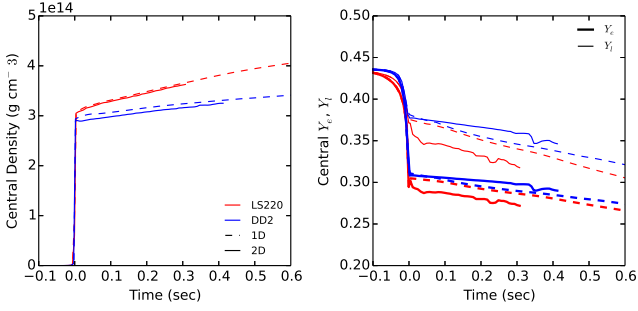}
\end{center}
\caption{\label{fig_eos_central_values} 
Similar to Figure~\ref{fig_2d_central_values} but for models 1D-DA15, 1D-LA15, 2D-DA15, and 2D-LA15 in Table~\ref{tab_simulations}.}
\end{figure}

The angle-averaged radial profiles of density, electron fraction, entropy, and radial velocity are shown in 
Figure~\ref{fig_eos_collapse} for the prebounce phase. 
As discussed in Section~\ref{sec_results}, the radial profiles in the collapse phase are nearly identical in both 1D and 2D. 
In addition, the density and radial velocity evolutions in models DA15 and LA15 are also similar during the collapse 
but models DA15 collapse slower than models LA15.  
Therefore, models DA15 reach core bounce about $\sim 30$~ms later than models LA15.  

%
\begin{figure}
\begin{center}
\epsscale{1.2}
\plotone{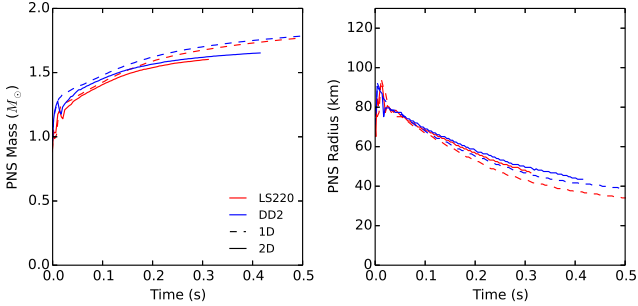}
\end{center}
\caption{\label{fig_eos_pns} 
Similar to Figure~\ref{fig_2d_pns} but for models 2D-DA15, and 2D-LA15 in Table~\ref{tab_simulations}.}
\end{figure}

While there are no significant differences of the neutrino luminosity 
and mean energy between models DA15 and LA15 (Figure~\ref{fig_eos_luminosity}),
models DA15 show a lower central density and higher PNS radius than LA15. 
The time evolutions of central density 
and electron fraction of models DA15 and LA15 are shown in Figure~\ref{fig_eos_central_values}.
Furthermore, models DA15 have a higher central electron and lepton fraction than models DL15. 
2D models also show a lower PNS mass and a higher PNS radius than 1D (Figure~\ref{fig_eos_pns}), 
but the PNS masses are similar in both DD2 and LS220 (see Table~\ref{tab_simulations}). 

%
\begin{figure}
\begin{center}
\epsscale{1.2}
\plotone{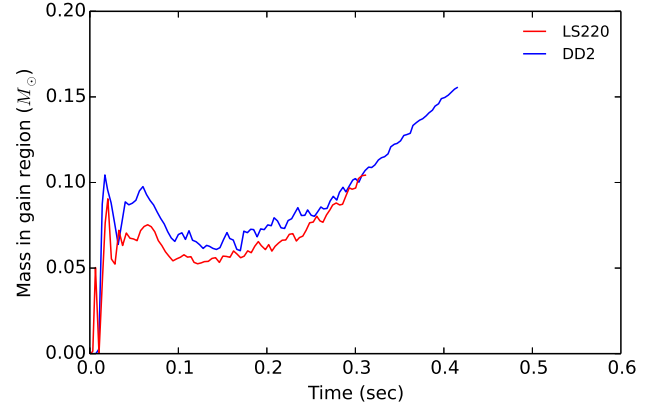}
\end{center}
\caption{\label{fig_eos_gain_mass} 
Similar to Figure~\ref{fig_2d_gain_mass} but for models 2D-DA15, and 2D-LA15 in Table~\ref{tab_simulations}.}
\end{figure} 

%
\begin{figure}
\begin{center}
\epsscale{1.2}
\plotone{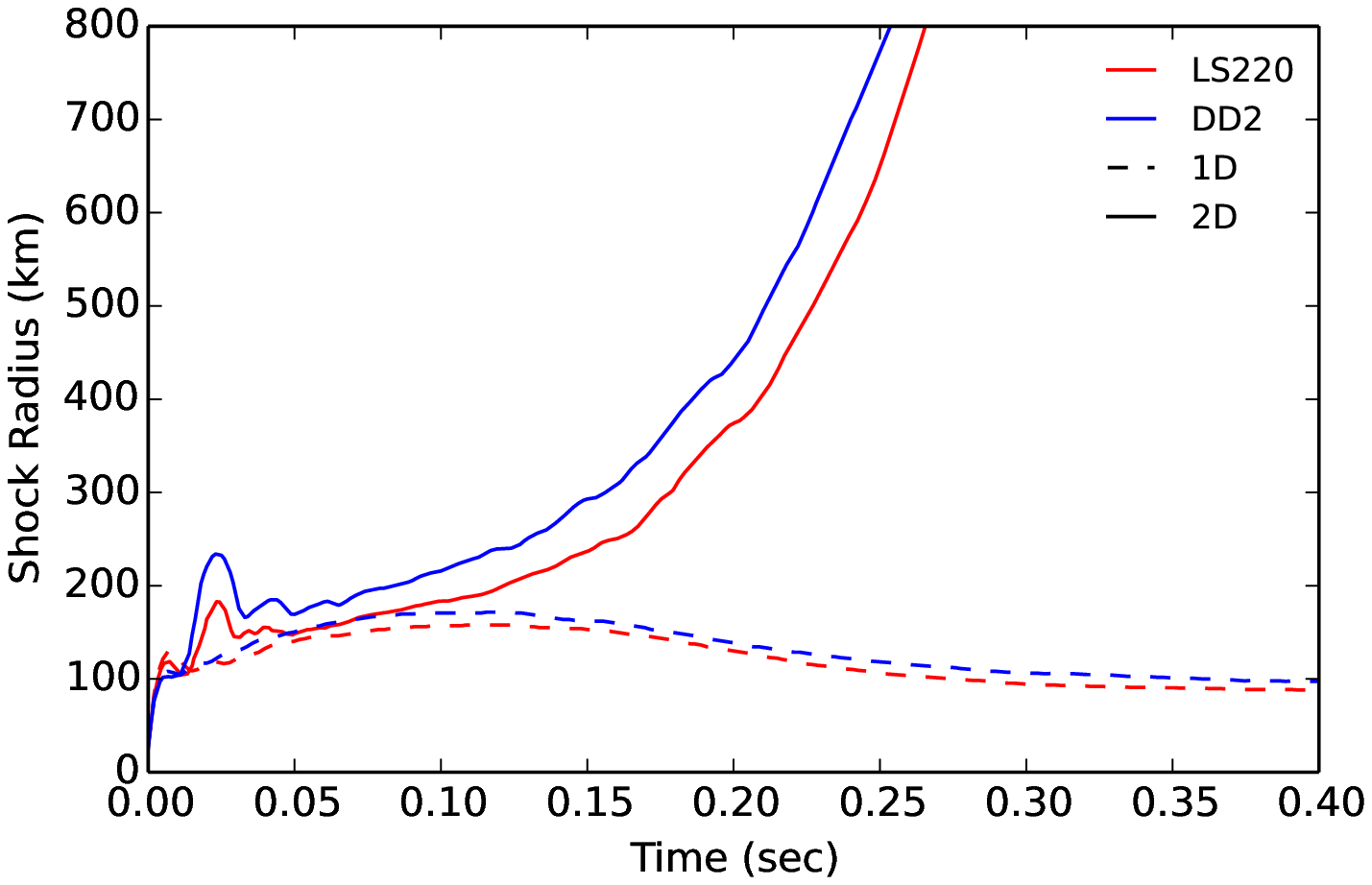}
\end{center}
\caption{\label{fig_eos_shock} 
Similar to Figure~\ref{fig_2d_shock_radius} but for models 1D-DA15, 1D-LA15, 2D-DA15, and 2D-LA15 in Table~\ref{tab_simulations}. }
\end{figure}

%
\begin{figure}
\begin{center}
\epsscale{1.2}
\plotone{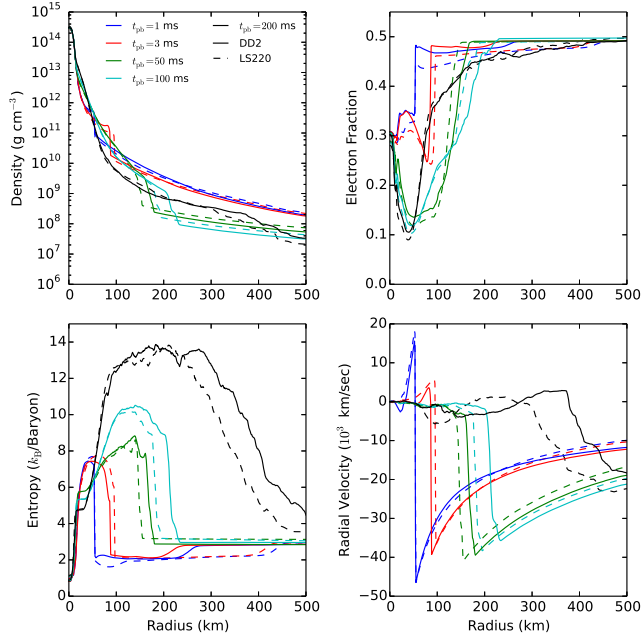}
\end{center}
\caption{\label{fig_eos_postbounce} 
Similar to Figure~\ref{fig_postbounce_s11} but for models 1D-DA15, 1D-LA15, 2D-DA15, and 2D-LA15 in Table~\ref{tab_simulations}.}
\end{figure}

Since the bounce time in models  DA15 take longer than models LA15, 
the Si/O interface reaches a smaller radius in models DA15 ($\sim 90$~km) 
than in LA15 ($\sim 120$~km, see Figure~\ref{fig_eos_collapse}).
The bounce shock hits the Si/O interface earlier in models DA15 than in LA15, 
generating a prompt convection and fast shock expansion at $\sim 20$~ms postbounce. 
The earlier interaction between the bounce shock with the Si/O interface in model 2D-DA15 
makes the shock radius in model 2D-DA15 larger than in simulation 2D-LA15. 
Furthermore, the larger shock radius in model 2D-DA15 produces a higher mass in the gain region (see Figure~\ref{fig_eos_gain_mass}),
suggesting that the models with the DD2 EOS explodes more easily than the models with the LS220 EOS.
The averaged shock radius evolution of models DA15 and LA15 can be seen in Figure~\ref{fig_eos_shock}.  
The explosion time ($t_{400}$) in 2D-DA15 is about $23$~ms earlier than in 2D-LA15. 
In Figure~\ref{fig_eos_postbounce}, we compare the radial density, electron fraction, entropy, and velocity profiles 
in 1D and 2D for DD2 and LS220 at 1, 3, 50, 100, and 200~ms postbounce. 
Before $\sim 20$~ms postbounce, models LS15 show a faster shock expansion than models DA15 
but the shock radius in DA15 become larger after the Si/O interface has reached the shock at $\sim 20$~ms. 

%
\begin{figure}
\begin{center}
\epsscale{1.2}
\plotone{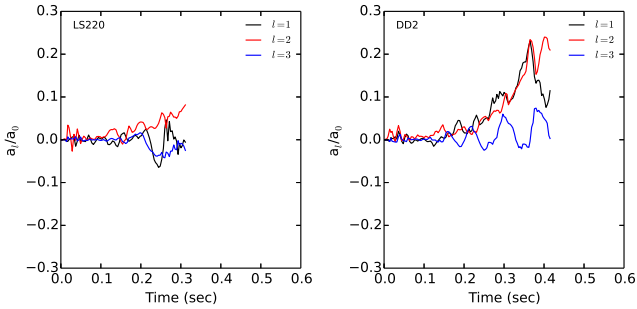}
\end{center}
\caption{\label{fig_eos_sasi} 
Similar to Figure~\ref{fig_2d_sasi} but for models 2D-LA15 (Left) and 2D-LD15 (Right) in Table~\ref{tab_simulations}. }
\end{figure}

%
\begin{figure*}
\begin{center}
\epsscale{1.2}
\plotone{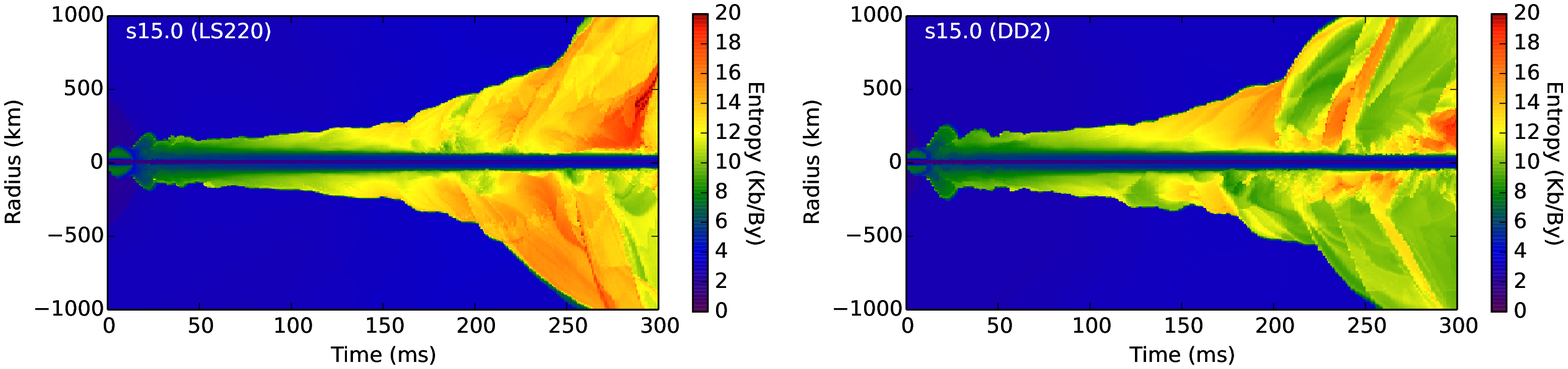}
\end{center}
\caption{\label{fig_eos_entr_timeevo} 
Similar to Figure~\ref{fig_2d_entr_timeevo} but for models 2D-LA15 (Left) and 2D-DA15 (Right) in Table~\ref{tab_simulations}.}
\end{figure*}

Figure~\ref{fig_eos_sasi} shows the normalized decomposition into Legendre polynomials 
(Equation~\ref{eq_sasi}) of the shock radius of models 2D-DA15 and 2D-LA15. 
The corresponding entropy distribution at the north and south pole is shown in Figure~\ref{fig_eos_entr_timeevo}.
The SASI activities start at $\sim 0.2$~s in both 2D-DA15 and 2D-LA15 after the Si/O interface reach the shock 
and the amplitude of the normalized coefficients $a_l/a_0$ do not show significant difference.  
We remark that a more through investigation of the EOS effects will be the subject of a future study.




\end{document}